\documentclass[fleqn,usenatbib,useAMS]{mnras}

\usepackage{newtxtext,newtxmath}

\usepackage[T1]{fontenc}
\usepackage{ae,aecompl}

\usepackage{graphicx}	
\usepackage{amsmath}	
\usepackage{amssymb}	
\usepackage{amsfonts}
\usepackage{gensymb}
\usepackage{soul}

\usepackage{float}
\usepackage{caption}
\usepackage{subcaption}

\def\ibid{{\it ibid.}}
\def\egg{{\it e.g.}}

\title[Dust Fans and Infrared Variability]{Infrared Variability due to Magnetic Pressure Driven Jets, Dust Ejection and Quasi-Puffed-Up Inner Rims 
}

\author[Liffman, Bryan, Hutchison, Maddison]{
Kurt Liffman$^1$,\thanks{E-mail: kliffman@swin.edu.au} 
Geoffrey Bryan$^1$, 
Mark Hutchison$^2$, 
Sarah T. Maddison$^1$
\\
$^{1}$Centre for Astrophysics and Supercomputing, Swinburne University of Technology, Victoria, AUSTRALIA\\
$^{2}$Physikalisches Institut, Universit\"at Bern, Switzerland; Institute for Computational Science, University of Zurich, Switzerland}

\date{Accepted XXX. Received YYY; in original form ZZZ}

\pubyear{2019}

\begin{document}
\label{firstpage}
\pagerange{\pageref{firstpage}--\pageref{lastpage}}
\maketitle
%
%
\begin{abstract}
The interaction between a YSO stellar magnetic field and its protostellar disc can result in stellar accretional flows and outflows from the inner disc rim.  Gas flows with a velocity component perpendicular to disc midplane subject particles to centrifugal acceleration away from the protostar, resulting in particles being catapulted across the face of the disc. The ejected material can produce a ``dust fan", which may be dense enough to mimic the appearance of a ``puffed-up" inner disc rim. 
We derive analytic equations for the time dependent disc toroidal field, the disc magnetic twist, the size of the stable toroidal disc region, the jet speed and the disc region of maximal jet flow speed. We show how the observed infrared variability of the pre-transition disc system LRLL~31 can be modelled by a dust ejecta fan from the inner-most regions of the disc whose height is partially dependent on the jet flow speed. The greater the jet flow speed, the higher is the potential dust fan scale height. 
An increase in mass accretion onto the star tends to increase the height and optical depth of the dust ejection fan, increasing the amount of 1--8~$\mu$m  radiation. The subsequent shadow reduces the amount of light falling on the outer disc and decreases the 8-- 40~$\mu$m radiation. A decrease in the accretion rate reverses this scenario, thereby producing the observed ``see-saw'' infrared variability.

\end{abstract}

\begin{keywords}
accretion discs -- protoplanetary discs -- magnetohydrodynamics -- radiative transfer -- stars: jets -- stars: variables: T Tauri, Herbig Ae/Be
\end{keywords}



\section{Introduction}

Protostars undergo a well-documented, although not fully understood, evolution from a collapsing cloud of gas and dust through to planets orbiting an evolving star \citep{2011ARA&A..49...67W}. The intervening stages are associated with the formation of a disc of gas and dust surrounding the protostar, where the disc may evolve from a continuous disc to one with an inner hole or an optically thin gap between optically thick inner and outer discs. These latter discs are known as pre-transition and transition discs respectively as they are in a transition phase between a continuous disc to a debris disc from which most of the gas has been removed \citep{2014prpl.conf..497E}.

\begin{figure}
	\includegraphics[width=\columnwidth]{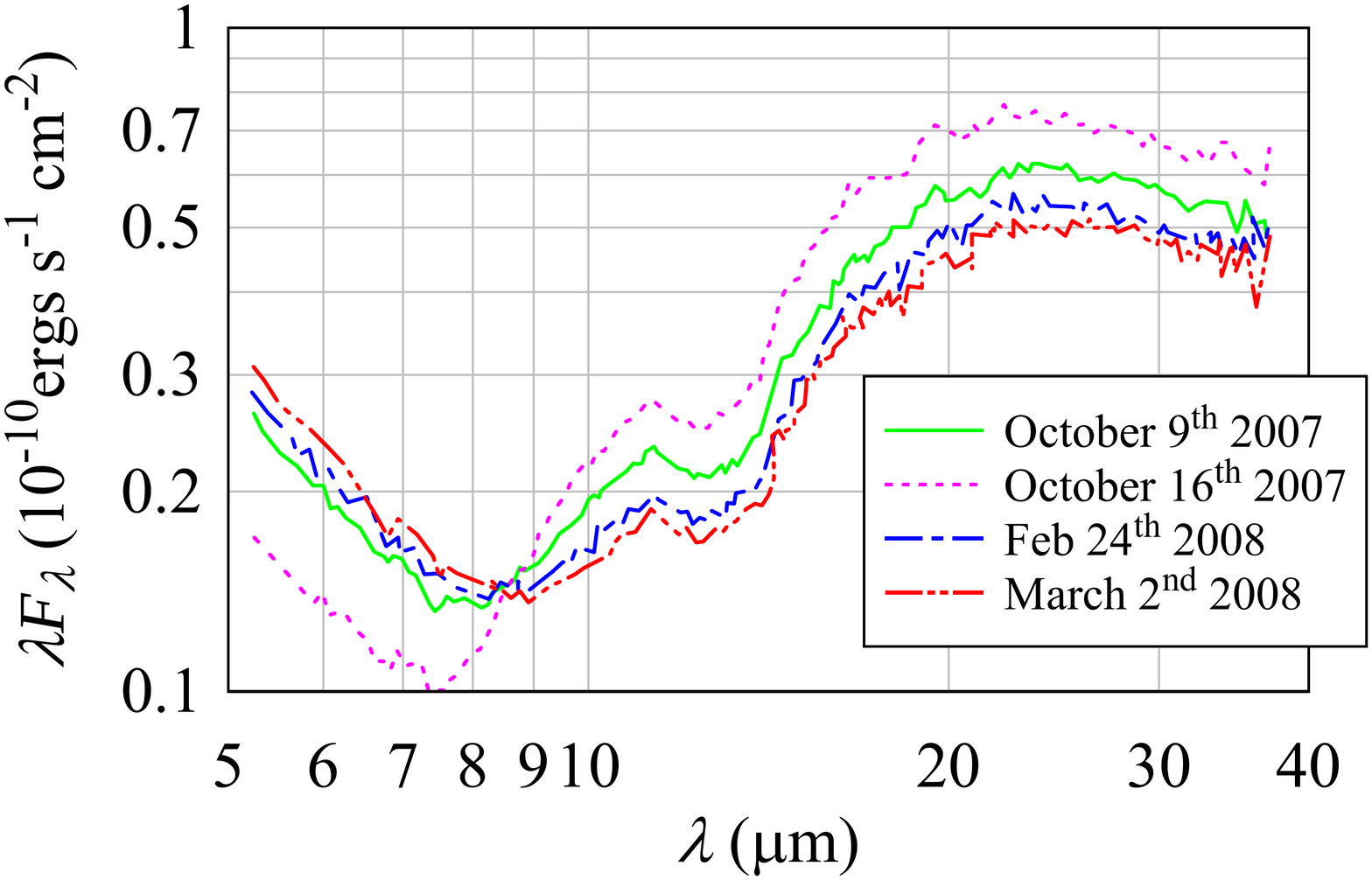}
    \caption{The temporal changes in infrared flux between 5 and 40~$\mu$m for LRLL 31. As the flux between 5 and 8.5~$\mu$m increases the flux between 8.5 and 40~$\mu$m decreases \citep{2009ApJ...704L..15M,2011ApJ...732...83F}.}
    \label{fig:figure1}
\end{figure}

The pre-transition disc LRLL 31 is located in the 2-3 Myr old star forming region IC 348 some 315~pc from the Earth. Current observations suggest that LRLL 31 is a G6 star with a rotation period of 3.4 days \citep{2011ApJ...732...83F}, a luminosity of 5.0~L$_{\odot}$, a radius of about 2.3~R$_{\odot}$, a mass of approximately 1.6~M$_{\odot}$, and an effective temperature of 5700~K \citep{2014A&A...564A..51P}. In addition, observations with the Spitzer Space Telescope show a ``see saw" oscillation in the system's infrared spectrum \citep{2009ApJ...704L..15M}. That is, when the flux from the 5 to 8.5~$\mu$m range increases then the flux from 8.5 to 40~$\mu$m decreases on timescales of weeks (Figure~\ref{fig:figure1}).  

\begin{figure}
	\includegraphics[width=\columnwidth]{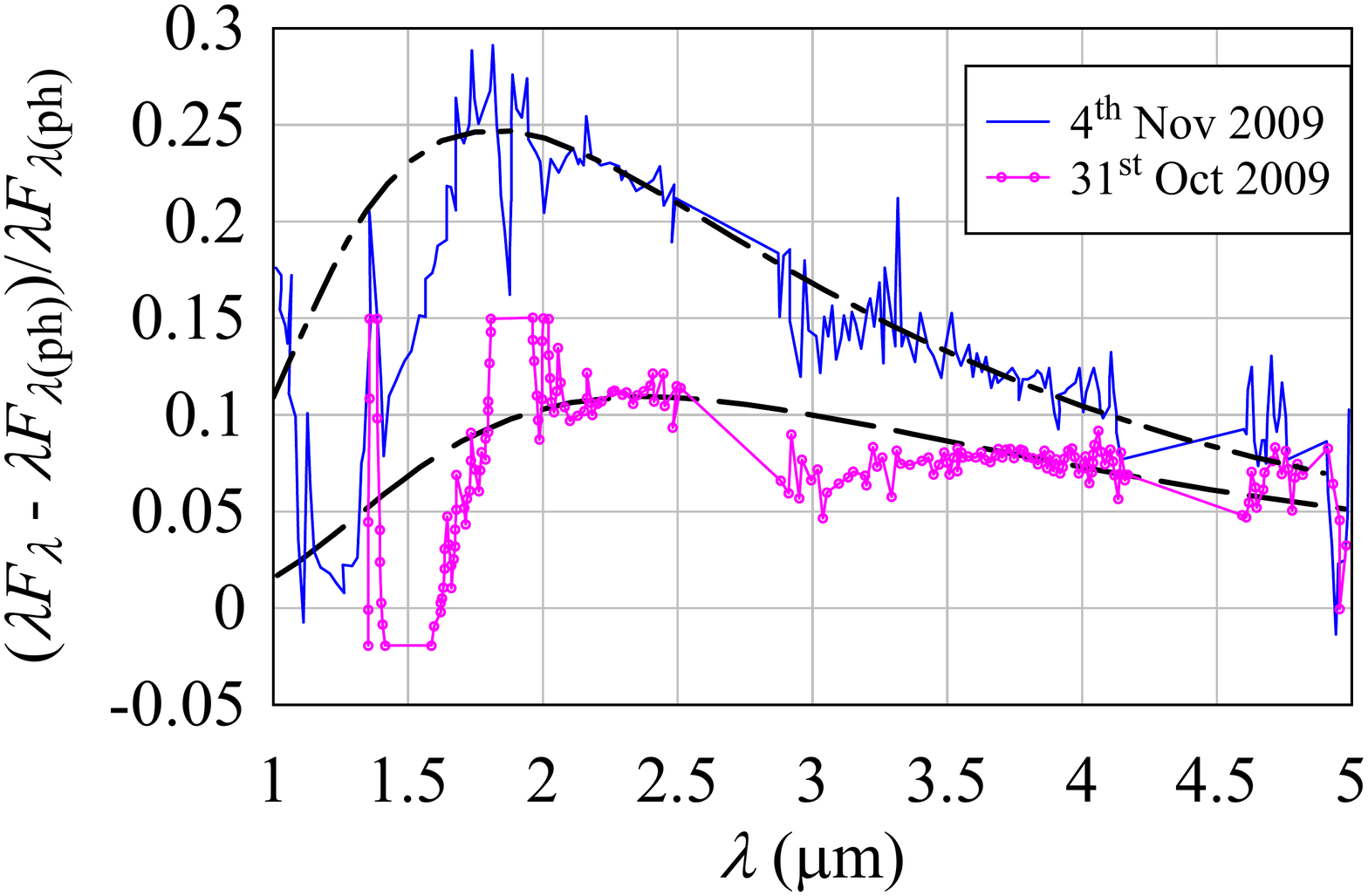}
    \caption{Infrared excesses obtained from the inner disc of LRLL 31 at two different dates. The plot shows the difference between LRLL 31 and the stellar photosphere, normalized to the photospheric flux at 2.15~$\mu$m. A blackbody best fit line is shown for each set of data \citep{2011ApJ...732...83F}.}
    \label{fig:figure2}
\end{figure}

At wavelengths between 1 and 5~$\mu$m, assumed to result from dust emission in the inner disc, a significant change in the LRLL 31 infrared excess can occur daily. As an example, Figure~\ref{fig:figure2} shows the approximate factor of two increase in infrared excess from the LRLL 31 inner disc wall between the 31st of October and the 4th of November 2009. 

The flux from the inner wall of a disc surrounding a star, $F_{\lambda \rm{rim}}$, is approximately given by 
\begin{equation}
    F_{\lambda \rm{rim}} \approx \frac{4\pi R_{\rm rim} H_{\rm rim} B_{\lambda}(T_{\rm rim})}{d^2} \sin(i) \, ,
	\label{eq:equation1}
\end{equation}
where $d$ is the distance between the disc and the observer, $R_{\rm rim}$ the distance from the inner disc wall to the centre of the star, $H_{\rm rim}$ the height of the inner rim wall measured from the disc midplane to the top of the wall,  $B_{\lambda}(T_{\rm rim})$ the blackbody radiation from the inner disc wall, which is at a temperature of $T_{\rm rim}$, and $i$ is the disc inclination angle. In the case of LRLL 31, the disc is thought to be nearly edge on to the observer with $i \sim 90^{\circ}$ \citep{2010ApJ...719.1733F}.

According to \citet{2011ApJ...732...83F}, analysis of the infrared excess indicates that $T_{\rm rim}$ and hence $R_{\rm rim}$ probably remained approximately constant. Thus, to explain the factor of two increase in the infrared excess, equation~\ref{eq:equation1} implies that $H_{\rm rim}$ increased by around a factor of two during the four days separating the 31st of October and the 4th of November 2009.  So again from \citet{2011ApJ...732...83F}, observations would imply that the covering fraction of the inner disc relative to the central star increased by approximately a factor of five over the course of a month: from $\sim 0.01$ (8th October 2009) to $\sim 0.054$ (8th November 2009), where a possible description of the covering fraction, $f_{\rm cover}$, is given by the approximate formula:
\begin{equation}
    f_{\rm cover} \approx \frac{2}{\pi} \tan^{-1}\left(\frac{H_{\rm rim}}{R_{\rm rim}}\right) \, .
	\label{eq:equation2}
\end{equation}
Thus, in that one month, the value of $H_{\rm rim}$ has possibly increased by about a factor of four. At the same time, the deduced mass accretion rate from the disc onto the star also increased by around a factor of four from $\sim 0.4 \times 10^{-8} {\rm M}_{\odot} {\rm yr}^{-1}$ (8th October 2009) to $\sim 1.6 \times 10^{-8} {\rm M}_{\odot} {\rm yr}^{-1}$  (8th November 2009) (\ibid).

Other authors have modelled transition disc systems and concluded that parts of the infrared variability might be explained by variation in the height of the inner disc rim, e.g., \citet{2007MNRAS.374.1242J} and \citet{2008ApJ...678.1070S}. \citet{2008ApJ...678.1070S} also examined disc winds as a possible explanation. 

As the inner 0.1 au inner regions of protostellar discs tend to be too small to be resolved with current capabilities (\egg, if 5 mas resolution interferometers such as the Atacama Large Millimetre Array could be used on LLRL 31 then the effective resolution would be $\sim$ 1.5 au). Consequently, only indirect data and modelling are available to understand the major physical mechanism that is producing the observed infrared variability in LRLL 31. At least nine separate models have been suggested to explain the deduced variation in scale height of the inner disc. These models range from higher accretion rates that increase the scale height of the inner disc, through to asymmetric, dynamic, warped inner discs, hidden planetary companions, inner disc radial fluctuations and magnetic field effects \citep{2011ApJ...732...83F}. 

While it is possible that some or all of these explanations may be applicable to particular young stellar systems, in this paper we will attempt to explain the observations as a by-product of the interaction of the stellar magnetic field with the inner disc. Other authors, e.g., \citet{2010ApJ...708..188T}, have considered magnetic fields within the disc associated with disc turbulence and the magnetorotational instability. However, they ignored their own jet flow results, so their subsequent deduced variations in the scale height are too small to account for the potential factor of four changes in scale height for LRLL 31. 

It is possible that accretional flows via the stellar magnetosphere are sufficiently opaque to produce shadowing on the outer disc. For example, \citet{2008MNRAS.386..673K} show very interesting numerical examples of such flows, but whether they can produce the observed infrared variability is uncertain as such flows tend to vary on a shorter timescale relative to the observed infrared variability. 

\citet{2008ApJ...683..949L} modelled the effect of a tilted, rotating stellar magnetic field on the inner region of the disc. They find that waves are produced in the disc, which produces semi-periodic changes in the disc height. They note that such a model may be applicable to neutron star systems, but, to date, infrared variability in LRLL 31 does not seem to be periodic. This may change with more observational data, but at this stage the Lai \& Zhang model does not appear to be an applicable mechanism. 

The interaction between a stellar magnetosphere and a surrounding accretion disc (Figure~\ref{fig:figure5}) produces a significant disc toroidal magnetic field  \citep{2005MNRAS.356..167M,2010ApJ...714..989M} on a probable timescale of hours to days (equation~\ref{eqn:15} and Appendix \ref{Toroidal_Field_Growth}).The toroidal field produces a magnetic pressure (Figure~\ref{fig:figure10}  and Appendix~\ref{MPD_Flow}) which may move material away from the disc surface to accrete onto the star or be ejected as an outflow \citep{2013A&A...550A..99Z} . Under certain circumstances \citep{2009MNRAS.399.1802R,2018NewA...62...94R}, the outflow is produced within a small region at the inner edge of the accretion disc (equation~\ref{eqn:25}), a result that is consistent with observations \citep{2018NatCo...9.4636L}.

We derive an analytic formula for the jet flow speed near the surface of the inner disc (equation~\ref{eqn:35}). The jet flow speed tends to increase with decreasing distance to the star with the exception of the region near the co-rotation radius where the jet flow turns off (Figure~\ref{fig:figure11}). 

Numerical simulations for protostellar systems \citep{2013A&A...550A..99Z,2018NewA...62...94R} and collapsing cloud cores \citep{2012MNRAS.423L..45P} show that the resulting magnetohydrodynamic (MHD) jet flows move at an angle relative to the disc midplane. In this study we are interested in the motion of particles that are launched by the jet fows, so for the purposes of establishing a base case scenario, we assume that the jet flow initially moves perpendicular to the disc midplane. By computing the motion of dust grains in the flow, we find that the dust can decouple from the flow and move radially across the face of the disc (Figure~\ref{fig:figure16}). A result that is consistent with observations from the Spitzer Space Telescope \citep{2012ApJ...744..118J}. The resulting dust fan may mimic the appearance of a puffed up inner rim (Figure~\ref{fig:figure17}) and possibly account for the observed behaviour of LLRL 31 (\S \ref{sec:dust_fan_sed}).

Observations, however, also indicate that the inner edge of the protostellar accretion disc is often within the dust sublimation radius \citep{2007IAUS..243..135C}. This poses a problem for our model, since there may be no dust particles to be entrained in the flow. Again, observations strongly suggest that dust is entrained in outflows from young stars \citep{2019MNRAS.483..132P}. So some dust or macroscopic particles may still be present in the inner disc regions and/or dust condenses in the flow just like dust formation in stellar winds once the flow has moved past the dust sublimation distance \citep{1995SSRv...73..211S}. Either way, our model requires dust to be present at some point in the outflow as it moves away from the inner edge of the accretion disc.

\section{LRLL 31}

In attempting to understand the behaviour of the LRLL 31 inner disc rim, it is helpful to obtain some length scales of the inner LRLL 31 disc and star. An important disc length scale is the dust temperature radius of the inner rim, $R_{\rm d}$. This is the distance between the inner disc rim and the centre of the protostar required to obtain a dust temperature. It has the approximate formula \citep{2010ApJ...717..441E}:
\begin{equation}
   R_{\rm d} \approx \sqrt{\frac{3(L_{\star}+L_{\rm a})}{16 \pi \sigma_{\rm SB} T^4_{\rm d}}} \approx 0.13 \, {\rm au} \sqrt{\frac{(L_{\star}+L_{\rm a})/5{\rm L}_{\odot}}{(T_{\rm d} / 1500\,{\rm K})^4}} 
  \label{eq:equation3}
\end{equation} 
where $\sigma_{\rm SB}$ is the Stefan-Boltzmann constant, $T_{\rm d}$ is the temperature of the dust, $L_{\star}$ is the luminosity of the star (in this case $5 {\rm L}_{\odot}$) and $L_{\rm a}$ is the accretion luminosity given by:
\begin{align}
    L_{\rm a} &= \frac{GM_{\star}\dot{M}_{\rm a}}{R_{\star}}\left(1-\frac{R_{\star}}{2R_{\rm t}} \right)  \nonumber \\
    &\approx 0.15 {\rm L}_{\odot} \frac{ (M_{\star}/{\rm M}_{\odot}) (\dot{M}_{\rm a}/10^{-8} {\rm M_{\odot}yr^{-1}}) }{(R_{\star}/2{\rm R}_{\odot})} \left(1-\frac{R_{\star}}{2R_{\rm t}}\right)
 \label{eq:equation4}
\end{align}
From equation~\ref{eq:equation4}, for LRLL 31, $L_{\rm a} \ll L_{\star}$.

\citet{2011ApJ...732...83F} suggest an average dust temperature of $\sim 1800$~K at the inner rim of the LRLL 31 disc. This implies that $R_{\rm d}$ is approximately equal to 0.09~au. 

Another relevant length scale is the truncation radius, $R_{\rm t}$, of the inner disc, which is the distance between the star and the inner edge of the disc as a function of mass accretion rate and stellar magnetic field strength. The inner truncation radius is produced by the approximate pressure balance between the infalling accretion disc and the stellar magnetosphere \citep{1978ApJ...223L..83G}, given by:
\begin{align}
    R_{\rm t} &\approx \left(\frac{4\pi}{\mu_0} \frac{B^2_\star R^6_\star}{\dot{M}_{\rm a} \sqrt{GM_\star}} \right)^{2/7}  \nonumber\\
    &\approx 0.067\,{\rm au} \left( \frac{(B_{\star}(R_{\star})/0.1T)^2 (R_{\star}/2{\rm R}_{\odot})^6} {(\dot{M}_{\rm a}/10^{-8} {\rm \, M_\odot yr^{-1}})(M_\star/{\rm M}_\odot)^{1/2} } \right)^{2/7}
    \label{eq:equation5}
\end{align}
where $R_\star$ is the radius of the star, $\dot{M}_{\rm a}$ the mass accretion rate onto the star, $M_\star$ the mass of the star, $\mu_0$ the permeability of free space, $G$ the universal gravitational constant, and $B_\star(R_\star)$ the magnetic field strength at the surface of the star. 
 
For LRLL 31, the magnetic field strength at the surface of the star is unknown. However, the average magnetic field strength for protostellar systems is in the kilogauss range \citep{2007prpl.conf..479B}. As such, if we set $B_\star(R_\star) \approx 0.15$~T, $R_{\rm t} \approx 2.3 {\rm R}_\odot$, and $M_\star \approx 1.6 M_\odot$, which implies that a mass accretion rate of $\dot{M}_{\rm a}\sim 0.4\times10^{-8}$~M$_\odot$yr$^{-1}$ gives $R_{\rm t} \approx 0.13$~au, while $\dot{M}_{\rm a}\sim 1.6\times10^{-8}$~M$_\odot$yr$^{-1}$ gives $R_{\rm t} \approx 0.09$~au. The latter distance is also the deduced dust temperature radius for a mass accretion of $\dot{M}_{\rm a} \sim 1.6 \times 10^{-8}$~M$_\odot$yr$^{-1}$ \citep{2011ApJ...732...83F}. 

\begin{figure*}
	\includegraphics[width=0.8\textwidth]{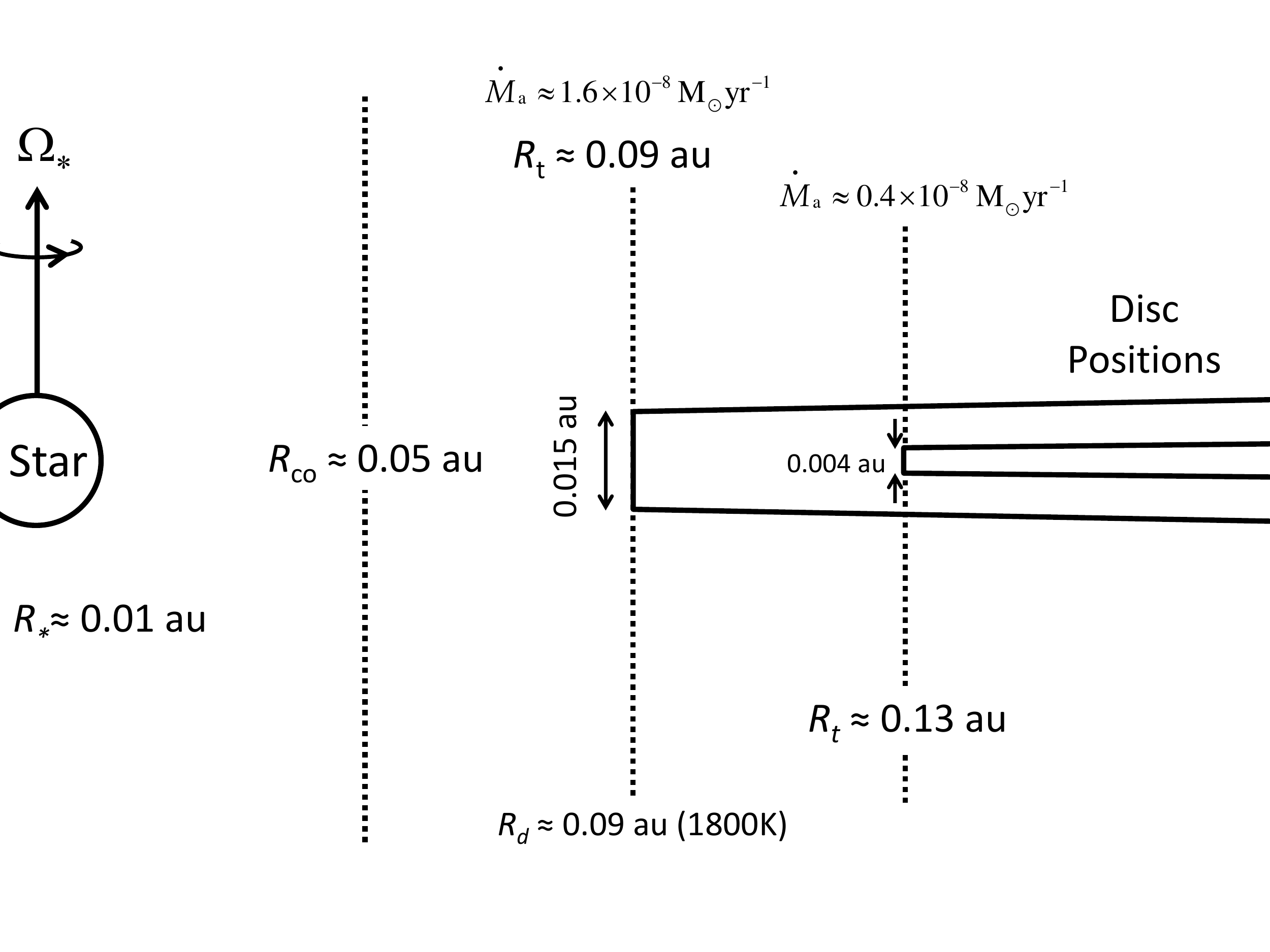}
    \caption{Scale diagram of the inner LRLL 31 system showing the disc truncation radius, $R_{\rm t}$,  the stellar radius, $R_\star$, the  co-rotation radius, $R_{\rm co}$, the dust radius for a particular dust temperature, $R_{\rm d}$, and  the angular velocity, $\Omega_\star$, of the protostar. Two truncation radii are shown at 0.9~au and 0.13~au, which correspond, respectively, to the high ($\dot{M}_{\rm a}\sim 1.6 \times 10^{-8}$~M$_\odot$ yr$^{-1}$ on Nov. 8th 2009) and low ($\dot{M}_{\rm a}\sim 0.4 \times 10^{-8}$~M$_\odot$ yr$^{-1}$  on Oct. 8th 2009) observed mass accretion rates. The very approximate scale heights of the inner rims are calculated from the deduced covering fractions. The stellar radius, disc inner rim heights and positions are approximately to scale, although the disc scale heights away from the inner disc are represented only for schematic purposes.}
    \label{fig:figure3}
\end{figure*}

The rotational angular velocity of the star, $\Omega_\star$, sets the co-rotation radius, which is the distance from the centre of the star where the Keplerian angular velocity, $\Omega_{\rm K}(r)$, equals the stellar rotational angular velocity: 
\begin{equation}
    \Omega_\star = \Omega_{\rm K}(R_{\rm co}),
    	\label{eq:equation6}
\end{equation}
 with
 \begin{equation}
     \Omega_{\rm K}(r) = \sqrt{ \frac{GM_\star}{r^3}}, 
     \label{eq:equation7}
 \end{equation}
$r$ being the cylindrical radial distance from the star. These equations imply:
\begin{equation}
    R_{\rm co} = \left(\frac{GM_\star}{\Omega_\star^2}\right)^{1/3} \approx 0.05\left( \left(\frac{M_\star}{1.6 \ {\rm M_\odot}} \right)\left(\frac{P_\star}{3.4 \ {\rm days}}\right)^2 \right)^{1/3} \, {\rm au} \, .
	\label{eq:equation8}
\end{equation}
Thus for LRLL 31, the co-rotation radius is $\sim 0.05$~au from the star. The stellar radius is 2.3 R$_\odot \approx 0.01$~au. The distance length scales are summarised in Figure~\ref{fig:figure3}, where the inner disc scale heights are calculated from the deduced covering fraction (equation~\ref{eq:equation2}).

\begin{figure*}
	\includegraphics[width=0.8\textwidth]{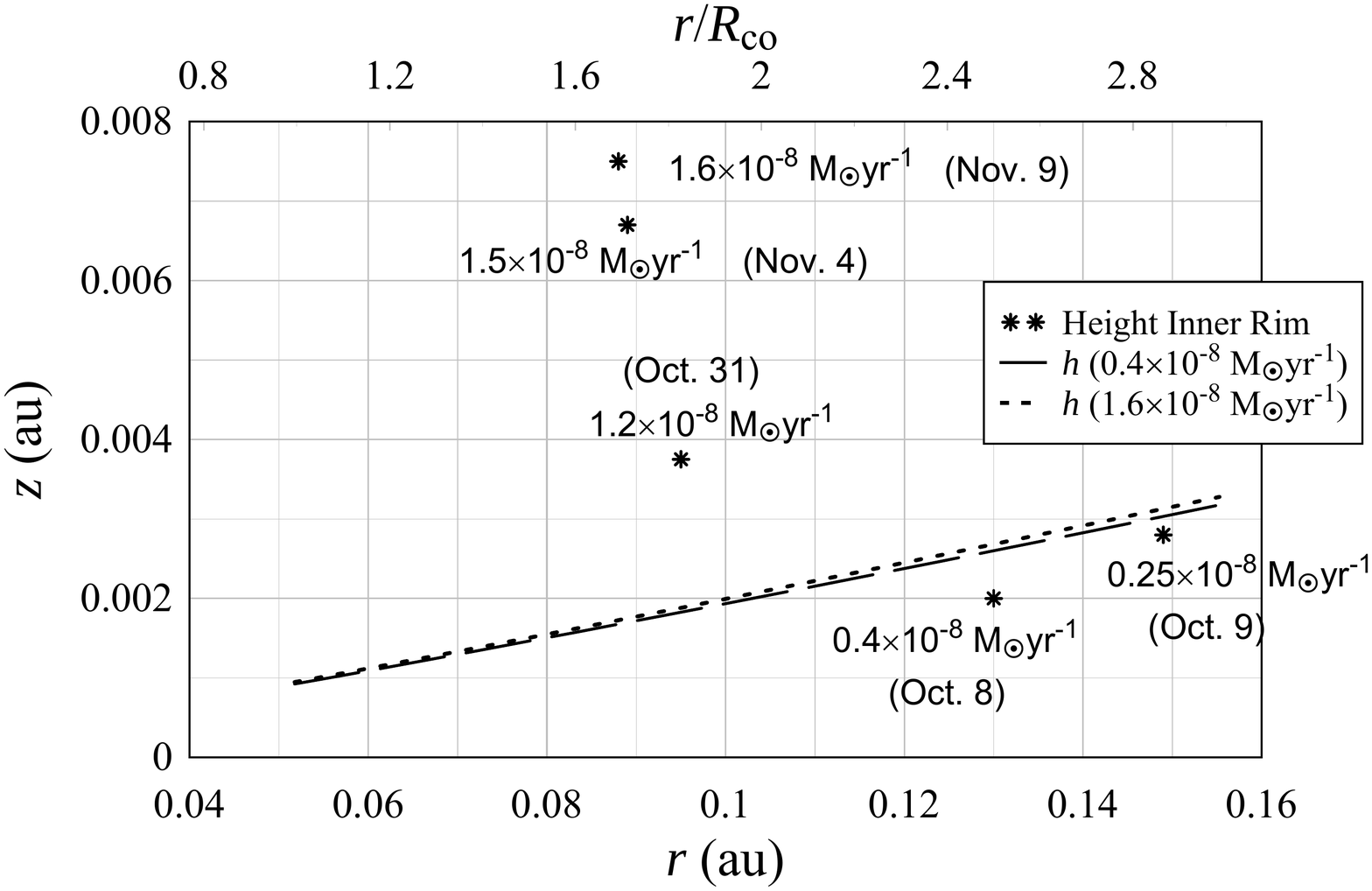}
    \caption{Two theoretical isothermal scale heights for two accretion rates are compared to the deduced heights of the inner rim of LRLL~31 for five different mass accretion rates \citep{2011ApJ...732...83F}. The radial extent of the disc is set from approximately 0.05~au to 0.15~au for illustrative purposes. The observed inner rim is significantly larger or ``puffed up'' relative to the expected isothermal scale heights for the higher accretion rates. The height of the inner rim appears to be somewhat proportional to the accretion rate. The highest inner rim occurs at the largest accretion rate at a distance of $\sim$ 0.09 au or $\sim$ 1.7 $R_{\rm co}$ The dates are all in the year 2009.}
    \label{fig:figure4}
\end{figure*}

It is of interest to compare the deduced heights of the LRLL 31 inner rim with the standard, isothermal scale height, $h(r)$, of an accretion disc: 
\begin{equation}
    h(r) = \sqrt{\frac{2k_{\rm B} T_{\rm g} r^3}{GM_\star \bar{m}}} \approx 0.0043 \, {\rm au} \sqrt{\frac{(T_{\rm g}/1000\,{\rm K})(r/0.1\,{\rm au})^3}{(M_\star / {\rm M}_\odot)(\bar{m}/{\rm m_H})}}
\end{equation}
where $k_{\rm B}$ is Boltzmann's constant, $T_{\rm g}$ the gas temperature, ${\rm m_H}$the mass of the hydrogen atom, and $\bar{m}$ the mean molecular mass of the gas. 
This comparison is shown in Figure~\ref{fig:figure4}, where the observed inner rim heights for the lower mass accretion rates \citep[data from][]{2011ApJ...732...83F} are smaller, but comparable to the expected isothermal scale height. However, the inner rim heights for the higher mass accretion rates are significantly higher (over a factor of four in one case) relative to the expected isothermal scale heights, and LRLL~31 has a puffed up inner rim. Such puffed up rims appear to be common in young stellar systems, but a comprehensive explanation for how they are produced has eluded researchers \citep{2014A&A...566A.117V}. As such, LRLL~31 infrared variability could be linked to the production of puffed up inner disc rims.

In calculating the isothermal scale height, the gas temperature is assumed to be approximately the same as the disc surface temperature, $T_{\rm disc}(r)$, where the temperature of an optically thick, flat disc subject to stellar radiation \citep{1985A&A...146..366F,1998apsf.book.....H}  and differential friction in the accretion disc \citep{2002apa..book.....F} is
\begin{align}
    T_{\rm disc}(r) &\approx \left( \frac{(L_\star+L_{\rm a})}{4\pi^2 \sigma_{\rm SB} R_\star^2} \biggl(\sin^{-1}\left(\frac{R_\star}{r}\right) - \frac{R_\star}{r}\sqrt{1-(R_\star/r)^2} \right)  \nonumber \\
    & \quad \quad \quad \quad + \frac{3GM_\star \dot{M}_{\rm a}}{8\pi r^3 \sigma_{\rm SB}}  \left(1-\sqrt{R_\star/r}\right) \biggr)^{1/4} \, . 
\end{align}
The parameters used for the LRLL~31 calculations are shown in Table~\ref{tab:table1}.

\begin{table}
    \centering
    \begin{tabular}{l|l} \\ \hline
    stellar parameter & value \\ \hline
       $M_\star$, stellar mass & 1.6 M$_\odot$ \\
       $R_\star$, stellar radius & 2.3 R$_\odot$\\ 
       $L_\star$, stellar luminosity & 5 L$_\odot$\\
       $P_\star$, stellar rotation period & 3.4 days \\
       $B_\star(R_\star)$, magnetic field strength & 0.15 T\\
       $\dot{M}_{\rm a}$, mass accretion rate & $0.4-1.6\times10^{-8}$ M$_\odot$yr$^{-1}$ \\
       $d$, stellar distance & 315 pc \\
       $\bar{m}$, mean molecular mass of disc gas & 2.3 amu \\ \hline
    \end{tabular}
    \caption{LRLL 31 parameters used in this paper.}
    \label{tab:table1}
\end{table}

To produce a physical model for puffed-up inner rims, we first consider a model for the interaction between the stellar magnetosphere and the surrounding disc.

\section{STELLAR MAGNETOSPHERE-DISC INTERACTION}

\subsection{Magnetic Disc Height and Scale Height}

If a co-rotating, stellar magnetosphere interacts with a surrounding disc of gas and dust then a radial disc current, $j_{\rm r}$, is generated in the disc (Figure 5), where the current has the steady state form \citep{1999MNRAS.309..443L}:
\begin{equation}
    j_{\rm r} = -\sigma_{\rm D}(r) \, r (\Omega_\star - \Omega_{\rm K}(r)) \, B_{\star z}(r) \widehat{\mathbf r} \, ,
\end{equation}
where
\begin{equation}
    |B_{\star z}(r)| \approx B_\star(R_\star)\left(R_
    \star/r\right)^3 \, ,
\end{equation}
with $\sigma_{\rm D}$ is the disc electrical conductivity, $B_{\star z}$ the $z$ component of the stellar magnetic field at the midplane of the disc, and $\widehat{\mathbf r}$ the unit vector in the $r$ direction.

\begin{figure}
	\includegraphics[width=\columnwidth]{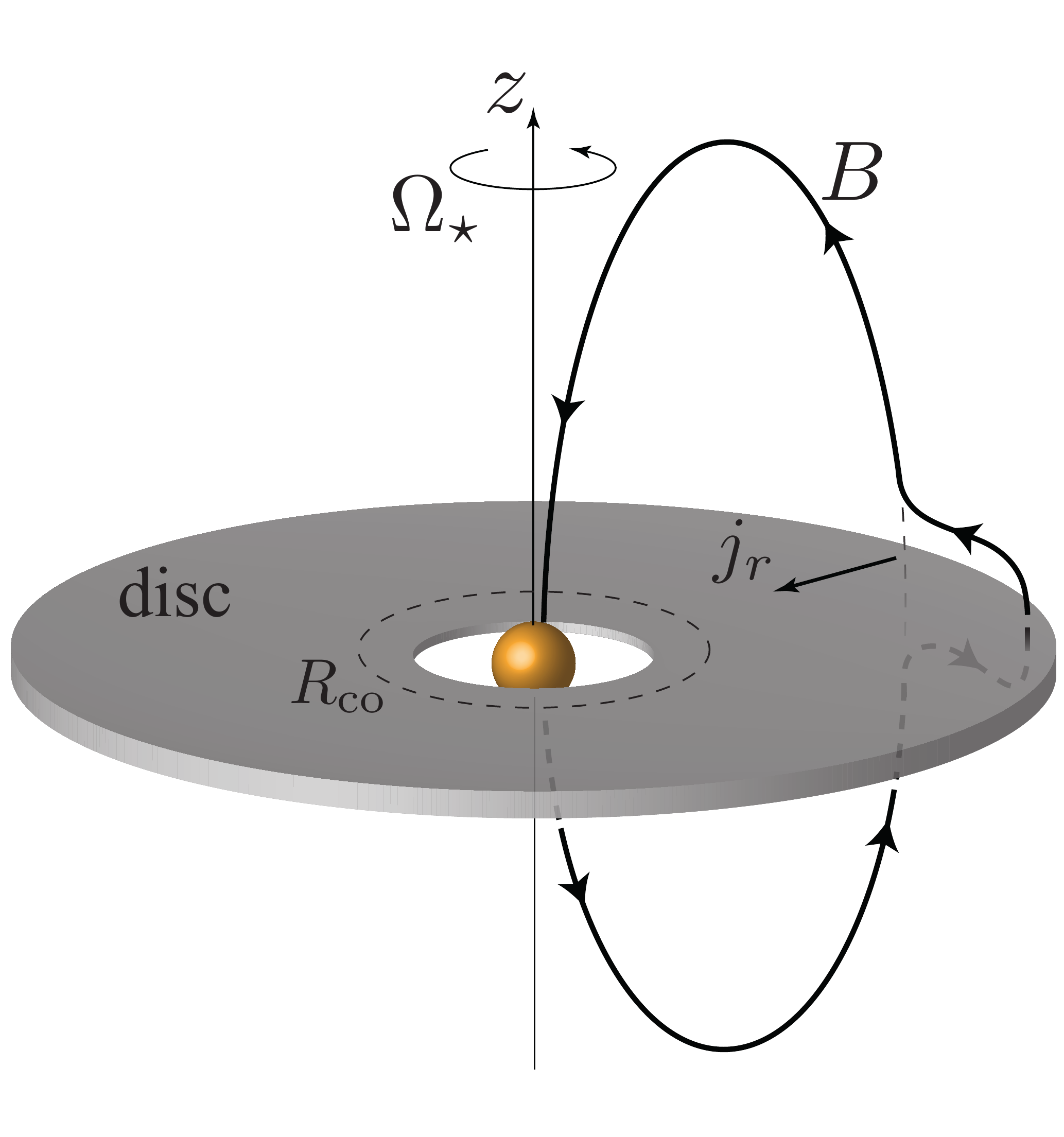}
    \caption{Interaction of a stellar magnetic field and an accretion disc. The difference in rotation between the central object and the disc creates a radial current, $j_{\rm r}$, and, thereby, toroidal magnetic fields, $B_\phi$, within the disc. We show the case $B_{\rm z} > 0$. At the co-rotation radius, $R_{\rm co}$, the angular velocity of the magnetospheric field is equal to the angular velocity of the disc. As the toroidal field increases in strength, the magnetic lines will tend to expand in the $z$ direction \citep{1995MNRAS.275..244L}.}
    \label{fig:figure5}
\end{figure}

The radial disc current generates a toroidal magnetic field in the disc with the steady state form (\ibid): 
\begin{equation}
    \mathbf{B}_\phi (r,z) = \mu_o \sigma_{\rm D}(r) r z (\Omega_\star - \Omega_{\rm K}(r))B_{\rm z}(r)\widehat{\mathbf{\phi}} \, ,
    \label{eqn:14}
\end{equation}
with $z$ the perpendicular distance from the midplane of the disc. In equation~\ref{eqn:14}, $z = 0$ is located at the midplane of the disc. For this equation, $z$ has a magnitude that is less than or equal to the height of the disc. 

One can show (Appendix \ref{Toroidal_Field_Growth}) that, in principle, the disc toroidal field grows to its steady state value on a timescale, $\tau_{{\rm B}_\phi}$, which has the approximate form 
\begin{equation}
    \tau_{{\rm B}_\phi} \approx \frac{\mu_0\sigma_{\rm D}(r)h(r)^2}{2} \approx 1.6 \left(\frac{\sigma_{\rm D}}{10^{-5} \, {\rm Sm}^{-1}}\right)\left(\frac{h}{0.001 \, {\rm au}}\right)^2 {\rm days.} 
    \label{eqn:15}
\end{equation}
In practice, the toroidal field may not reach its steady state value due to field instabilities - for example the inflation of the field that arises when the torioidal field and poloidal field strengths become comparable \citep{1992ApJ...392..622N,1995MNRAS.275..244L}.

The interaction between the radial disc current and the generated toroidal field produces a Lorentz compressive $\mathbf{j}\times\mathbf{B}$ force on the disc in the $z$ direction that is directed towards the midplane of the disc. For a disc that is approximately isothermal in the z direction, this Lorentz compression changes the standard isothermal density profile to \citep{1999MNRAS.309..443L}
\begin{equation}
    \rho(r,z) = \rho_{\rm c}(r) \exp\left(-\left(\frac{z}{h(r)}\right)^2\right)
    -\rho_\infty\left(1-\exp{\left(-\left(\frac{z}{h(r)}\right)^2\right)}\right)\, ,
    \label{eqn:16}
\end{equation}
with $\rho_{\rm c}(r)$the midplane mass density of the disc gas, $h(r)$ the standard isothermal scale height, and
\begin{equation}
    \rho_\infty = \mu_0\sigma_{\rm D}^2r^2B^2_{\rm z}(r) \left(\frac{\Omega_\star}{\Omega_{\rm K}(r)}-1\right)^2 \, .
        \label{eqn:17}
\end{equation}
The first term in equation~\ref{eqn:16} is the standard isothermal density profile, while the second term introduces the magnetic compression of the disc. The combination of the two terms produces a magnetic disc height due to the sharp cut off in the disc at a distance, $H_{\rm B}$, from the midplane of the disc, where
 \begin{equation}
     H_{\rm B}(r) = h(r) \sqrt{\ln{\left(1+\frac{\rho_{\rm c}(r)}{\rho_\infty}\right)}} \, .
 \end{equation}
From equation~\ref{eqn:17}, if $\sigma_{\rm D} \rightarrow 0$ and/or $B_{\rm z}\rightarrow 0$, then $\rho_\infty \rightarrow 0$ and equation~\ref{eqn:16} returns to the standard isothermal disc density profile with $H_{\rm B} \rightarrow \infty$. The latter property, while physically correct, is mathematically inconvenient and it is useful to define a magnetic scale height, $h_{\rm B}$, where $\rho(r,h_{\rm B})=\rho_{\rm c}(r)e^{-1}$ with
\begin{equation}
    h_{\rm B}(r) = h(r) \sqrt{\ln{\left(1+\frac{1-1/e}{1/e+\rho_\infty/\rho_{\rm c}(r)}\right)}} \, .
\end{equation}

\subsection{Magnetic Compression and Disc Conductivity}

\begin{figure}
	\includegraphics[width=\columnwidth]{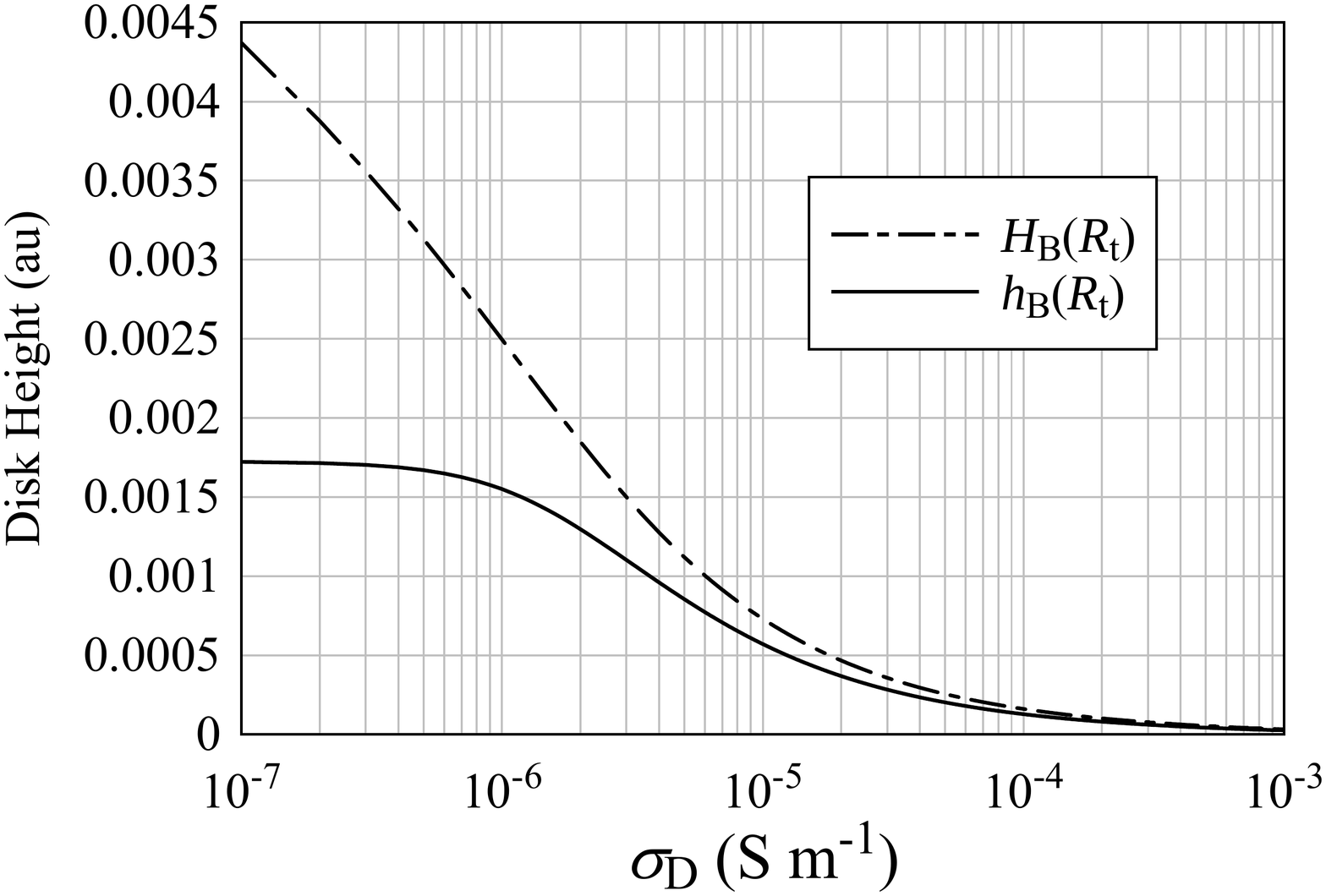}
    \caption{The magnetically compressed disc height as a function of the inner disc electrical conductivity in units of au. $H_{\rm B}$ is the magnetic disc height from the midplane of the disc, while $h_{\rm B}$ is the e-folding magnetic scale height.}
    \label{fig:figure6}
\end{figure}

This magnetic compression effect (obtained, independently, via different derivations by \citet{1986ApJS...62....1L,1998MNRAS.299...31C} and \citet{1999MNRAS.309..443L}) is dependent on the disc conductivity. In Figure~\ref{fig:figure6}, we show the decrease in values of $H_{\rm B}$ and $h_{\rm B}$ as a function of disc conductivity at the truncation radius of the disc, $R_{\rm t}$, for the mass accretion rate of $1.6\times10^{-8}$~M$_\odot$yr$^{-1}$ (i.e., for $R_{\rm t} \approx 0.09$~au).  
A magnetically confined disc can suffer significant compression with increasing disc electrical conductivity.
%

For small disc electrical conductivities, $h_{\rm B}$ approaches the standard isothermal disc height. $H_{\rm B}$ is the cut-off of the disc density profile due to magnetic compression, which increases as $\sigma_{\rm D} \rightarrow 0$. 

As discussed in the previous sections, the inner disc of LRLL~31 appears to be ``puffed up'', so discussion of magnetised disc compression would appear to be not relevant. However, there is also a wind-up of the toroidal field, (Appendix \ref{Toroidal_Field_Growth}) which, via magnetic pressure, may power an outflow from the compressed disc. It is of interest to understand how such contradictory behaviour may arise.

\subsection{Twist and Interaction Region} 

From equation~\ref{eqn:14}, the steady state twist of the disc magnetic field, $\gamma_{\rm B}$, is 
\begin{equation}
    \gamma_{\rm B}(r,z) = \frac{B_\phi(r,z)}{B_{\rm z}(r,z)} = \mu_0\sigma_{\rm D}(r) rz(\Omega_\star - \Omega_{\rm K}(r)) \, .
    \label{eqn:20}
\end{equation}
An estimate of the maximum twist value as a function of distance from a star is
\begin{align}
\gamma_{B_{\rm max}}(r,z) &\sim \gamma_{\rm B}(r,h_{\rm B}(r)) = \mu_0\sigma_{\rm D}(r)rh_{\rm B}(r)(\Omega_\star - \Omega_{\rm K}(r)) \nonumber\\
& = \mu_0\sigma_{\rm D}(r)rh_{\rm B}(r)\Omega_\star\left(1 - \left(\frac{R_{\rm co}}{r}\right)^{3/2} \right) \nonumber\\
& = 1.0 \left( \frac{\sigma_{\rm D}}{10^{-7}{\rm Sm^{-1}}}\right)\left( \frac{r}{0.05 \ {\rm au}}\right)\left( \frac{h_{\rm B}}{0.001 \ {\rm au}}\right)
\left(\frac{10\ {\rm days}}{P_\star} \right) \nonumber\\
&\times\left(1 - \left(\frac{R_{\rm co}}{r}\right)^{3/2} \right)
\,.
     \label{eqn:21}
\end{align}
From equation~\ref{eqn:21} we see that the twist at the co-rotation radius, $R_{\rm co}$, is zero, but for the chosen representative values, it quickly increases as a function of distance $r$ from LRLL~31 to a value much larger than one. Thus, the wound-up toroidal field strength may be orders of magnitude greater than the magnetospheric poloidal field strength. This implies that that the stellar dipole field may have expanded, opened and disconnected from the disc  \citep[e.g.][]{2002ApJ...565.1191U}. Alternatively, wound up, strong toroidal fields may produce collimated jet flows that are perpendicular to the disc midplane \citep{2012MNRAS.423L..45P}.  

It is of interest to note the particular region where the twist is possibly stable, i.e. $\gamma_{\rm B} < 1$. To do this, we set
\begin{equation}
    \gamma_{B_{\rm max}}(r,z) \sim \gamma_{\rm B}(r,h(r)) = \mu_0\sigma_{\rm D}(r)rh(r) |\Omega_\star - \Omega_{\rm K}(r)| \, ,
    \label{eqn:22}
\end{equation}
and let
\begin{equation}
    r = R_{\rm co} + \Delta r \, ,
     \label{eqn:23}
\end{equation}
where we assume that $\Delta r \ll R_{\rm co}$.  Substituting equation~\ref{eqn:23} into equation~\ref{eqn:22} gives
\begin{equation}
    \gamma_{\rm B}(\Delta r,h(r)) \approx \frac{3\mu_0\sigma_{\rm D}|\Delta r|}{2} \sqrt{\frac{2k_{\rm B} T_{\rm g}}{\bar{m}}} \, .
\end{equation}

Setting $\gamma_{\rm B} = \gamma_{\rm c} \approx 1$, where $\gamma_{\rm c}$ is the `critical' twist (that is the toroidal disc field becomes comparable to the stellar magnetospheric field), gives 
\begin{align}
       |\Delta r_{\rm c}| &\approx \frac{2\gamma_{\rm c}}{3\mu_0\sigma_{\rm D}} \sqrt{\frac{\bar{m}}{2k_{\rm B} T_{\rm g}}} \nonumber \\
    & \approx 1.0\times10^{-4} {\rm au} 
    \left(\frac{10^{-5}{\rm Sm}^{-1}}{\sigma_{\rm D}}\right) \sqrt{\frac{\bar{m}/{\rm m_{H_2}}}{T_{\rm g}/1400~{\rm K}}} \, .
    \label{eqn:25}
\end{align}
So  $\Delta r \ll R_{\rm co}$, as was assumed.

For significantly lower values of $\sigma_D$, a much larger region of the inner disc will have a stable twist, with the field lines opening up at larger distances away from the star. This leads to some interesting disc twist stability regions which are discussed in \citet{2005MNRAS.356..167M}.

\begin{figure}
	\includegraphics[width=\columnwidth]{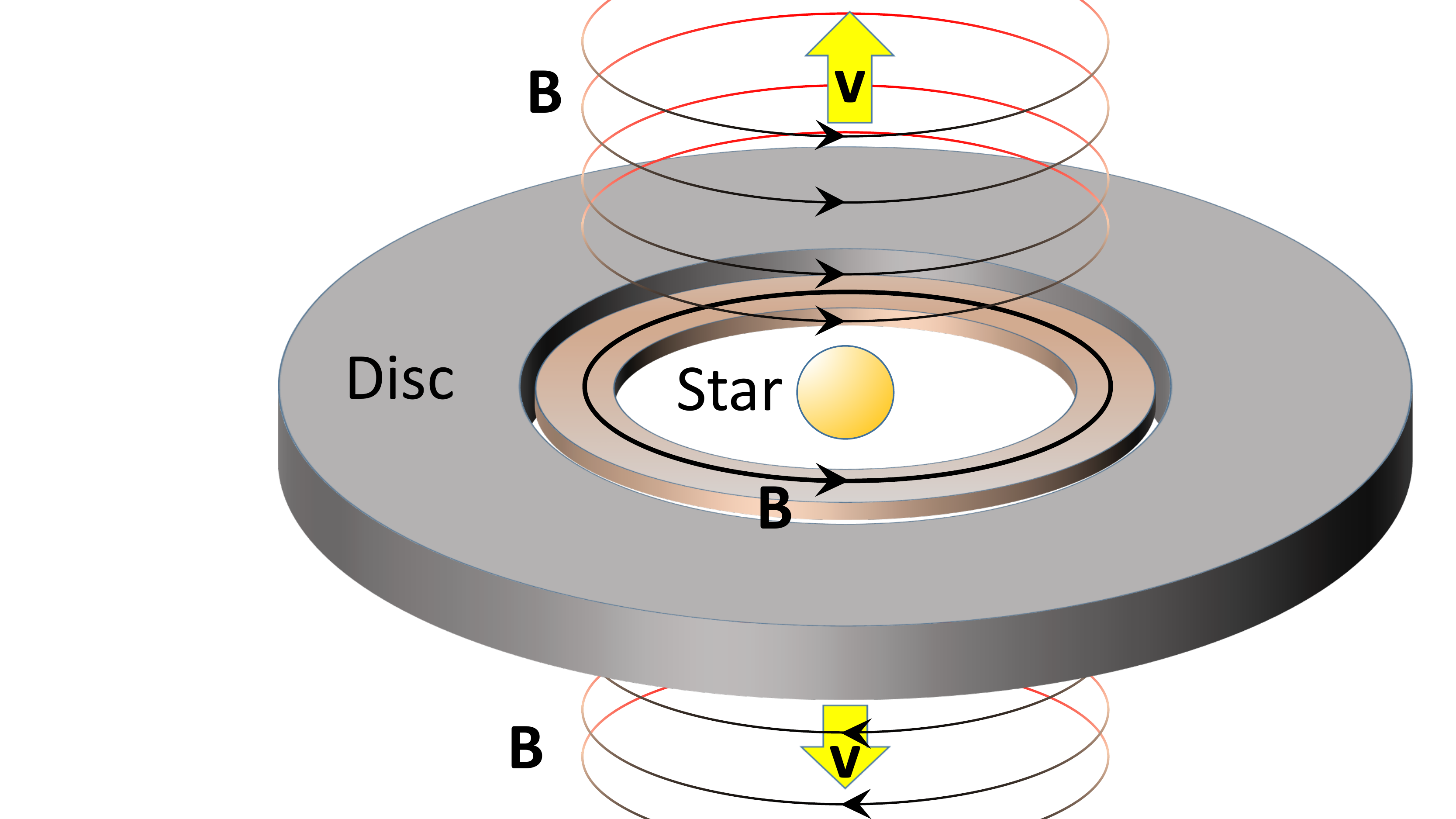}
    \caption{A schematic of a wind flow and possible driving magnetic field. The interaction of a stellar magnetic field with an accretion disc produces a toroidal field, $\mathbf{B}$, that compresses the inner disc and may, simultaneously, produce a magnetic pressure driven outflow with a velocity $\mathbf{v}$. Our jet flow model implicitly assumes that the jet is produced at or near the surface of the disc and that a small toroidal magnetic  field would be carried away with the flow. As the jet moves away from the disc, it is likely to expand in the radial direction \citep{1991ApJ...379..696L}. The jet speed will also decrease as it moves away from the star.}
    \label{fig:figure8}
\end{figure}

So far, in this section, we have shown that a dipole stellar magnetic field interacting with an inner accretion disc will generate a toroidal field that can compress the disc and, at the same time, potentially generate an outflow from the surface of the disc (Figure~\ref{fig:figure8}). 
%

\subsection{Jet Speed} 
\label{subsec:jet_exhaust_speed}

\begin{figure}
 \begin{subfigure}{\linewidth}\centering
 \includegraphics[width=\linewidth]{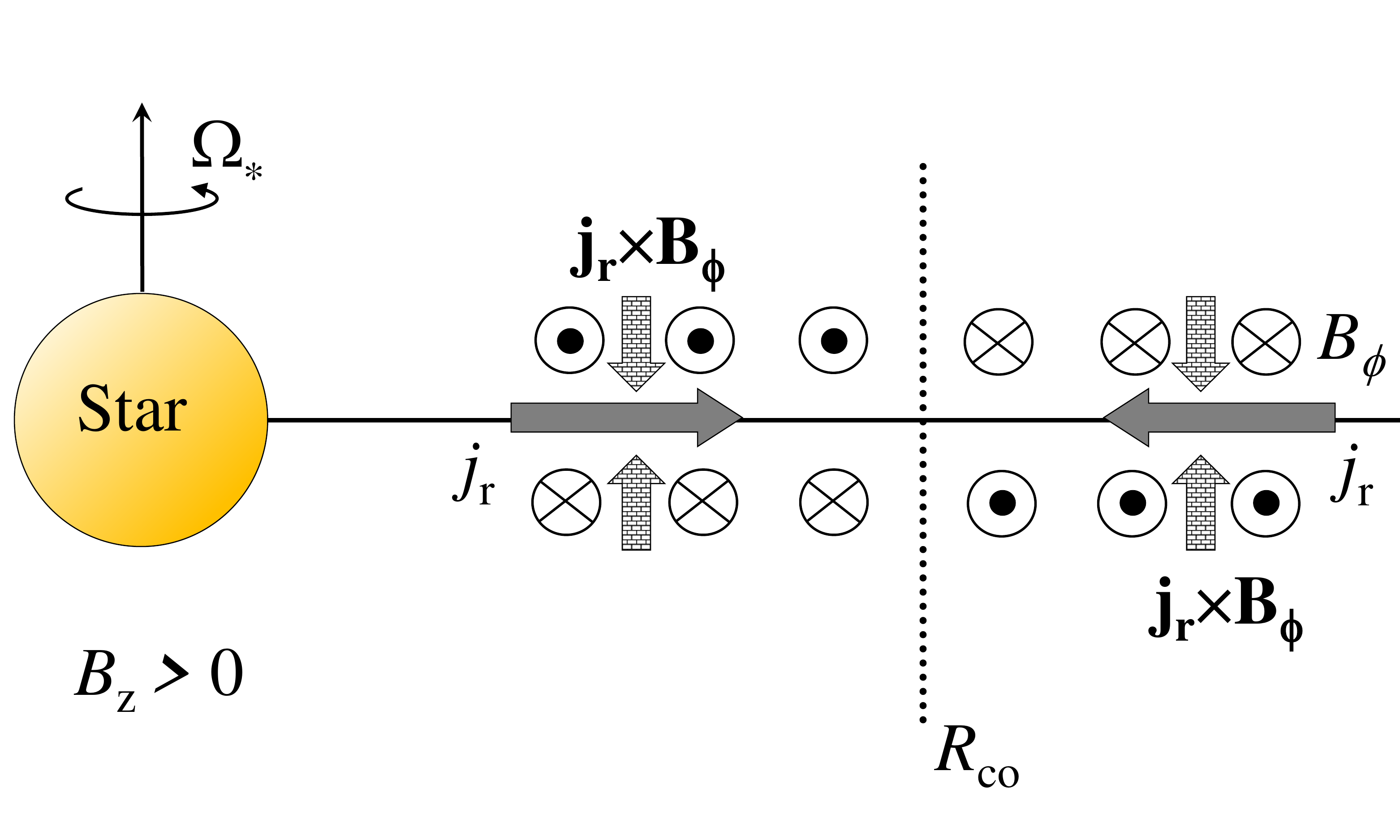}\caption{$B_{\rm z} > 0$}
  \end{subfigure}
 \begin{subfigure}{\linewidth}\centering
 \includegraphics[width=\linewidth]{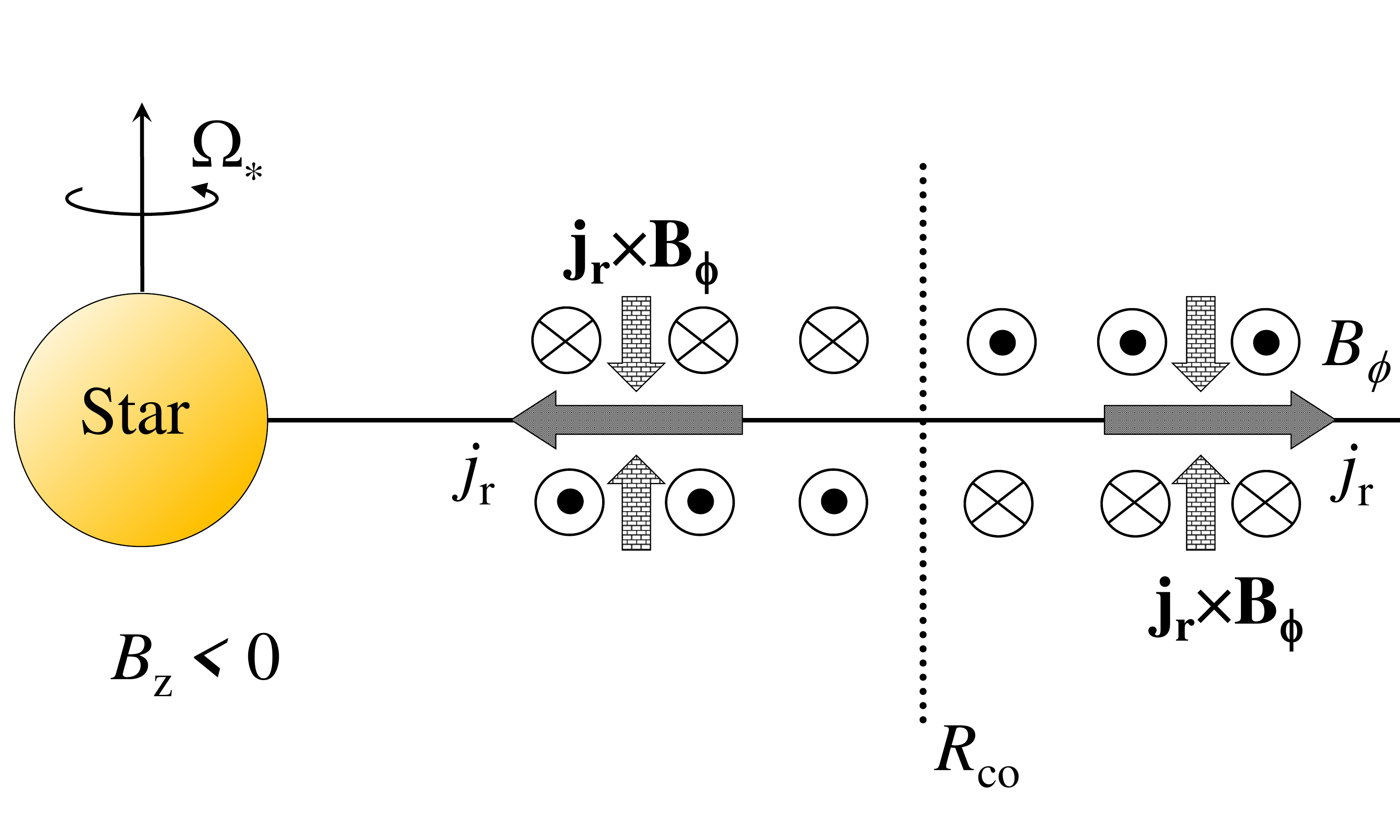}\caption{$B_{\rm z} < 0$}
   \end{subfigure}
    \caption{The interaction of a stellar magnetic field with an accretion disc produces a radial disc current, $j_{\rm r}$, and a toroidal field, $B_\phi$, that compresses the inner disc. The resulting Lorentz Force  $\mathbf{j}\times \mathbf{B}$ (or $\mathbf{j_{\rm r}}\times \mathbf{B_\phi}$)  compresses the disc. The compression effect is independent of the orientation of the magnetic field direction of the stellar magnetosphere, as seen when $B_{\rm z}>0$ in (a) and $B_{\rm z}<0$ in (b). }
    \label{fig:figure9}
\end{figure}

Given our stellar magnetosphere$+$disc model as shown in Figures \ref{fig:figure5} and \ref{fig:figure8}, it is instructive to obtain an intuitive idea of how this toroidal field may form a jet flow. From Figure~\ref{fig:figure5}, which shows the case $B_{\rm z} > 0$, the resulting radial disc currents and fields have the directions shown in Figure~\ref{fig:figure9}(a), where we have taken a slice through the disc. Here we see that the radial disc currents and toroidal magnetic fields produce compressive forces on the disc. In Figure~\ref{fig:figure9}(b), the stellar magnetic field points in the opposite direction, i.e., $B_{\rm z} < 0$. In this case the directions of the magnetic fields and currents are reversed, but the Lorentz compressive force remains as shown in Figure~\ref{fig:figure9}(b). So, the compressive Lorentz force is independent of the direction of the stellar magnetic field.
%

\begin{figure}
 \begin{subfigure}{\linewidth}\centering
 \includegraphics[width=\linewidth]{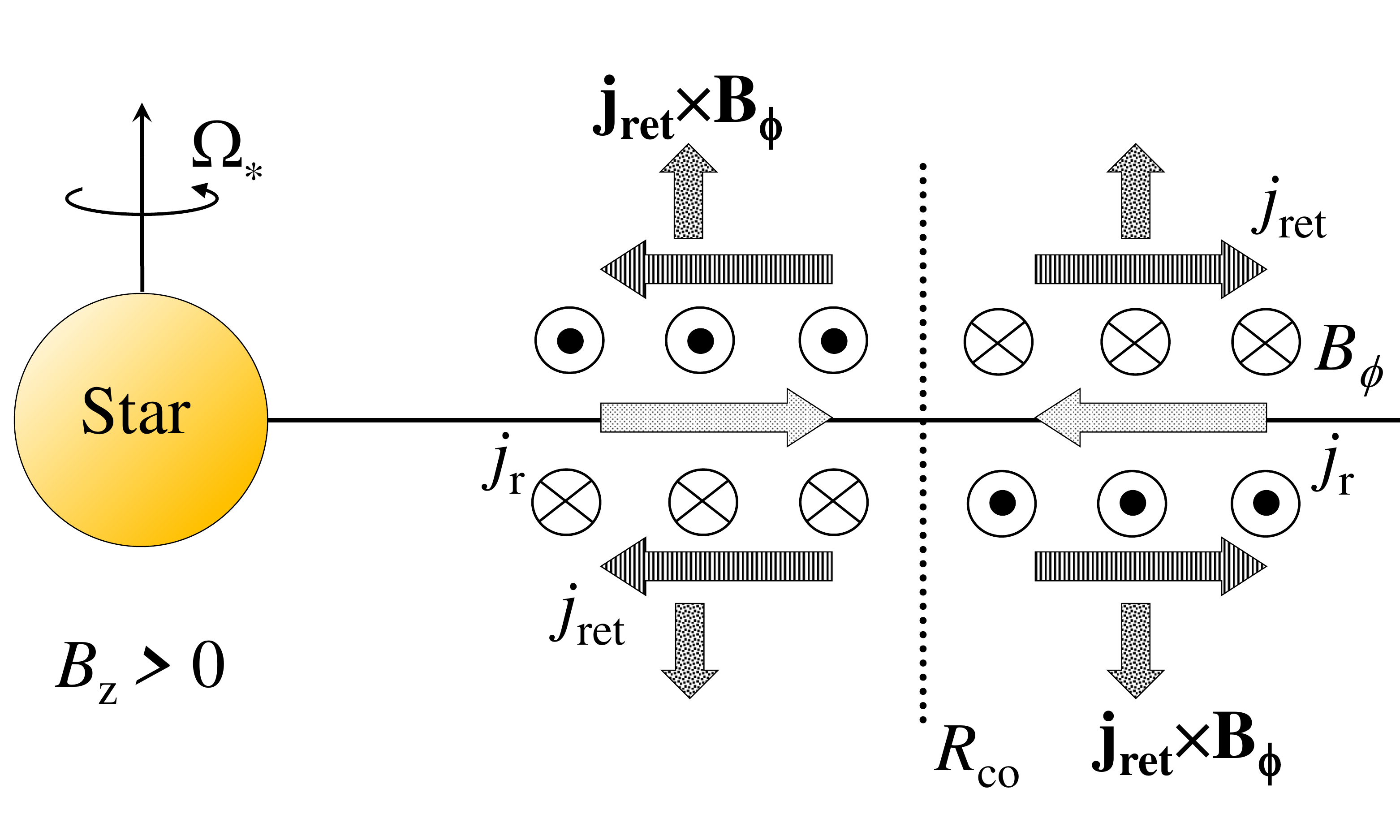}\caption{$B_{\rm z} > 0$}
  \end{subfigure}
 \begin{subfigure}{\linewidth}\centering
 \includegraphics[width=\linewidth]{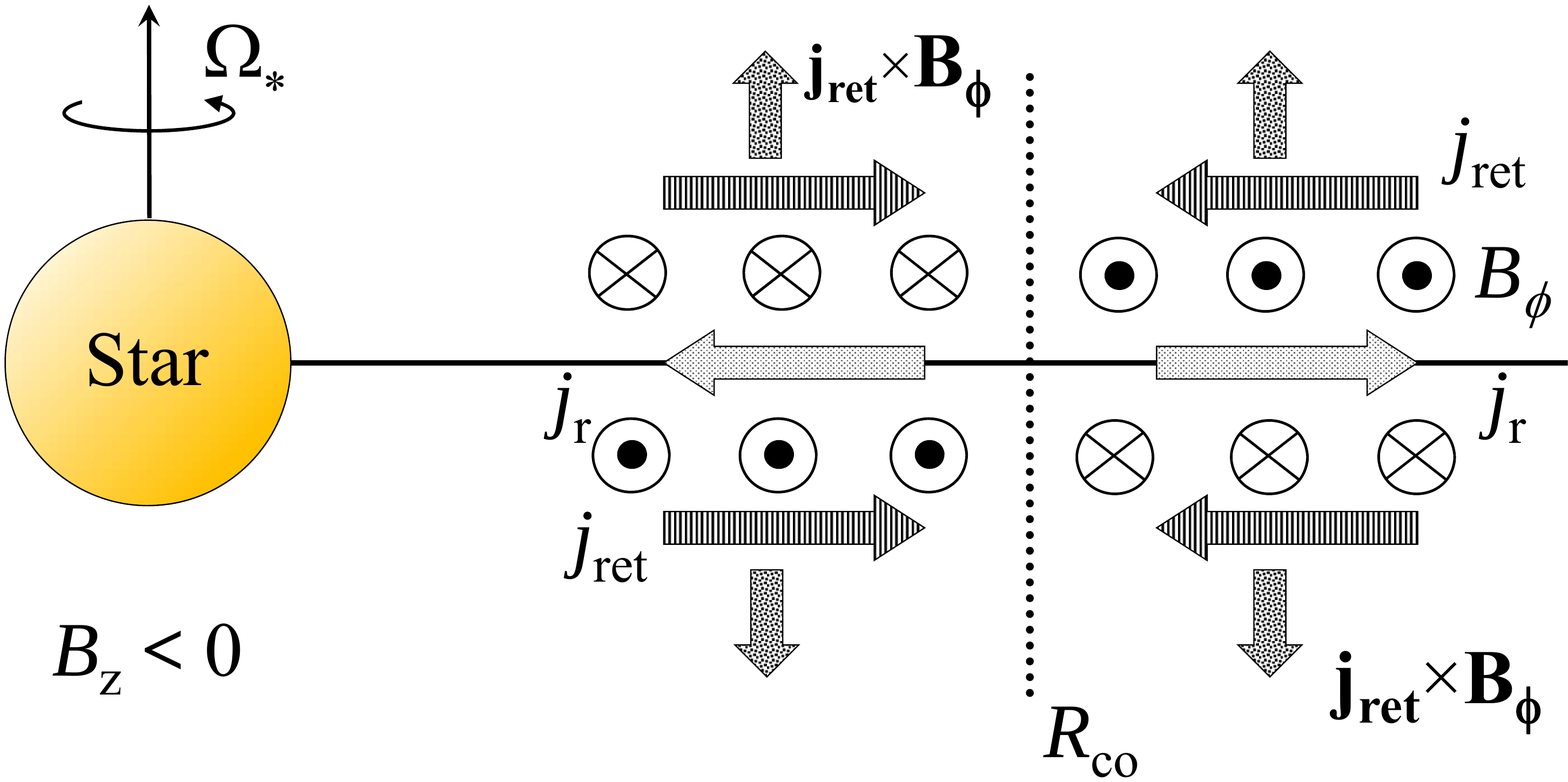}\caption{$B_{\rm z} < 0$}
   \end{subfigure}
    \caption{The return current $j_{\rm ret}$ of the radial disc current $j_{\rm r}$ interacts with the toroidal field, $B_\phi$, to produce a Lorentz force, $\mathbf{j_{\rm ret}}\times \mathbf{B_\phi}$, away from the disc. The outflow effect occurs near the surface of the disc and is independent of the orientation of the magnetic field direction of the stellar magnetosphere (with (a) showing $B_{\rm z}>0$ and (b) $B_{\rm z} < 0$).}
    \label{fig:figure10}
\end{figure}

We now suppose that the upper disc atmosphere allows a return current to flow. As discussed in Appendix \ref{MPD_Flow}, this implies there exist separated layers of peak Pedersen or Hall conductivity that allow trans magnetic field currents to flow. One maximum of conductivity occurs within the disc, approximately at the disc midplane, while the other conductivity maxima occur on the top and bottom disc surfaces. In such a circumstance, the Lorentz force driven by the return currents, $j_{\rm ret}$, and the toroidal fields, $B_\phi$,  points away from the disc midplane and so material is forced to move away from the disc (Figure \ref{fig:figure10}). 

%

As derived in Appendix \ref{MPD_Flow}, the speed, $\varv_{\rm ex}$, of the outflow produced by the return current at or near the surface of the disc is approximately 
\begin{equation}
    \varv_{\rm ex}(r,z_{\rm T}) \approx \sqrt{\frac{\mu_0}{\bar{\rho}}} \sigma_{\rm D}(r) rz_0 \Omega_\star |1-(R_{\rm co}/r)^{3/2}| \, |B_{\rm z}(r)| \, ,
    \label{eqn:35}
\end{equation}
where $z_0$ is the distance (or altitude) from the disc midplane to the entrance of the jet flow, $z_{\rm T}$ is the altitude of the jet flow exit, $\bar{\rho}$ is the average gas mass density within the jet flow, while the $r$ in this case is, approximately, the inner edge of the disc, i.e., $r\approx R_{\rm t}$. The derivation of equation~\ref{eqn:35} implicitly assumes that $z_0$ and $z_{\rm T}$  are $\ll r$. As a consequence, this is the expression for the jet speed close to the top and bottom surfaces of the inner disc.

The form of equation~\ref{eqn:35} tells us that the jet flow speed goes to zero at $r=R_{\rm co}$, because the stellar field does not wind up into a toroidal disc field at this point. Trivially, the jet flow speed goes to zero when $r \rightarrow \infty$. If we assume that the disc conductivity $\sigma_{\rm D}(r)$ is approximately constant for the regions of interest, then for $r> R_{\rm co}$ the jet flow has a maximum speed of
\begin{equation}
    \varv_{\rm ex}(r_{\rm m},z_{\rm T}) \approx \frac{3}{7}\left(\frac{4}{7}\right)^{4/3} \sqrt{\frac{\mu_0}{\bar{\rho}}}\sigma_{\rm D} z_0 \Omega_\star R_{\rm co} |B_{\rm z}(R_{\rm co})| \, ,
    \label{eqn:36}
\end{equation}
where $(3/7) (4/7)^{4/3} \approx 0.203$ and $r_{\rm m} = (7/4)^{2/3}R_{\rm co} \approx 1.45 R_{\rm co}$.

For $r<R_{\rm co}$ then the jet speed approaches infinity as $r \rightarrow 0$. As such the maximum practical jet speed is when the disc is touching the stellar surface
\begin{align}
     \varv_{\rm ex}&(R_\star,z_{\rm T}) \approx 
    \nonumber \\
    &\sqrt{\frac{\mu_0}{\bar{\rho}}}\sigma_{\rm D}(R_\star) R_\star z_0 \Omega_\star |1-(R_{\rm co}/R_\star)^{3/2}| \, |B_\star(R_\star)| . 
    \label{eqn:37}
\end{align}

\begin{figure}
	\includegraphics[width=\columnwidth]{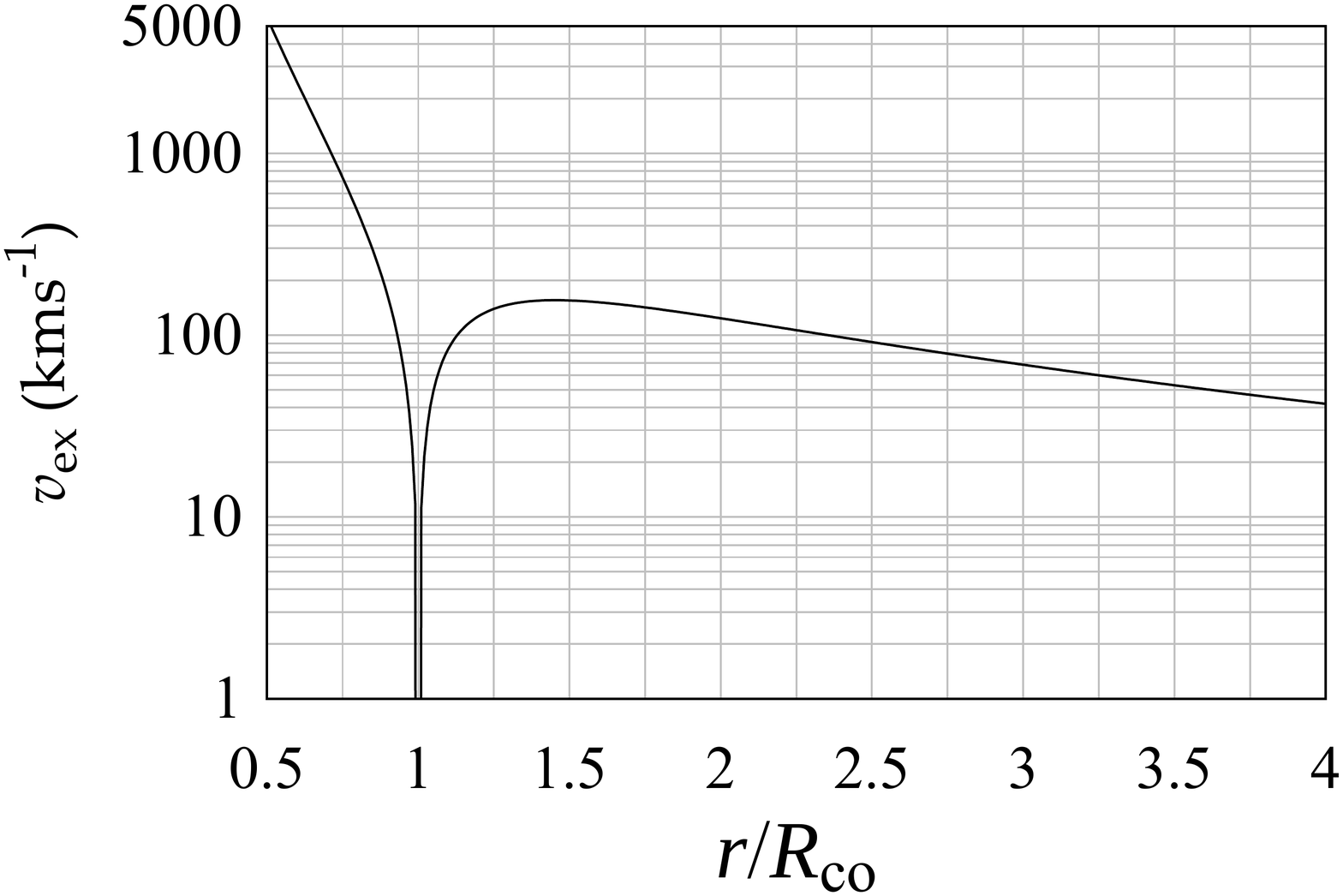}
    \caption{The speed of a (bipolar) jet flow from a young stellar system at or near the surface of the accretion disc as calculated from equation~\ref{eqn:35}  as a function of distance  from the star (in units of the co-rotation radius). Although the flow speed tends to increase with decreasing distance from the star, the jet flow shuts down near the co-rotation radius. 
    }
    \label{fig:figure11}
\end{figure}

We plot equation~\ref{eqn:35} as a function of $r$ in Figure~\ref{fig:figure11}, which shows the jet flow speed tends to increase with decreasing distance from the star. 
Here we have assumed the following parameters:  $z_0=0.001$~au, $P_\star = 3.4$~days, $\sigma_{\rm D}=10^{-5}$~Sm$^{-1}$, $\bar{\rho}=10^{-7}$~kgm$^{-3}$, $B_\star(R_\star)=0.15$~T, and $R_{\rm co}=0.052$~au. These representative values give flow speeds of order 10 to 1000 kms$^{-1}$. For $r/R_{\rm co} > 1$ the maximum speed occurs at $r_{\rm m} = (7/4)^{2/3}R_{\rm co} \approx 1.45 R_{\rm co}$.
In the region  $R_{\rm co} < r < r_{\rm m}$, the jet flow speed decreases for decreasing $r$, while for the region $R_\star < r < R_{\rm co}$, the jet flow speed increases as $r$ decreases.
%

\begin{figure}
	\includegraphics[width=\columnwidth]{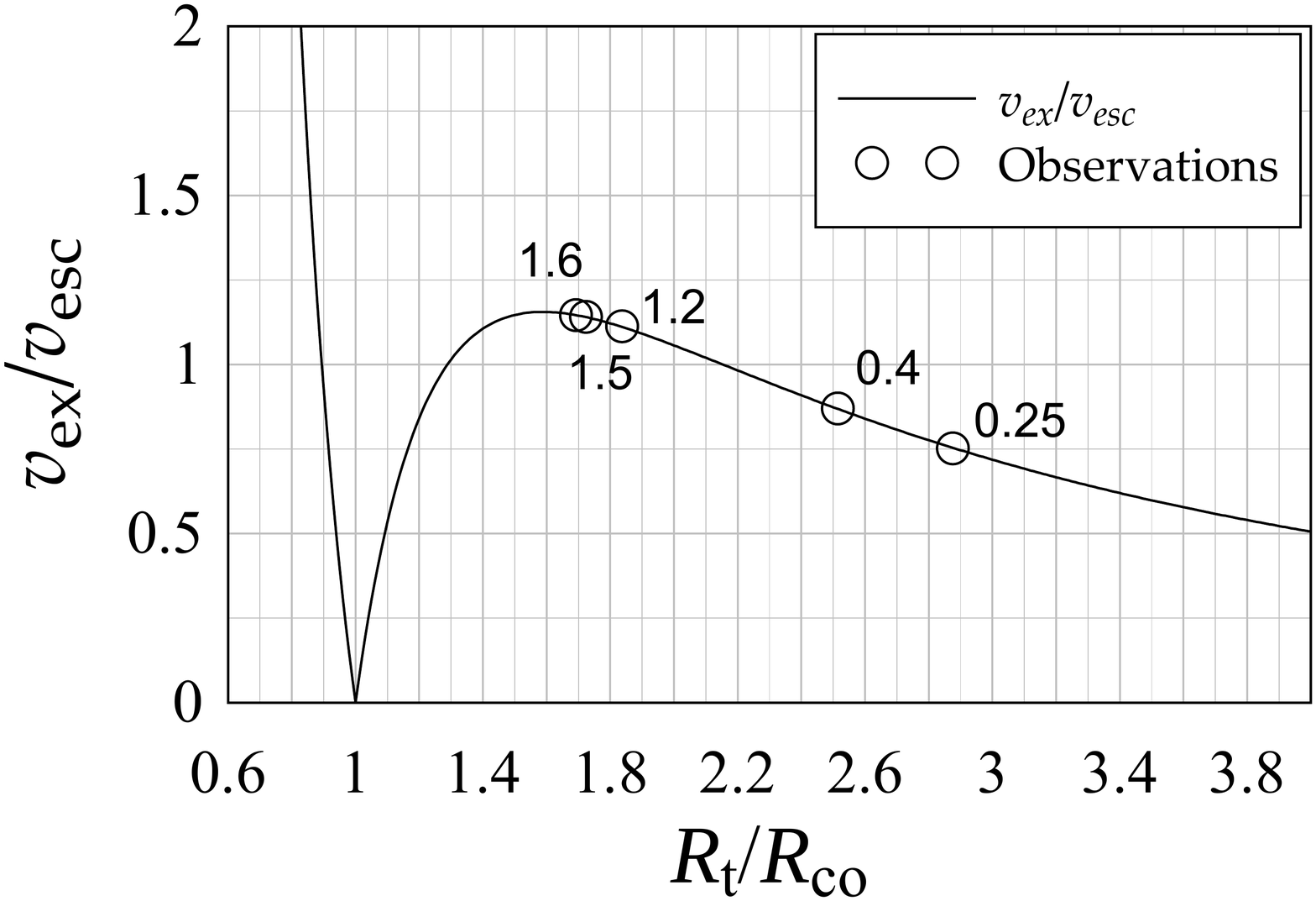}
    \caption{The speed of a (bipolar) jet flow from the surface of an accretion disc in a young stellar system (equation~\ref{eqn:35}) divided by the escape speed (equation~\ref{eqn:38}) as a function of distance away from the star (distance in units of the co-rotation radius). In this case, the jet flow exceeds the escape speed for the range $1.3 < R_{\rm t}/R_{\rm co} < 2.1$. The numbers on the observation points represent the mass accretion rate in units of $10^{-8}$~Myr$^{-1}$. Accretion rates between 1.2 and $1.6\times 10^{-8}$~Myr$^{-1}$ are in the range where the jet speed is greater than the escape speed.  In \S~\ref{sec:dust_fan} we show that jet flows with higher speeds eject dust to higher altitudes, where the resulting dust cloud may have the appearance of a puffed-up inner rim. This may explain the behaviour shown in Figure~\ref{fig:figure4}, where increasing mass accretion rates produce higher jet flow speeds which, in turn, produce higher inner disc rims. 
    }
    \label{fig:figure12}
\end{figure}

We plot in Figure~\ref{fig:figure12} the jet speed as a ratio of the escape speed, $\varv_{esc}$, where for material in a Keplerian orbit around a star:
\begin{equation}
    \varv_{esc}(r) = \sqrt{\frac{GM_\star}{r}} \, ,
    \label{eqn:38}
\end{equation}
which is simply the Keplerian speed. 

The same parameters are used as for Figure~\ref{fig:figure11}, so for the given values of $z_0$, $P_\star$, $\sigma_{\rm D}$, $\bar{\rho}$, and $B_\star$, the flow in Figure~\ref{fig:figure12} reaches escape speed for the approximate range $1.3 < R_{\rm t}/R_{\rm co} < 2.1$. Indeed, it can be shown that for $r > R_{\rm co}$, the ratio of the jet speed to the escape speed has a maximum at $r_{\rm m} = 2^{2/3}R_{\rm co} \approx 1.59 R_{\rm co}$. In Figure~\ref{fig:figure12}, we also plot the observed values of mass accretion rates in LLRL~31: 0.25, 0.4, 1.2, 1.5 and $1.6\times 10^{-8}$~Myr$^{-1}$. These accretion rates give approximate values of the inner accretion disc radius, $R_{\rm t}$, via equation~\ref{eq:equation5}. As can be seen from Figure~\ref{fig:figure12}, the deduced values of $R_{\rm t}/R_{\rm co}$ from the observed accretion rates move towards the maximum flow speed at $1.59 R_{\rm co}$. 

In terms of puffed up inner discs, it can be seen from Figure~\ref{fig:figure4}, that the maximum inner rim height occurs near $1.7 R_{\rm co}$, this is approaching the region where the jet speed relative to the escape speed is a local maximum. 

As we will show in the next section, increasing jet ejection speeds forces dust particles to reach higher altitudes and produces regions of raised dust above and below the inner accretion disc that may have the appearance of puffed-up inner rims. So, higher mass accretion rates force the inner rim closer to the star thereby increasing the jet flow speed, which then increases the height of the ejected dust fan and increases the perceived height of the disc inner rim.
%

\section{DUST FANS AND THE CATAPULT EFFECT}
\label{sec:dust_fan}

In this section we examine the motion of particles that are entrained with a disc outflow or accretional inflow. As mentioned in the introduction, we assume that dust particles are present in the disc at or near the base of the flow, and that dust may condense in the flow in regions where the temperature and gas densities allow dust to nucleate from the gas. 

 Numerical simulations of MHD outflows show that the outflows tend to leave the disc at an angle which is not perpendicular to the disc midplane \citep{2012MNRAS.423L..45P,2013A&A...550A..99Z,2018NewA...62...94R}. However, as a base case, we assume that the initial direction of the entrained particles is perpendicular to the disc midplane. This is done to illustrate the potential effect of centrifugal acceleration moving the particles away from the flow direction.

\subsection{Particle Ejection Model}

In our model, the dust particles are initially in a circular Keplerian orbit at or near the inner truncation radius of the disc. As discussed in  \S~\ref{subsec:jet_exhaust_speed}, we assume that the accretional inflow onto the star and/or the protostellar jet will tend to flow away from the disc in a direction approximately perpendicular to the disc midplane. This will give the particles an initial `boost' velocity that is assumed to be primarily in the $z$ direction. The subsequent motion of the particles is described by the equations given in Appendix~\ref{Particle_Motion}. Although the particles start with a Keplerian azimuthal velocity, the azimuthal velocity of the dust particles as they move above the disc may change due to the gas, in the stellar magnetosphere, co-rotating with the star.

\begin{figure}
	\includegraphics[width=\columnwidth]{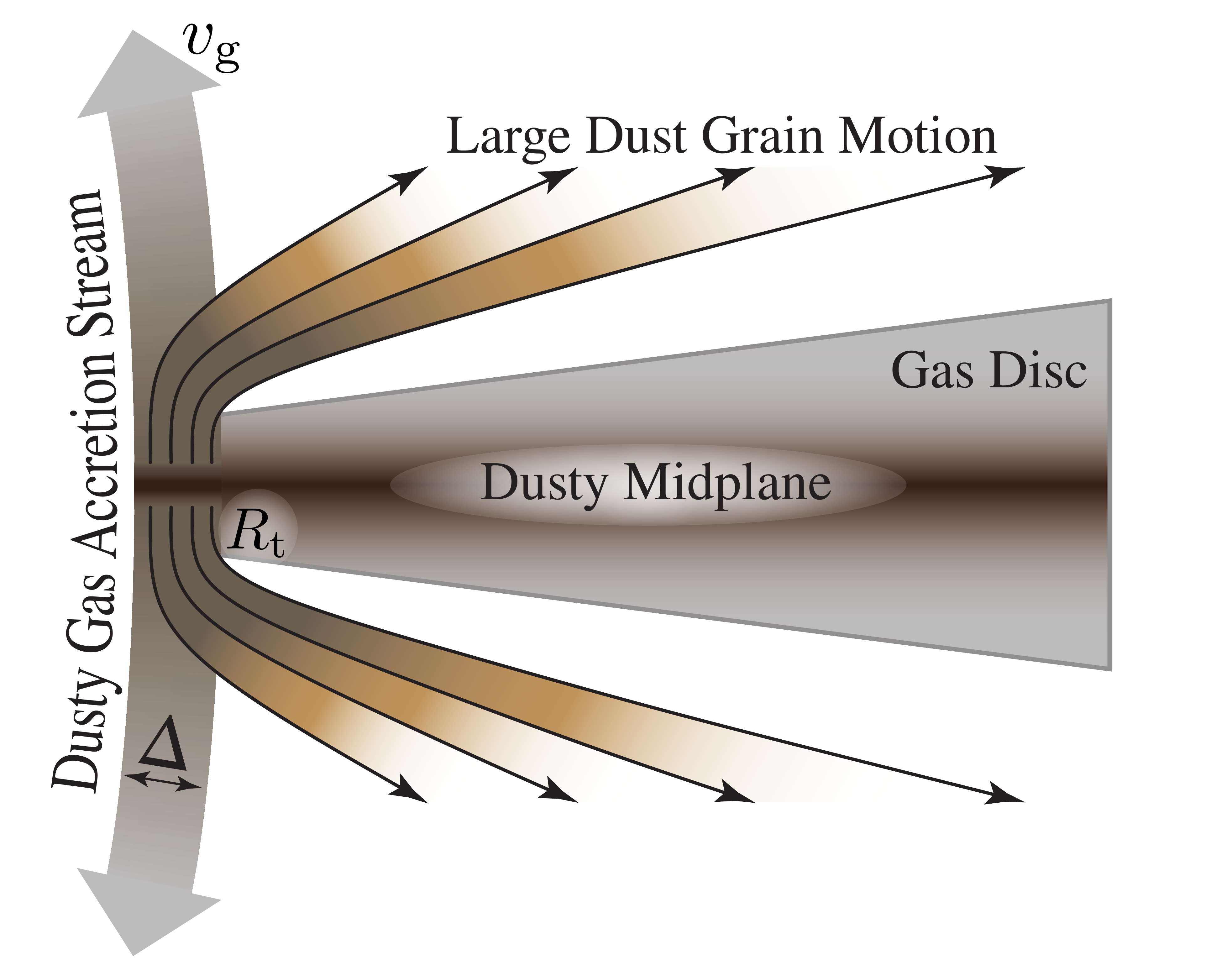}
    \caption{A schematic of the dust fan model: a gas flow of thickness $\Delta$ and speed $\varv_{\rm g}$ is generated at the inner edge of the accretion disc that ejects dust particles, either  from the disc's dusty midplane or  dust particles that condense in the flow. Dust particles with sufficiently high angular momentum can escape  the flow and are ejected across the face of the disc. Note that in the  launch area, the disc is assumed to be magnetically compressed.
    }
    \label{fig:figure13}
\end{figure}

At the truncation radius, the ejected gas and dust will initially tend to flow along the stellar magnetic field lines with speed $\varv_{\rm g}$ in the $z$ direction in an (assumed) axisymmetric channel of initial width $\Delta$ ( Figure~\ref{fig:figure13}, and equation~\ref{eqn:52}). As the particles move away from the disc midplane, the radial gravitational force will decrease, but their angular momentum remains constant. The resulting centrifugal force potentially flings the particles on a ballistic trajectory across the face of the disc, as is shown schematically in Figure~\ref{fig:figure13}.

In our simulations, as discussed in Appendix~\ref{Particle_Motion}, the dust particles are placed at the inner edge of the gas flow, so they have to travel through the entire width of the  outflow  before they can escape the flow. The particles are subject to gas drag (equation~\ref{eq:C_D}) when they are embedded in the gas flow. We set the gas drag to zero when or if the dust particles leave the initial gas flow and are moving out across the face of the accreton disc.

%

\subsection{Potential Projectile Motion}

\begin{figure}
 \begin{subfigure}{\linewidth}\centering
 \includegraphics[width=\linewidth]{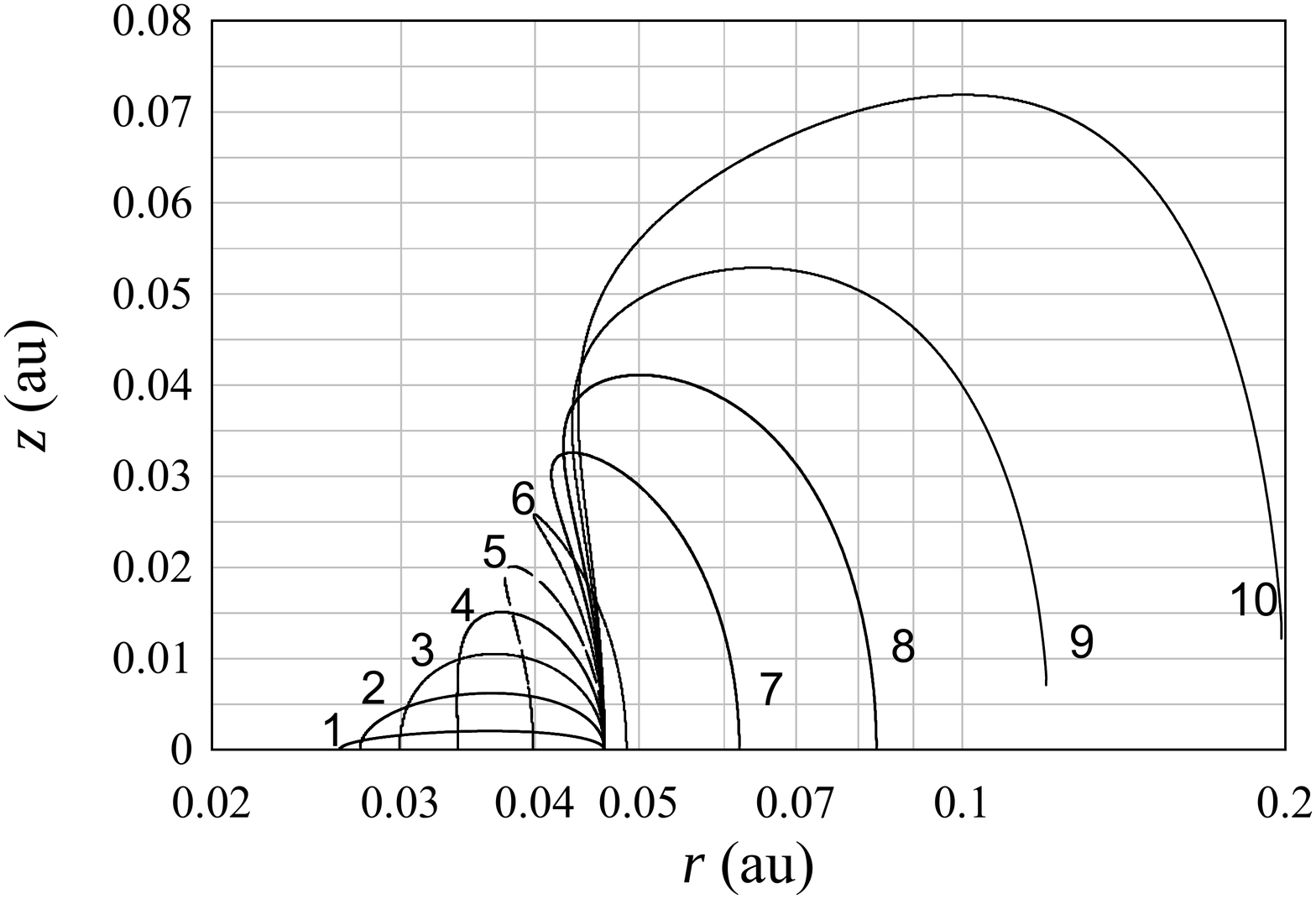}\caption{$r < R_{\rm co}$. Flight paths of particles ejected from $r = 0.9R_{\rm co}$.}
  \end{subfigure}
 \begin{subfigure}{\linewidth}\centering
 \includegraphics[width=\linewidth]{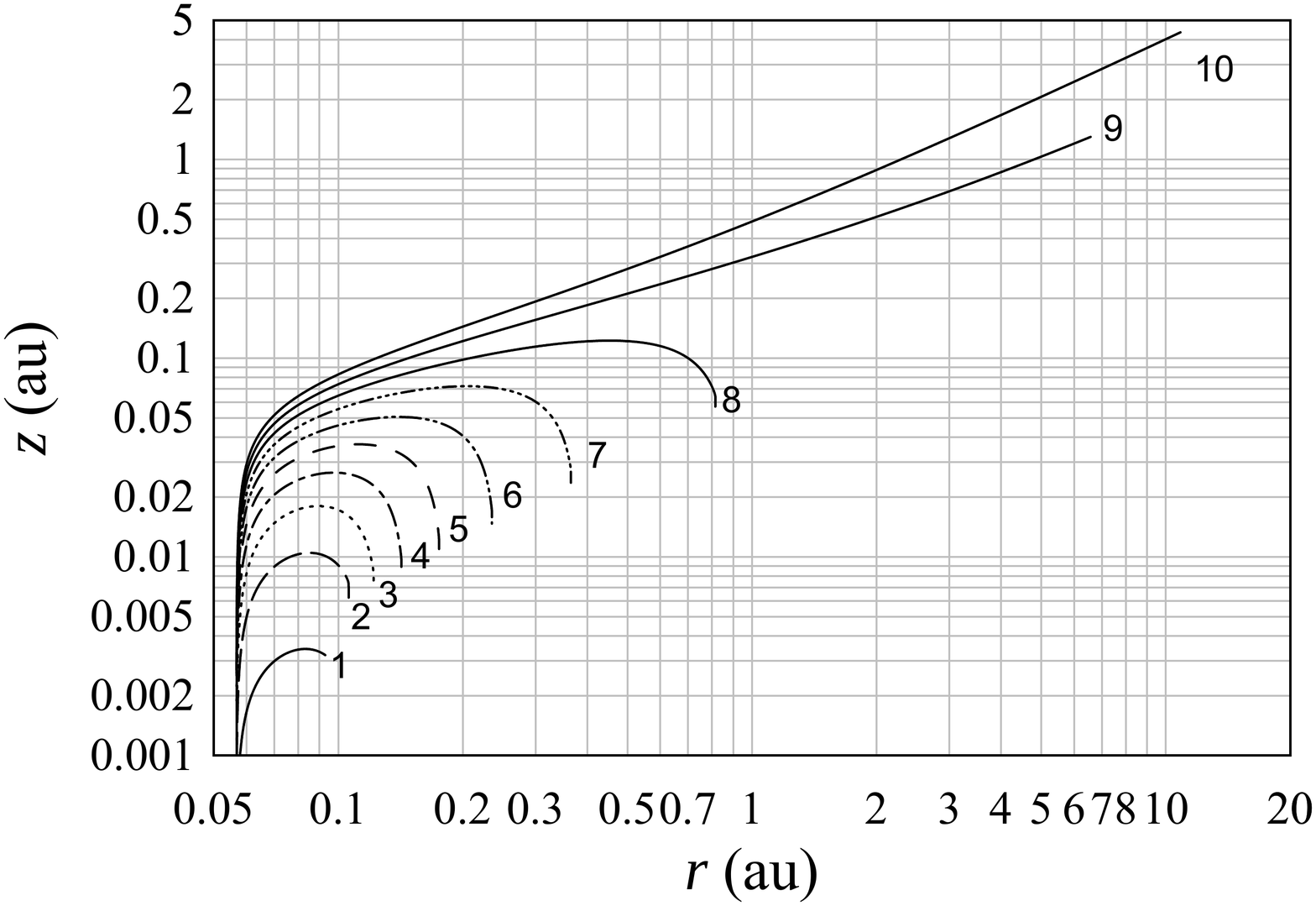}\caption{$r > R_{\rm co}$. Flight paths of particles ejected from $r = 1.1R_{\rm co}$}
   \end{subfigure}
    \caption{Trajectories for 1 mm diameter, silicate-like particles that are ejected perpendicular to the midplane. (a) Ten particles are ejected from $r = 0.9R_{\rm co} = $0.0464 au. Particles 1 to 5 have vertical ejections speeds of 8.75, 26.2, 43.7, 61.2 and 78.7 kms$^{-1}$, respectively, and these particles subsequently fall back towards the star. Particles 6 to 10 have vertical ejection speeds of 96.2, 113.7, 131.2, 148.6 and 166 kms$^{-1}$, respectively, and they reach altitudes comparable to the observed puffed up inner rims before they fall back towards the disc midplane away from the star. (b)  Ten particles are ejected from $r = 1.1R_{\rm co} = $0.0567 au with speeds ranging from 7.9 to 150 kms$^{-1}$. For this case all the particles subsequently move away from the star. 
    }
    \label{fig:figure16}
\end{figure}

As an example of the potential projectile paths, Figure~\ref{fig:figure16} shows the generic paths of ten, 1 mm diameter, silicate-like particles ejected with initial vertical speeds, $\varv_{\rm pz}$, of 0.05, 0.15, 0.25 ... 0.95 the local Keplerian orbital speed.

Figure~\ref{fig:figure16}(a) shows the case where the initial launch distance from LRLL~31 is assumed to be 0.0464 au, which is 90\% of the co-rotation radius $R_{\rm co} \approx 0.0516$ au. The particles are given a stellar co-rotation azimuthal velocity of 148.5 kms$^{-1}$ which is 85\% of the local Keplerian orbital speed of 174.9 kms$^{-1}$. The first five particles (1 to 5), with vertical ejection speeds of 8.75, 26.2, 43.7, 61.2 and 78.7 kms$^{-1}$, fall back towards the star. The remaining five particles (ejection speeds: 96.2, 113.7, 131.2, 148.6 and 166 kms$^{-1}$) reach altitudes comparable to the observed puffed up inner rims and fall back to the disc at distances further away from the star. 

It is unknown whether stellar outflows or accretional infall can produce such high particle ejection speeds, but there is an obvious direct proportionality between the height of the projectile path above the disc midplane and the magnitude of the vertical ejection speed, $\varv_{\rm pz}$.

Figure~\ref{fig:figure16}(b) shows the case where the initial launch distance from the LRLL~31 is now set to 0.0567 au which is 110\% of the co-rotation radius $R_{\rm co}$. The particles are now given a stellar co-rotation azimuthal velocity of 181.5 kms$^{-1}$ or about 114\% of the local Keplerian orbital speed of 158.2 kms$^{-1}$. In this case, all the particles (1 through 10) are ejected to larger distances with initial ejection speeds of 7.9, 23.7, 39.5, 55.4, 71.2, 87, 103, 119, 134 and 150 kms$^{-1}$.

In principle, it is relatively easy to obtain dust particle trajectories (Figure~\ref{fig:figure16}. Figure~\ref{fig:figurecatpult}) that have similar heights to the maximum puffed-up inner rim heights shown in Figures~\ref{fig:figure3} and \ref{fig:figure4}. The height of the dust particle trajectory is simply dependent on the initial $z$ speed of the particle moving away from the disc midplane.

\subsection{Dust Fan as a Puffed-Up Inner Rim}

\begin{figure}
	\includegraphics[width=\columnwidth]{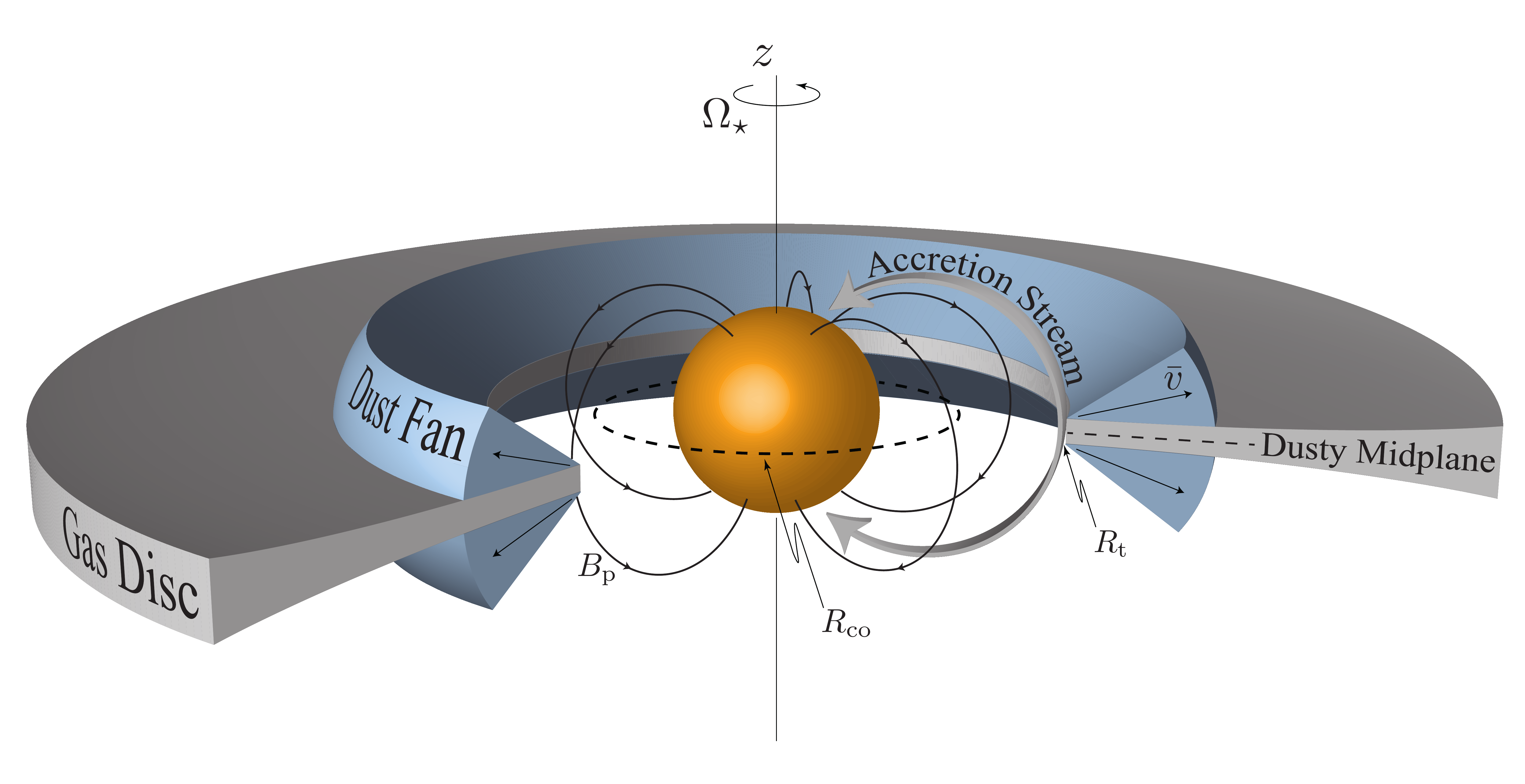}
    \caption{Dust fan produced from accretional gas flow onto a star or by a protostellar jet wind. Dust particles are initially entrained in the flow, but are then catapulted out of the flow due to centrifugal force. The dust may be initially in the disc and/or may condense in the outflow. The outward average speed of the dust is denoted by $\bar{\varv}$. Also shown are the corotation radius, $R_{\rm co}$, the inner truncation radius of the disc, $R_{\rm t}$, the poloidal magnetic field of the star, $B_{\rm p}$, and the angular velocity of the star, $\Omega_\star$}
    \label{fig:figure17}
\end{figure}

The particle paths given in Figure~\ref{fig:figure16} are akin to a fan of dust particles emerging from the inner edge of the disc. This idea is displayed schematically in Figure~\ref{fig:figure17}. If the number density and size of the ejected dust grains is suitably high and large respectively, then the dust fan can become optically thick and may appear to a distant observer like a ``puffed-up'' inner rim of the disc. 

\citet{2012ApJ...758..100B} have suggested a similar scenario, where the puffed-up inner rim is produced by dust entrained in a protostellar jetflow from the inner regions of a disc surrounding a young star.  \citet{2011ApJ...733L..32P} provided observational evidence that is consistent with the Bans and K\"{o}nigl model when they observed crystalline forsterite dust in the near neighbourhood of the young stellar system HOPS~68, where they suggest this dust was transported from the inner disc regions of HOPS~68 into the surrounding molecular cloud via a jet flow.

The model discussed here is different from the Bans and K\"{o}nigl model as we have the dust initially entrained in an accretion flow onto a star and/or a jet flow emerging from a disc surrounding a star, where the dust subsequently leaves either flow due to centrifugal forces and follows a ballistic path across the face of the disc. Dust in the Bans and and K\"{o}nigl model tends to stay entrained in the flow.
 
 Our "catapult'' model is consistent with the observational results given in \citet{2012ApJ...744..118J}, who deduced that silicate dust was moving away from the protostar Ex~Lup across the face of the disc at radial speeds of about 38 km s$^{-1}$. As discussed in Appendix~\ref{Particle_Motion}, our model can produce such radial speeds for ejected dust particles. For example, Figure~\ref{fig:figure15} shows an extreme case where a radial speed of over 200 km s$^{-1}$ is obtained. It is also possible to obtain an approximate analytic equation that explains how a particular radial speed may be obtained from our particle ejection process (equation~\ref{eqn:63}). 
%

\section{DUST FAN SEDS}
\label{sec:dust_fan_sed}
It is useful to examine whether a dust fan/puffed up inner disc is a feasible explanation for the ``see saw'' oscillation in the mid-infrared spectrum of LLRL~31  (Figure\ref{fig:figure1}). To study this idea, we use the Monte Carlo radiative transfer code Hochunk3D \citep{2003ApJ...591.1049W, 2003ApJ...598.1079W, 2013ApJS..207...30W} to simulate a protostellar system with a puffed-up inner rim.

In a later paper, we hope to simulate a bipolar outflow with a full dust fan. However, as a start we use a schematic replication of the dust fan effect by assuming that the dust is ejected in a thin channel or thin wall located at the inner edge of the disc, where the channel is perpendicular to the midplane of the disc, which is a slight modification of Figure~\ref{fig:figure17}. 

\begin{figure}
	\includegraphics[width=\columnwidth]{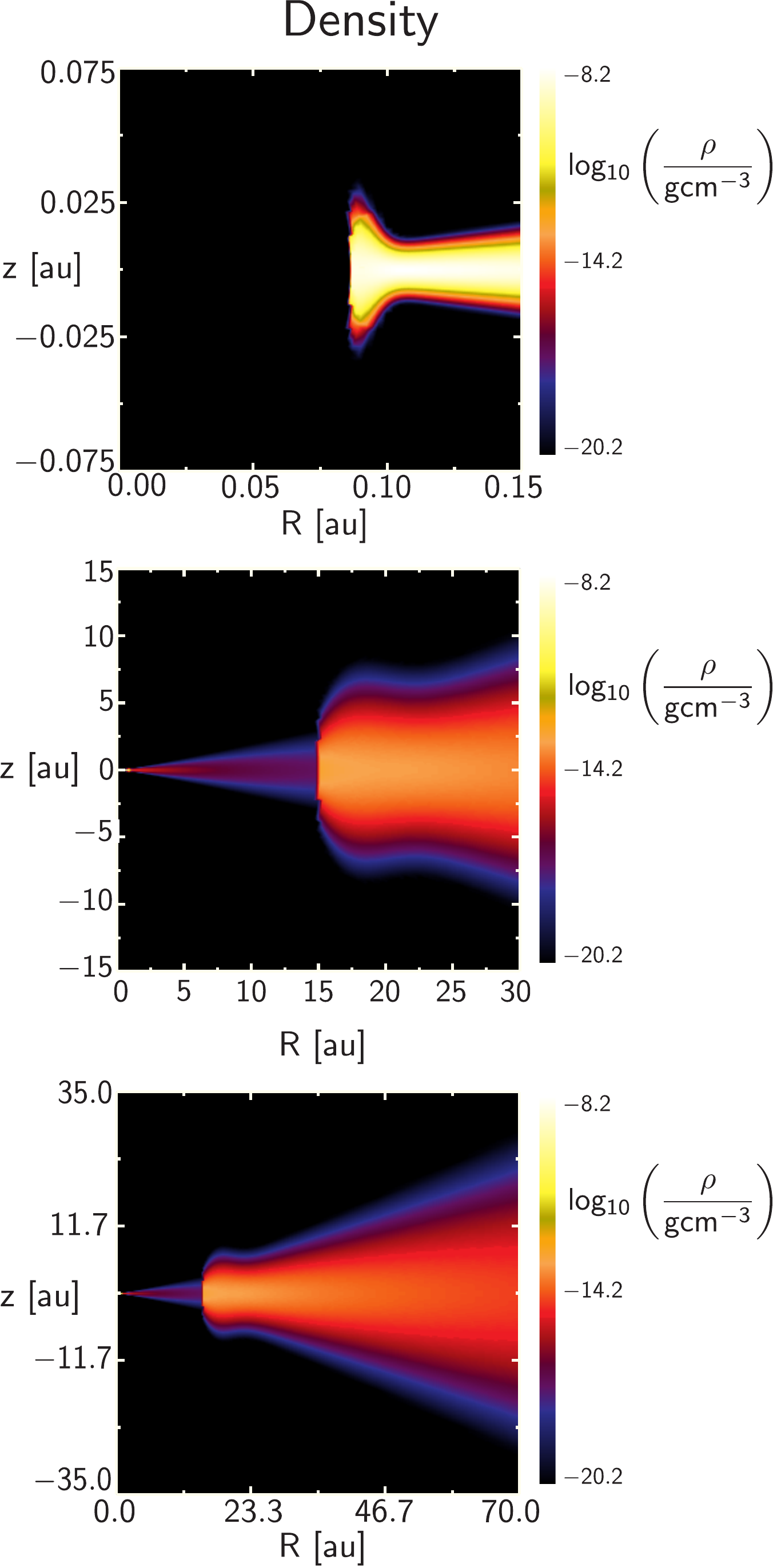}
    \caption{Density cross sections of our model protostellar disc surrounding LLRL~31. The three frames show (a)  the inner disc with a puffed up inner rim of 3.6 scale heights, (b) the inner region of the disc with a gap between 1 and 15~au, and (c) the large scale outer regions of the disc. Length units are in au and the  colour bar shows the density scale in units of $\log_{10}$ g cm$^{-3}$.
    }
    \label{fig:figure19}
\end{figure}

Hochunk3D has an inbuilt model of a protostellar disc and makes provision for puffed inner rim walls through an additive scale height. This is added to the isothermal scale height:
\begin{equation}
    h_{\rm rim}(r) = h_{\rm fid} \left(H_{\rm rim} \exp{\left(-\left(\frac{r-R_{\rm t}}{L_{\rm rim}}\right)^2\right)} \right) \, ,
    \label{eqn:65}
\end{equation}
where $H_{\rm rim}$ is the puffed-rim scale factor, $h_{\rm fid}$ is the fiducial scale height (i.e. the isothermal scale height at radius $r$), and $L_{\rm rim}$ is the radial scale length for the puffed inner rim. In our case, we have $R_{\rm t} = 0.086$~au, $h_{\rm fid} = 0.00137$~au and $L_{\rm rim} = 0.01$~au. The first two values are representative for a LLRL~31 high accretion scenario, while the latter value is a tentative minimum thickness for a puffed up inner rim based on the simulations shown in Figure~\ref{fig:figure16}. These parameters and equation~\ref{eqn:65} provide us with a channel-like, puffed-up inner rim, as shown in Figure~\ref{fig:figure19}.

 Figure~\ref{fig:figure19} shows the density cross section of our model protostellar disc that surrounds LLRL~31. The  inner disc is shown in Figure~\ref{fig:figure19}(a), while Figure~\ref{fig:figure19}(b) shows the gap in the disc between 1 and 15~au. The mass density  ranges from $10^{-8.3}$ g cm$^{-3}$ to $10^{-20.3}$ g cm$^{-3}$. 
%

\begin{figure}
	\includegraphics[width=\columnwidth]{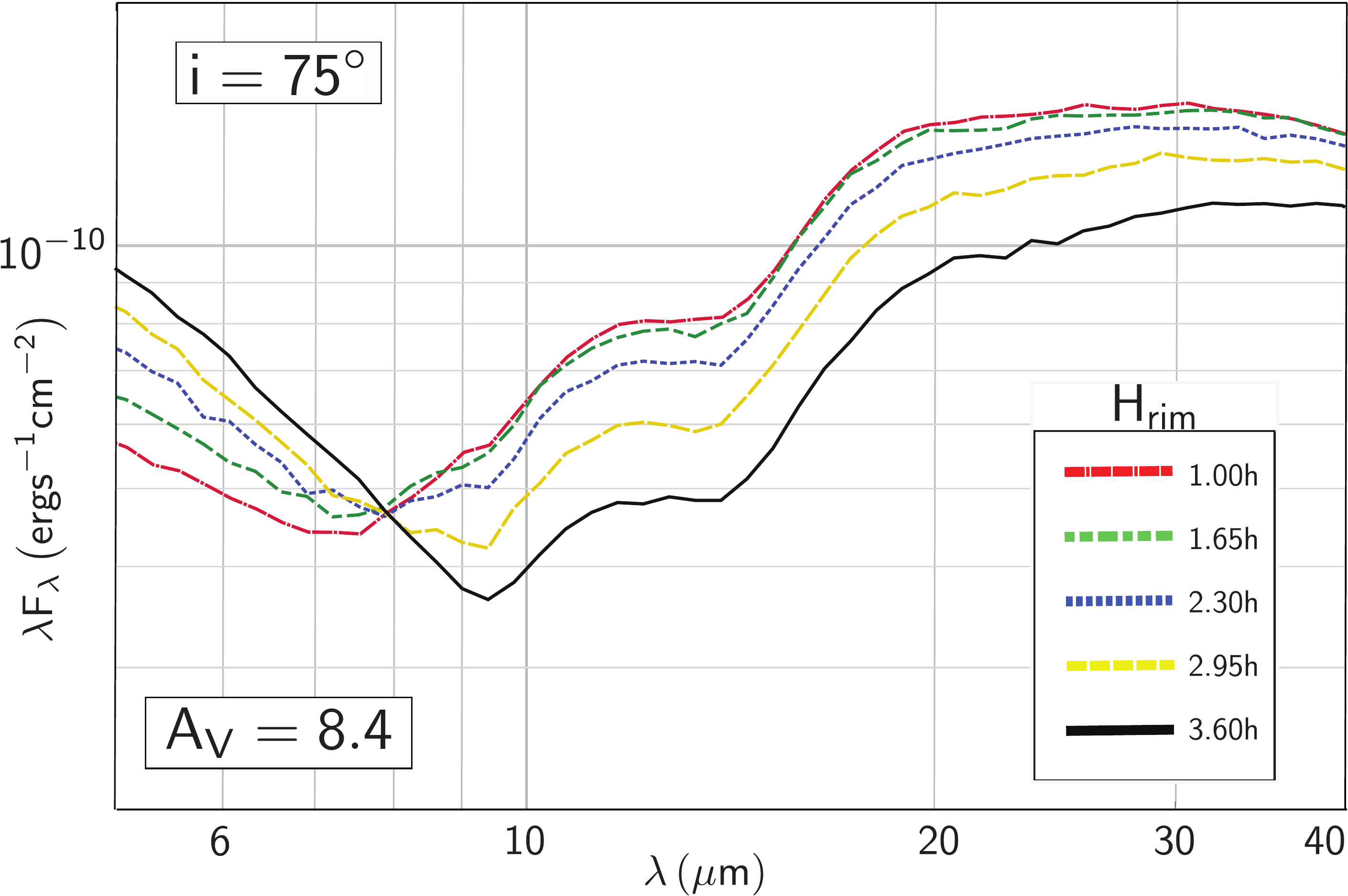}
    \caption{Resulting spectral energy distributions (SEDs) for a protostellar disc that is modelled on LRLL~31 with varying inner him height. The SEDs show similar behaviour to observations as displayed in Figure 1, where the higher puffed up inner rims produce more flux in the 5 to 8~$\mu$m range and less flux in the 8 to 40~$\mu$m range. The SED pivot point is around 8~$\mu$m. The disc is inclined at an angle of 75\degree with an assumed extinction of $A_V$ = 8.4. The inset displays the heights of the puffed-up inner rim in terms of natural scale height, $h$, at the inner rim.
    }
    \label{fig:figure20}
\end{figure}

To a first approximation, the puffed up inner rim may be produced by a gas flow with a speed given by the parameterised form of equation~\ref{eqn:53}
\begin{align}
    \varv_{\rm g} &= \frac{\dot{M}_{\rm a}}{4\pi R_{\rm t} \rho_{\rm g} \Delta} \nonumber \\
    & = 22.4 \, {\rm ms}^{-1} \frac{(\dot{M}_{\rm a})/10^{-8}{\rm M}_\odot{\rm yr}^{-1}}{(R_{\rm t}/0.1\,{\rm au})(\rho_{\rm g}/10^{-7}\,{\rm kg m}^{-3})(\Delta/0.01\,{\rm au})} \, .
\end{align}
To compute the relevant spectral energy distribution, we consider a range of inner rim heights, $H_{\rm rim}$, from 1 to 3.6 scale heights, where we note that the higher rim heights correspond to higher outflow ejection speeds. The resulting SEDs are shown in Figure~\ref{fig:figure20}, a puffed up inner rim has a greater radiative flux in the 5 to 8~$\mu$m range and a lower flux in the 8 to 40~$\mu$m region relative to smaller inner rims. 

The results in Figure~\ref{fig:figure20} are qualitatively similar to those shown in Figure~\ref{fig:figure1}. The pivot point in both figures is around 8~$\mu$m. The correspondence between model and observations is close, but not exact since we wish only to demonstrate that a puffed rim, as inspired by our dust fan model, can approximately reproduce observations. 

In our model, the changing value of $H_{\rm rim}$ would be directly due to the changing speed of the jet flow generated at the inner disc rim. As discussed in \S~\ref{subsec:jet_exhaust_speed} there are two local maxima for the jet flow speed: a  maximum when the inner disc rim is touching the stellar surface and another maximum, relative to the escape speed, when the inner disc rim is at the approximate distance of 1.59$R_{\rm co}$ from the centre of the star. 

As the accretion rate changes and the inner rim approaches the 1.59$R_{\rm co}$ point then the jet flow speed increases. As a consequence, as is shown in \S~\ref{sec:dust_fan}, ejected dust particles can reach higher altitudes after the particles are catapulted from the flow and subsequently move radially away from the star across the face of the accretion disc. It is the increase in the jet flow speed that produces the increase in the height of the dust fan and the consequent, perceived increase in the height of the puffed-up inner rim. We show more detailed radiative transfer modelling of LRLL~31 in \citet{2019arXiv190808703B}. 
%

\section{CONCLUSIONS}

In this study, we have derived a model for the mid-infrared variability of the young stellar system LLRL~31, which displays a decrease in the 8 to 40~$\mu$m flux when there is an increase in the 1 to 8~$\mu$m flux and vice versa (Figure 1). We have concluded, as have other authors, that this variability is primarily due to the perceived change in the rim height of the inner disc surrounding the central star. 

When the  inner rim height is perceived to increase, the inner wall is heated by stellar radiation and there is an increase in the 1 to 8~$\mu$m flux. A puffed inner rim also produces a shadow that obscures the outer disc, thereby resulting in the decrease in the 8 to 40~$\mu$m flux. Similarly, the opposite occurs when the perceived inner disc rim decreases in height. We say, ``perceived inner disc", because this deduced change in height may not actually be occurring to the disc itself, but may be produced by an optically thick fan of dust that is ejected from the disc due to the accretion of dust and gas onto the star.

As accreting gas in the disc moves towards the star, there is an interaction between the poloidal, approximately dipole, stellar magnetic field and the disc. The resulting toroidal disc field can produce an outflow such that gas and dust is ejected with a component of the flow that is perpendicular to the disc midplane. As this dusty gas moves away from the disc midplane, the dust may centrifugally decouple from the gas flow and move on a ballistic trajectory across the face of the disc. We suggest that the resulting inner disc dust fan may produce a shadow over the outer disc and provide the distant perception of a ``puffed-up" inner rim. The dust may be resident in the inner disc rim and/or it may have condensed in the outflow.

This model has allowed us to derive a number of analytic formulae: the speed of the jet flow produced from a toroidal magnetic disc field at or near the surface on an accretion disc (equation~\ref{eqn:35}), the time dependent disc toroidal field (equation~\ref{eqn:A14}), the disc magnetic twist (equation~\ref{eqn:20}), the size of the disc region where the magnetic twist is likely to be stable (equation~\ref{eqn:25}), the distance from the star for the maximal jet flow speed (equations~\ref{eqn:36} and \ref{eqn:37}) plus the radial speeds of particles ejected from the jet flow (equations~\ref{eqn:60} and \ref{eqn:63}).

This theoretical work indicates that the major timescale for this process is the magnetic diffusion time scale of the inner disc (equation~\ref{eqn:15}) due to the wind up of the stellar magnetosphere into a disc toroidal field. This timescale is dependent on the conductivity of the inner disc, but plausible conductivity values suggest a timescale of days, which is consistent with observations (Figure~\ref{fig:figure2}).

As such, puffed-up inner rims may be symptomatic of the magnetic interaction between a star and surrounding accretion disc. They are also indicative of the radial transport of processed dust from the inner regions of an accretion disc to the outer regions. Such a result is consistent with the Stardust mission results, where the dust obtained from Comet Wild~2 had been exposed to temperatures greater than 1000~K \citep{2014AREPS..42..179B}. The ballistic radial transport of dust has also been observed via the Spitzer Space Telescope \citep{2011ApJ...733L..32P, 2012ApJ...744..118J}. Puffed-up inner rims and the subsequent radial transport of dust may be an intimately intertwined process that is applicable to many young stellar systems including the some of the very first radial transport processes in the early Solar System.

\section*{Acknowledgements}

 The SED modelling work was performed on the gSTAR national supercomputing facility at Swinburne University of Technology. gSTAR is funded by Swinburne and the Australian Government\textquotesingle s Education Investment Fund. GRB acknowledges the  support of  a Swinburne University Postgraduate Research Award (SUPRA). We gratefully acknowledge the constructive suggestions and criticisms from the anonymous reviewers which were very helpful in improving the quality of this paper.




\bibliographystyle{mnras}
\bibliography{lighthouse} 

\begin{thebibliography}{}
\makeatletter
\relax
\def\mn@urlcharsother{\let\do\@makeother \do\$\do\&\do\#\do\^\do\_\do\%\do\~}
\def\mn@doi{\begingroup\mn@urlcharsother \@ifnextchar [ {\mn@doi@}
  {\mn@doi@[]}}
\def\mn@doi@[#1]#2{\def\@tempa{#1}\ifx\@tempa\@empty \href
  {http://dx.doi.org/#2} {doi:#2}\else \href {http://dx.doi.org/#2} {#1}\fi
  \endgroup}
\def\mn@eprint#1#2{\mn@eprint@#1:#2::\@nil}
\def\mn@eprint@arXiv#1{\href {http://arxiv.org/abs/#1} {{\tt arXiv:#1}}}
\def\mn@eprint@dblp#1{\href {http://dblp.uni-trier.de/rec/bibtex/#1.xml}
  {dblp:#1}}
\def\mn@eprint@#1:#2:#3:#4\@nil{\def\@tempa {#1}\def\@tempb {#2}\def\@tempc
  {#3}\ifx \@tempc \@empty \let \@tempc \@tempb \let \@tempb \@tempa \fi \ifx
  \@tempb \@empty \def\@tempb {arXiv}\fi \@ifundefined
  {mn@eprint@\@tempb}{\@tempb:\@tempc}{\expandafter \expandafter \csname
  mn@eprint@\@tempb\endcsname \expandafter{\@tempc}}}

\bibitem[\protect\citeauthoryear{{Adams} \& {Gregory}}{{Adams} \&
  {Gregory}}{2012}]{2012ApJ...744...55A}
{Adams} F.~C.,  {Gregory} S.~G.,  2012, \mn@doi [\apj]
  {10.1088/0004-637X/744/1/55}, \href
  {https://ui.adsabs.harvard.edu/\#abs/2012ApJ...744...55A} {744, 55}

\bibitem[\protect\citeauthoryear{{Bans} \& {K{\"o}nigl}}{{Bans} \&
  {K{\"o}nigl}}{2012}]{2012ApJ...758..100B}
{Bans} A.,  {K{\"o}nigl} A.,  2012, \mn@doi [\apj]
  {10.1088/0004-637X/758/2/100}, \href
  {https://ui.adsabs.harvard.edu/\#abs/2012ApJ...758..100B} {758, 100}

\bibitem[\protect\citeauthoryear{{Bouvier}, {Alencar}, {Harries}, {Johns-
  Krull}  \& {Romanova}}{{Bouvier} et~al.}{2007}]{2007prpl.conf..479B}
{Bouvier} J.,  {Alencar} S.~H.~P.,  {Harries} T.~J.,  {Johns- Krull} C.~M.,
  {Romanova} M.~M.,  2007, in {Reipurth} B.,  {Jewitt} D.,   {Keil} K.,  eds,
  Protostars and Planets V. p.~479 (\mn@eprint {arXiv} {astro-ph/0603498})

\bibitem[\protect\citeauthoryear{{Brownlee}}{{Brownlee}}{2014}]{2014AREPS..42..179B}
{Brownlee} D.,  2014, \mn@doi [Annual Review of Earth and Planetary Sciences]
  {10.1146/annurev-earth-050212-124203}, \href
  {https://ui.adsabs.harvard.edu/\#abs/2014AREPS..42..179B} {42, 179}

\bibitem[\protect\citeauthoryear{{Bryan}, {Maddison}  \& {Liffman}}{{Bryan}
  et~al.}{2019}]{2019arXiv190808703B}
{Bryan} G.~R.,  {Maddison} S.~T.,   {Liffman} K.,  2019, arXiv e-prints, \href
  {https://ui.adsabs.harvard.edu/abs/2019arXiv190808703B} {p. arXiv:1908.08703}

\bibitem[\protect\citeauthoryear{{Campbell} \& {Heptinstall}}{{Campbell} \&
  {Heptinstall}}{1998}]{1998MNRAS.299...31C}
{Campbell} C.~G.,  {Heptinstall} P.~M.,  1998, \mn@doi [\mnras]
  {10.1046/j.1365-8711.1998.01576.x}, \href
  {https://ui.adsabs.harvard.edu/abs/1998MNRAS.299...31C} {299, 31}

\bibitem[\protect\citeauthoryear{{Carr}}{{Carr}}{2007}]{2007IAUS..243..135C}
{Carr} J.~S.,  2007, in {Bouvier} J.,  {Appenzeller} I.,  eds,  IAU Symposium
  Vol. 243, Star-Disk Interaction in Young Stars. pp 135--146,
  \mn@doi{10.1017/S1743921307009490}

\bibitem[\protect\citeauthoryear{{Espaillat} et~al.,}{{Espaillat}
  et~al.}{2010}]{2010ApJ...717..441E}
{Espaillat} C.,  et~al., 2010, \mn@doi [\apj] {10.1088/0004-637X/717/1/441},
  \href {https://ui.adsabs.harvard.edu/#abs/2010ApJ...717..441E} {717, 441}

\bibitem[\protect\citeauthoryear{{Espaillat} et~al.,}{{Espaillat}
  et~al.}{2014}]{2014prpl.conf..497E}
{Espaillat} C.,  et~al., 2014, in {Beuther} H.,  {Klessen} R.~S.,  {Dullemond}
  C.~P.,   {Henning} T.,  eds, Protostars and Planets VI. p.~497 (\mn@eprint
  {arXiv} {1402.7103}), \mn@doi{10.2458/azu_uapress_9780816531240-ch022}

\bibitem[\protect\citeauthoryear{{Flaherty} \& {Muzerolle}}{{Flaherty} \&
  {Muzerolle}}{2010}]{2010ApJ...719.1733F}
{Flaherty} K.~M.,  {Muzerolle} J.,  2010, \mn@doi [\apj]
  {10.1088/0004-637X/719/2/1733}, \href
  {https://ui.adsabs.harvard.edu/#abs/2010ApJ...719.1733F} {719, 1733}

\bibitem[\protect\citeauthoryear{{Flaherty}, {Muzerolle}, {Rieke}, {Gutermuth},
  {Balog}, {Herbst}, {Megeath}  \& {Kun}}{{Flaherty}
  et~al.}{2011}]{2011ApJ...732...83F}
{Flaherty} K.~M.,  {Muzerolle} J.,  {Rieke} G.,  {Gutermuth} R.,  {Balog} Z.,
  {Herbst} W.,  {Megeath} S.~T.,   {Kun} M.,  2011, \mn@doi [\apj]
  {10.1088/0004-637X/732/2/83}, \href
  {https://ui.adsabs.harvard.edu/#abs/2011ApJ...732...83F} {732, 83}

\bibitem[\protect\citeauthoryear{{Frank}, {King}  \& {Raine}}{{Frank}
  et~al.}{2002}]{2002apa..book.....F}
{Frank} J.,  {King} A.,   {Raine} D.~J.,  2002, {Accretion Power in
  Astrophysics: Third Edition}.
Cambridge University Press

\bibitem[\protect\citeauthoryear{{Friedjung}}{{Friedjung}}{1985}]{1985A&A...146..366F}
{Friedjung} M.,  1985, \aap, \href
  {https://ui.adsabs.harvard.edu/\#abs/1985A&A...146..366F} {146, 366}

\bibitem[\protect\citeauthoryear{{Ghosh} \& {Lamb}}{{Ghosh} \&
  {Lamb}}{1978}]{1978ApJ...223L..83G}
{Ghosh} P.,  {Lamb} F.~K.,  1978, \mn@doi [\apj] {10.1086/182734}, \href
  {https://ui.adsabs.harvard.edu/#abs/1978ApJ...223L..83G} {223, L83}

\bibitem[\protect\citeauthoryear{{Hartmann}}{{Hartmann}}{1998}]{1998apsf.book.....H}
{Hartmann} L.,  1998, {Accretion Processes in Star Formation}.
Cambridge University Press

\bibitem[\protect\citeauthoryear{Hayes \& Probstein}{Hayes \&
  Probstein}{1959}]{hayes_probstein_1959}
Hayes W.~D.,  Probstein R.~F.,  1959, Hypersonic Flow Theory.
Academic Press

\bibitem[\protect\citeauthoryear{{Juh{\'a}sz}, {Prusti}, {{\'A}brah{\'a}m}  \&
  {Dullemond}}{{Juh{\'a}sz} et~al.}{2007}]{2007MNRAS.374.1242J}
{Juh{\'a}sz} A.,  {Prusti} T.,  {{\'A}brah{\'a}m} P.,   {Dullemond} C.~P.,
  2007, \mn@doi [\mnras] {10.1111/j.1365-2966.2006.11208.x}, \href
  {https://ui.adsabs.harvard.edu/#abs/2007MNRAS.374.1242J} {374, 1242}

\bibitem[\protect\citeauthoryear{{Juh{\'a}sz} et~al.,}{{Juh{\'a}sz}
  et~al.}{2012}]{2012ApJ...744..118J}
{Juh{\'a}sz} A.,  et~al., 2012, \mn@doi [\apj] {10.1088/0004-637X/744/2/118},
  \href {https://ui.adsabs.harvard.edu/\#abs/2012ApJ...744..118J} {744, 118}

\bibitem[\protect\citeauthoryear{{Kulkarni} \& {Romanova}}{{Kulkarni} \&
  {Romanova}}{2008}]{2008MNRAS.386..673K}
{Kulkarni} A.~K.,  {Romanova} M.~M.,  2008, \mn@doi [\mnras]
  {10.1111/j.1365-2966.2008.13094.x}, \href
  {https://ui.adsabs.harvard.edu/#abs/2008MNRAS.386..673K} {386, 673}

\bibitem[\protect\citeauthoryear{{Lai} \& {Zhang}}{{Lai} \&
  {Zhang}}{2008}]{2008ApJ...683..949L}
{Lai} D.,  {Zhang} H.,  2008, \mn@doi [\apj] {10.1086/589822}, \href
  {https://ui.adsabs.harvard.edu/#abs/2008ApJ...683..949L} {683, 949}

\bibitem[\protect\citeauthoryear{{Lee}, {Hwang}, {Ching}, {Hirano}, {Lai},
  {Rao}  \& {Ho}}{{Lee} et~al.}{2018}]{2018NatCo...9.4636L}
{Lee} C.-F.,  {Hwang} H.-C.,  {Ching} T.-C.,  {Hirano} N.,  {Lai} S.-P.,  {Rao}
  R.,   {Ho} P. T.~P.,  2018, \mn@doi [Nature Communications]
  {10.1038/s41467-018-07143-8}, \href
  {https://ui.adsabs.harvard.edu/\#abs/2018NatCo...9.4636L} {9, 4636}

\bibitem[\protect\citeauthoryear{{Liffman} \& {Bardou}}{{Liffman} \&
  {Bardou}}{1999}]{1999MNRAS.309..443L}
{Liffman} K.,  {Bardou} A.,  1999, \mn@doi [\mnras]
  {10.1046/j.1365-8711.1999.02852.x}, \href
  {https://ui.adsabs.harvard.edu/\#abs/1999MNRAS.309..443L} {309, 443}

\bibitem[\protect\citeauthoryear{{Liffman} \& {Siora}}{{Liffman} \&
  {Siora}}{1997}]{1997MNRAS.290..629L}
{Liffman} K.,  {Siora} A.,  1997, \mn@doi [\mnras] {10.1093/mnras/290.4.629},
  \href {https://ui.adsabs.harvard.edu/\#abs/1997MNRAS.290..629L} {290, 629}

\bibitem[\protect\citeauthoryear{{Lovelace}, {Mehanian}, {Mobarry}  \&
  {Sulkanen}}{{Lovelace} et~al.}{1986}]{1986ApJS...62....1L}
{Lovelace} R.~V.~E.,  {Mehanian} C.,  {Mobarry} C.~M.,   {Sulkanen} M.~E.,
  1986, \mn@doi [\apjs] {10.1086/191132}, \href
  {https://ui.adsabs.harvard.edu/abs/1986ApJS...62....1L} {62, 1}

\bibitem[\protect\citeauthoryear{{Lovelace}, {Berk}  \&
  {Contopoulos}}{{Lovelace} et~al.}{1991}]{1991ApJ...379..696L}
{Lovelace} R.~V.~E.,  {Berk} H.~L.,   {Contopoulos} J.,  1991, \mn@doi [\apj]
  {10.1086/170544}, \href
  {https://ui.adsabs.harvard.edu/abs/1991ApJ...379..696L} {379, 696}

\bibitem[\protect\citeauthoryear{{Lovelace}, {Romanova}  \&
  {Bisnovatyi-Kogan}}{{Lovelace} et~al.}{1995}]{1995MNRAS.275..244L}
{Lovelace} R.~V.~E.,  {Romanova} M.~M.,   {Bisnovatyi-Kogan} G.~S.,  1995,
  \mn@doi [\mnras] {10.1093/mnras/275.2.244}, \href
  {https://ui.adsabs.harvard.edu/abs/1995MNRAS.275..244L} {275, 244}

\bibitem[\protect\citeauthoryear{{Matt} \& {Pudritz}}{{Matt} \&
  {Pudritz}}{2005}]{2005MNRAS.356..167M}
{Matt} S.,  {Pudritz} R.~E.,  2005, \mn@doi [\mnras]
  {10.1111/j.1365-2966.2004.08431.x}, \href
  {https://ui.adsabs.harvard.edu/abs/2005MNRAS.356..167M} {356, 167}

\bibitem[\protect\citeauthoryear{{Matt}, {Pinz{\'o}n}, {de la Reza}  \&
  {Greene}}{{Matt} et~al.}{2010}]{2010ApJ...714..989M}
{Matt} S.~P.,  {Pinz{\'o}n} G.,  {de la Reza} R.,   {Greene} T.~P.,  2010,
  \mn@doi [\apj] {10.1088/0004-637X/714/2/989}, \href
  {https://ui.adsabs.harvard.edu/abs/2010ApJ...714..989M} {714, 989}

\bibitem[\protect\citeauthoryear{{Muzerolle} et~al.,}{{Muzerolle}
  et~al.}{2009}]{2009ApJ...704L..15M}
{Muzerolle} J.,  et~al., 2009, \mn@doi [\apj] {10.1088/0004-637X/704/1/L15},
  \href {https://ui.adsabs.harvard.edu/#abs/2009ApJ...704L..15M} {704, L15}

\bibitem[\protect\citeauthoryear{{Newman}, {Newman}  \& {Lovelace}}{{Newman}
  et~al.}{1992}]{1992ApJ...392..622N}
{Newman} W.~I.,  {Newman} A.~L.,   {Lovelace} R. V.~E.,  1992, \mn@doi [\apj]
  {10.1086/171462}, \href
  {https://ui.adsabs.harvard.edu/abs/1992ApJ...392..622N} {392, 622}

\bibitem[\protect\citeauthoryear{{Petrov} et~al.,}{{Petrov}
  et~al.}{2019}]{2019MNRAS.483..132P}
{Petrov} P.~P.,  et~al., 2019, \mn@doi [\mnras] {10.1093/mnras/sty3066}, \href
  {https://ui.adsabs.harvard.edu/\#abs/2019MNRAS.483..132P} {483, 132}

\bibitem[\protect\citeauthoryear{{Pinilla} et~al.,}{{Pinilla}
  et~al.}{2014}]{2014A&A...564A..51P}
{Pinilla} P.,  et~al., 2014, \mn@doi [\aap] {10.1051/0004-6361/201323322},
  \href {https://ui.adsabs.harvard.edu/#abs/2014A&A...564A..51P} {564, A51}

\bibitem[\protect\citeauthoryear{{Poteet} et~al.,}{{Poteet}
  et~al.}{2011}]{2011ApJ...733L..32P}
{Poteet} C.~A.,  et~al., 2011, \mn@doi [\apj] {10.1088/2041-8205/733/2/L32},
  \href {https://ui.adsabs.harvard.edu/\#abs/2011ApJ...733L..32P} {733, L32}

\bibitem[\protect\citeauthoryear{{Price}, {Tricco}  \& {Bate}}{{Price}
  et~al.}{2012}]{2012MNRAS.423L..45P}
{Price} D.~J.,  {Tricco} T.~S.,   {Bate} M.~R.,  2012, \mn@doi [\mnras]
  {10.1111/j.1745-3933.2012.01254.x}, \href
  {https://ui.adsabs.harvard.edu/\#abs/2012MNRAS.423L..45P} {423, L45}

\bibitem[\protect\citeauthoryear{{Probstein}}{{Probstein}}{1968}]{1968Probstein}
{Probstein} R.~F.,  1968, in {Lavret\textquotesingle ev} M.~A.,  ed., Problems
  of Hydrodynamics and Continuum Mechanics. SIAM.
pp 568--583

\bibitem[\protect\citeauthoryear{{Romanova}, {Ustyugova}, {Koldoba}  \&
  {Lovelace}}{{Romanova} et~al.}{2009}]{2009MNRAS.399.1802R}
{Romanova} M.~M.,  {Ustyugova} G.~V.,  {Koldoba} A.~V.,   {Lovelace} R.~V.~E.,
  2009, \mn@doi [\mnras] {10.1111/j.1365-2966.2009.15413.x}, \href
  {https://ui.adsabs.harvard.edu/abs/2009MNRAS.399.1802R} {399, 1802}

\bibitem[\protect\citeauthoryear{{Romanova}, {Blinova}, {Ustyugova}, {Koldoba}
  \& {Lovelace}}{{Romanova} et~al.}{2018}]{2018NewA...62...94R}
{Romanova} M.~M.,  {Blinova} A.~A.,  {Ustyugova} G.~V.,  {Koldoba} A.~V.,
  {Lovelace} R.~V.~E.,  2018, \mn@doi [\na] {10.1016/j.newast.2018.01.011},
  \href {https://ui.adsabs.harvard.edu/abs/2018NewA...62...94R} {62, 94}

\bibitem[\protect\citeauthoryear{{Rosenqvist} et~al.,}{{Rosenqvist}
  et~al.}{2009}]{2009P&SS...57.1828R}
{Rosenqvist} L.,  et~al., 2009, \mn@doi [Planetary and Space Science]
  {10.1016/j.pss.2009.01.007}, \href
  {https://ui.adsabs.harvard.edu/\#abs/2009P&SS...57.1828R} {57, 1828}

\bibitem[\protect\citeauthoryear{{Sedlmayr} \& {Dominik}}{{Sedlmayr} \&
  {Dominik}}{1995}]{1995SSRv...73..211S}
{Sedlmayr} E.,  {Dominik} C.,  1995, \mn@doi [\ssr] {10.1007/BF00751238}, \href
  {https://ui.adsabs.harvard.edu/\#abs/1995SSRv...73..211S} {73, 211}

\bibitem[\protect\citeauthoryear{{Sitko} et~al.,}{{Sitko}
  et~al.}{2008}]{2008ApJ...678.1070S}
{Sitko} M.~L.,  et~al., 2008, \mn@doi [\apj] {10.1086/529003}, \href
  {https://ui.adsabs.harvard.edu/#abs/2008ApJ...678.1070S} {678, 1070}

\bibitem[\protect\citeauthoryear{{Turner}, {Carballido}  \& {Sano}}{{Turner}
  et~al.}{2010}]{2010ApJ...708..188T}
{Turner} N.~J.,  {Carballido} A.,   {Sano} T.,  2010, \mn@doi [\apj]
  {10.1088/0004-637X/708/1/188}, \href
  {https://ui.adsabs.harvard.edu/#abs/2010ApJ...708..188T} {708, 188}

\bibitem[\protect\citeauthoryear{{Uzdensky}, {K{\"o}nigl}  \&
  {Litwin}}{{Uzdensky} et~al.}{2002}]{2002ApJ...565.1191U}
{Uzdensky} D.~A.,  {K{\"o}nigl} A.,   {Litwin} C.,  2002, \mn@doi [\apj]
  {10.1086/324720}, \href
  {https://ui.adsabs.harvard.edu/\#abs/2002ApJ...565.1191U} {565, 1191}

\bibitem[\protect\citeauthoryear{{Vinkovi{\'c}}}{{Vinkovi{\'c}}}{2014}]{2014A&A...566A.117V}
{Vinkovi{\'c}} D.,  2014, \mn@doi [\aap] {10.1051/0004-6361/201322008}, \href
  {https://ui.adsabs.harvard.edu/\#abs/2014A&A...566A.117V} {566, A117}

\bibitem[\protect\citeauthoryear{{Whitney}, {Wood}, {Bjorkman}  \&
  {Wolff}}{{Whitney} et~al.}{2003a}]{2003ApJ...591.1049W}
{Whitney} B.~A.,  {Wood} K.,  {Bjorkman} J.~E.,   {Wolff} M.~J.,  2003a,
  \mn@doi [\apj] {10.1086/375415}, \href
  {https://ui.adsabs.harvard.edu/\#abs/2003ApJ...591.1049W} {591, 1049}

\bibitem[\protect\citeauthoryear{{Whitney}, {Wood}, {Bjorkman}  \&
  {Cohen}}{{Whitney} et~al.}{2003b}]{2003ApJ...598.1079W}
{Whitney} B.~A.,  {Wood} K.,  {Bjorkman} J.~E.,   {Cohen} M.,  2003b, \mn@doi
  [\apj] {10.1086/379068}, \href
  {https://ui.adsabs.harvard.edu/\#abs/2003ApJ...598.1079W} {598, 1079}

\bibitem[\protect\citeauthoryear{{Whitney}, {Robitaille}, {Bjorkman}, {Dong},
  {Wolff}, {Wood}  \& {Honor}}{{Whitney} et~al.}{2013}]{2013ApJS..207...30W}
{Whitney} B.~A.,  {Robitaille} T.~P.,  {Bjorkman} J.~E.,  {Dong} R.,  {Wolff}
  M.~J.,  {Wood} K.,   {Honor} J.,  2013, \mn@doi [The Astrophysical Journal
  Supplement Series] {10.1088/0067-0049/207/2/30}, \href
  {https://ui.adsabs.harvard.edu/\#abs/2013ApJS..207...30W} {207, 30}

\bibitem[\protect\citeauthoryear{{Williams} \& {Cieza}}{{Williams} \&
  {Cieza}}{2011}]{2011ARA&A..49...67W}
{Williams} J.~P.,  {Cieza} L.~A.,  2011, \mn@doi [Annual Review of Astronomy
  and Astrophysics] {10.1146/annurev-astro-081710-102548}, \href
  {https://ui.adsabs.harvard.edu/#abs/2011ARA&A..49...67W} {49, 67}

\bibitem[\protect\citeauthoryear{{Zanni} \& {Ferreira}}{{Zanni} \&
  {Ferreira}}{2013}]{2013A&A...550A..99Z}
{Zanni} C.,  {Ferreira} J.,  2013, \mn@doi [\aap]
  {10.1051/0004-6361/201220168}, \href
  {https://ui.adsabs.harvard.edu/abs/2013A&A...550A..99Z} {550, A99}

\makeatother
\end{thebibliography}

\appendix

\section{Toroidal Field Growth}
\label{Toroidal_Field_Growth}

As illustrated in Figure~\ref{fig:figure5}, we make the plausible, first order approximation that the dipole component of the stellar magnetic field co-rotates with the star and this magnetic field interacts with the surrounding accretion disc. As the stellar magnetic field moves over the accretion disc, it will generate a toroidal field in the disc. To obtain a timescale for the development of the disc toroidal field and the subsequent changes in the inner disc, we require the induction equation:
\begin{equation}
    \frac{\partial \mathbf{B}}{\partial t} = \nabla \times (\mathbf{v} \times \mathbf{B}) + \eta \nabla^2\mathbf{B} \, ,
    \label{eqn:A1}
\end{equation}
where $\mathbf{B}$ is the magnetic vector field, $t$ the time and $\eta$ is the magnetic diffusivity with
\begin{equation}
    \eta = \frac{1}{\mu_0 \sigma_{\rm D}} \, .
    \label{eqn:A2}
\end{equation}

The assumption of co-rotation with the star of the stellar magnetic field implies that at the disc surface, the speed of the stellar field, $\varv_{\rm B}$ is
\begin{equation}
    \mathbf{v}_{{\rm B}_\star} = r \Omega \mathbf{\hat{\phi}} \, ,
\end{equation}
where $\hat{\mathbf{\phi}}$ is the unit vector in the cylindrical coordinate azimuthal direction. The azimuthal velocity of the disc relative to the co-rotating stellar field, $\varv_{\rm DB}$, is
\begin{equation}
   \mathbf{v}_{\rm DB}(r) = (\varv_{\rm K}(r) - r\Omega_\star) \hat{\mathbf{\phi}} \ ,
        \label{eqn:A3}
\end{equation}
where $\varv_{\rm K}(r)$ is the Keplerian azimuthal velocity ($\varv_{\rm K}(r) = r\Omega_{\rm K}(r)$). In equation~\ref{eqn:A1},  
\begin{equation}
    \mathbf{v} \times \mathbf{B} = (\varv_{\rm K}(r) - r\Omega_\star) B_{\rm z} \hat{\mathbf{r}}\, ,
    \label{eqn:A4}
\end{equation}
with $B \approx B_{\rm z}$ at or near the midplane of the accretion disc. Thus, assuming axisymmetry
\begin{equation}
    \nabla \times (\mathbf{v} \times \mathbf{B}) = \frac{\partial((\varv_{\rm K}(r)-r\Omega_\star)B_{\rm z})}{\partial z} \hat{\mathbf{\phi}}\, ,
     \label{eqn:A5}
\end{equation}

For this analysis, we are assuming that $\mathbf{B}=(0,B_\phi,B_{\rm z})$ and that the change in $B_\phi$ as a function of $r$ is small relative to the change of $B_\phi$ in $z$. So
\begin{equation}
    \nabla^2\mathbf{B}|_\phi \approx \frac{\partial^2 B_\phi}{\partial z^2} - \frac{B_\phi}{r^2}
       \label{eqn:A6}
\end{equation}
and equation~\ref{eqn:A1}  becomes
\begin{equation}
    \frac{\partial B_\phi}{\partial t} \approx \frac{\partial((\varv_{\rm K}(r) - r\Omega_\star) B_{\rm z})}{\partial z} + \eta \frac{\partial^2 B_\phi}{\partial z^2} - \eta \frac{B_\phi}{r^2}  .
       \label{eqn:A7}
\end{equation}

Integrating the components of equation~\ref{eqn:A7} with respect to $z$ gives
\begin{equation}
    \int_0^h B_\phi dz = \bar{B}_\phi h \, ,
      \label{eqn:A8}
\end{equation}
with $\bar{B}_\phi$ the height averaged value of $B_\phi$.
\begin{equation}
    \int_0^h \frac{\partial B_\phi}{\partial t} dz = h \frac{\partial \bar{B}_\phi }{\partial t}\, ,
      \label{eqn:A9}
\end{equation}
where we have assumed that $\partial h / \partial t$ can be neglected.
\begin{align}
    \int_0^h \frac{\partial (\varv_{\rm DB} B_{\rm z})}{\partial z} dz &= (\varv_{\rm DB}(h) - \varv_{\rm DB}(0)) B_{\rm z}(r) \, , \\
    \int_0^h \frac{\partial^2 B_\phi}{\partial z^2} dz &= \frac{\partial B_\phi}{\partial z}\biggm\lvert_h - \frac{\partial B_\phi}{\partial z}\biggm\lvert_0 \approx -\frac{\partial B_\phi}{\partial z}\biggm\lvert_0 \approx -\frac{2}{h}\bar{B}_\phi \,
      \label{eqn:A11}
\end{align}
where we have set $\partial B_\phi / \partial z |_h =0 $ as a boundary condition as we should expect that the generated toriodal field will decrease with increasing z as one moves away from the disc surface ($z \gtrsim h$). The boundary condition  $\partial B_\phi / \partial z|_0 \approx (2/h) \bar{B}_\phi$, is a semi-plausible ansatz.

Finally,
\begin{equation}
    \frac{\eta \int_0^h B_\phi dz}{r^2} = \frac{\eta h \bar{B}_\phi}{r^2} \, ,
      \label{eqn:A12}
\end{equation}
where we neglect this latter term as $(h/r) \ll 1$.

Putting all this together, our height averaged form of equation~\ref{eqn:A7}  is
\begin{equation}
    \frac{\partial \bar{B}_\phi}{\partial t} + \frac{2\eta}{h^2}\bar{B}_\phi \approx \frac{(\varv_{\rm DB}(h)-\varv_{\rm DB}(0)) B_{\rm z}(r)}{h} \, ,
    \label{eqn:A13}
\end{equation}
which, for constant $r$, is a first order differential equation with constant coefficients and has the solution 
\begin{equation}
    \bar{B}_\phi(t) \approx \bar{B}_\phi(0) \, {\rm e}^{\frac{-2\eta}{h^2}t} + \frac{h}{2\eta}(\varv_{\rm DB}(h)-\varv_{\rm DB}(0)) B_{\rm z}(r) \left(1-{\rm e}^{\frac{-2\eta}{h^2}t}\right) \, .
    \label{eqn:A14}
\end{equation}
Here the e-folding time scale for the build-up in the toroidal field has the expected dimensional form:
 \begin{equation}
     \tau_{{\rm B}_\phi} = \frac{h(r)^2}{2\eta} = \frac{\mu_0\sigma_{\rm D} h(r)^2}{2} \, .
        \label{eqn:A15}
 \end{equation}  
So as $t \rightarrow \infty$, we have the steady state form
\begin{equation}
    \bar{B}_\phi(t) \approx \frac{h \mu_0\sigma_{\rm D}(r)}{2} (\varv_{\rm DB}(h)-\varv_{\rm DB}(0)) B_{\rm z}(r) \, .
       \label{eqn:A16}
\end{equation}

By assumption the corona at the surface of the disc is corotating with the star:
\begin{equation}
   \varv_{\rm DB}(h) \approx 0 \, , 
   \label{eqn:A17}
\end{equation}
while at the midplane of the disc
\begin{equation}
   v_{\rm DB}(0) \approx r\Omega_{\rm K}(r) - r \Omega_\star \, , 
   \label{eqn:A18}
\end{equation}
So
\begin{equation}
    \bar{B}_\phi(r) \approx \frac{h \mu_0 r \sigma_{\rm D}(r)}{2} (\Omega_\star - \Omega_{\rm K}(r)) B_{\rm z}(r) \ , 
       \label{eqn:A19}
\end{equation}
which is the height-averaged integral ($\frac{1}{h} \int_0^h B_\phi(r,z)dz$) of equation~\ref{eqn:14}.



\section{MAGNETIC PRESSURE DRIVEN FLOW}
\label{MPD_Flow}

\subsection{MAGNETIC FIELD}

In Figure~\ref{fig:figure5}, we show a radial current generated by the relative motion of the stellar magnetic field and the disc. The assumed direction of the stellar field and the directions of rotation of the disc and star produce radial disc current flows that are within the disc and flow towards the star. However, for the current to exist then there must be a return current, otherwise, charge separation would occur in the disc and the current would shut down. We thereby assume that the disc surface is also conductive and there is a return current along the disc surface to complete the circuit. For this to occur, there must exist separated layers of peak Pedersen or Hall conductivity that allow trans magnetic field currents to flow. One maximum of conductivity occurs within the disc, while the other conductivity maxima occur on the separate disc surfaces.

Such altitude dependent, multiple conductivity maxima do not occur in the Earth\textquotesingle s ionosphere, but have been observed in the upper atmosphere of Titan \citep{2009P&SS...57.1828R}. Such a phenomenon may also occur in the inner discs around young stars, where the inner disc is interacting with a stellar magnetosphere.

\begin{figure}
	\includegraphics[width=\columnwidth]{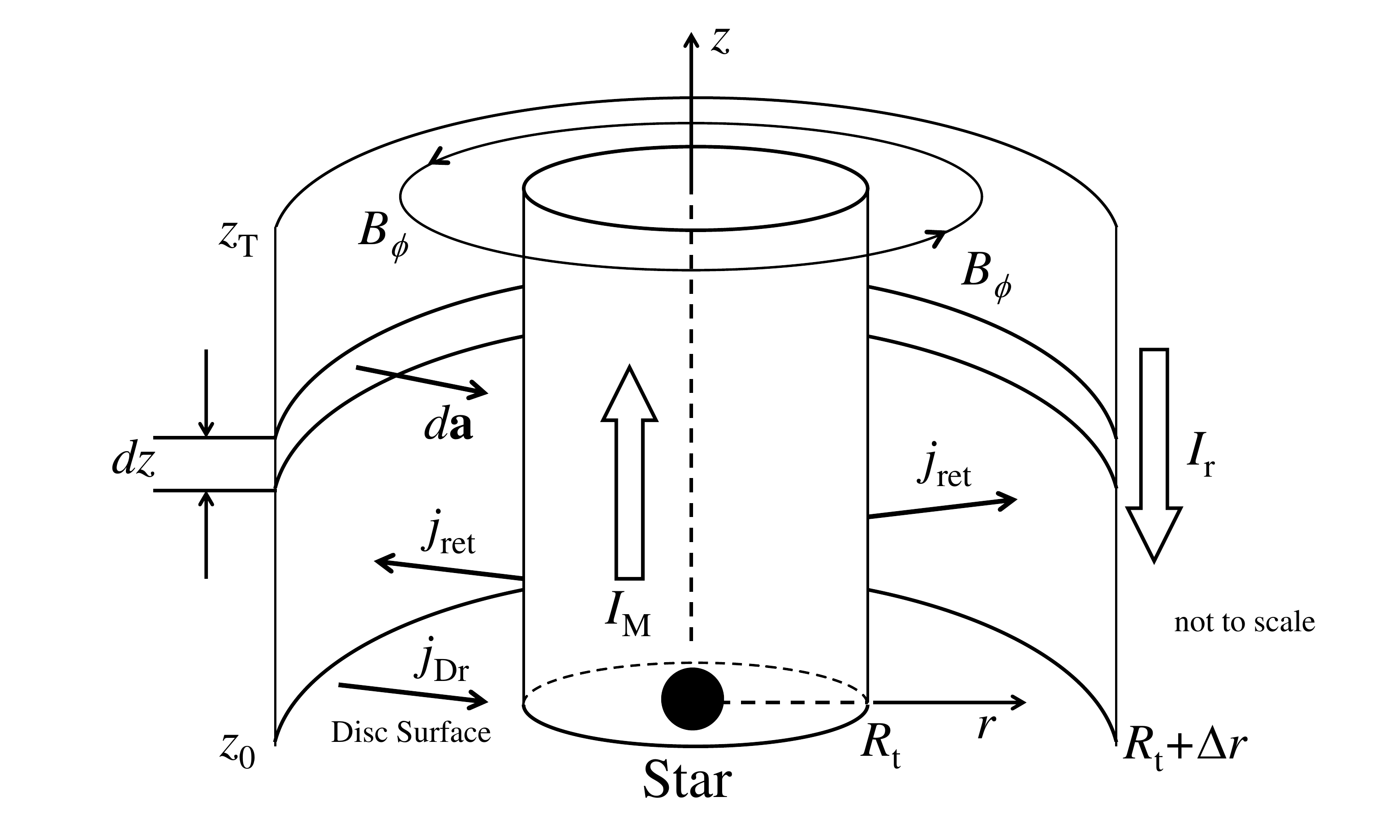}
    \caption{Geometry of the current flow in the jet acceleration region. $j_{\rm Dr}$ is the current density within the disc. $I_{\rm M}$ is the current between the inner edge of the disc and the star; $j_{\rm ret}$ is the radial transfield return current density between the stellar field lines; $I_{\rm r}$ is the total radial transfield return current at a distance $r$ from the star; $B_\phi$ is the toroidal magnetic field generated by $I_{\rm M}$. This picture is not to scale. In our model, the bottom of the jet flow acceleration region, $z_0$, is located at the disc surface, while the top of the jet acceleration region is very close to the surface of the disc, i.e., $z_0 < z_{\rm T} \ll r$.}
    \label{fig:figureB1}
\end{figure}

The inner disc current shown in Figure~\ref{fig:figure5}, interacts with the wrapped-up disc toroidal field to compress the inner disc (Figure~\ref{fig:figure9}). Conversely, because the surface current flows in the opposite direction then its interaction with the disc toroidal field pushes material away from the disc surface (Figure~\ref{fig:figure10}). The general current flows and magnetic fields are shown in Figure~\ref{fig:figureB1}. 

This figure is somewhat busy and complex, but we first concentrate on the current flows. The internal disc current density $j_{\rm Dr}$ flows through the disc towards the star, it then flows up the stellar magnetosphere as a total current, $I_{\rm M}$, and out across the upper surface of the disc with a return current density $j_{\rm ret}$. Finally, it flows back to the disc along the stellar magnetosphere, which we represent via the current $I_{\rm r}$. This jet acceleration region is assumed to take up only a small area of the inner disc starting from the inner truncation radius, $R_{\rm t}$, to a slightly larger radius of $R_{\rm t}+\Delta r$. The base of the jet acceleration region is located at height $z_0$, which is on or near the surface of the disc. The top of the acceleration jet occurs at $z_{\rm T}$, where $z_{\rm T}$ is also located in the upper regions of the disc. So $z_0 < z_{\rm T} \ll r$. The magnetospheric current, $I_{\rm M}$, generates a toroidal magnetic field $B_\phi$ in the acceleration region. The radial return Pedersen (or, possibly, Hall current), $j_{\rm ret}$, bleeds off the magnetospheric current and interacts with the toroidal field to produce a $\mathbf{j}_{\rm ret}\times \mathbf{B}_\phi$ Lorenz force that pushes the disc gas away from the disc plane, thereby producing the jet flow.

The toroidal magnetic field within the acceleration region is given by Ampere's Law:
\begin{equation}
    \nabla \times \mathbf{B} = \mu_o \mathbf{j} \, ,
      \label{eqn:B1}
\end{equation}
Integrating the right hand side of equation~\ref{eqn:B1}  over the area element $d\mathbf{a}$ shown in Figure~\ref{fig:figureB1} gives
\begin{equation}
    \int_A \mu_0 \mathbf{j} \cdot d\mathbf{a} =-2\pi \mu_0 r \int^z_{z_0} j_{\rm ret}(z) dz = -2 \mu_0 \pi  r j_{\rm ret}(z_{\rm c})(z-z_0) \, ,
    \label{eqn:B2}
\end{equation}
where the final equality is derived from the Mean Value Theorem with $z_{\rm c} \in [z_0,z]$.

The average radial return current $\bar{j}_{\rm ret}$ is given by the definition
\begin{equation}
    \bar{j}_{\rm ret}(r) = \frac{1}{z_{\rm T} - z_0} \int^{z_{\rm T}}_{z_0} j_{\rm ret}   (z) dz = \frac{I_{\rm r}}{2\pi r(z_{\rm T} - z_0)} \, ,
     \label{eqn:B3}
\end{equation}
so we can write
\begin{equation}
    \int_A \mu_0 \mathbf{j} \cdot d\mathbf{a} = -\mu_0 I_{\rm r} \frac{j_{\rm ret}(z_{\rm c})}{\bar{j}_{\rm ret}(r)}\frac{(z-z_0)}{(z_{\rm T} - z_0)} \, .
       \label{eqn:B4}
\end{equation}

Integrating the left hand side of equation~\ref{eqn:B1} over the boundary of the area
element $d\mathbf{a}$ shown in Figure~\ref{fig:figureB1} gives
\begin{equation}
    \int_A \nabla \times \mathbf{B} \cdot d\mathbf{a} = \oint_{\partial A} \mathbf{B} \cdot d\mathbf{I} = 2\pi r (B_\phi(z)-B_\phi(z_0)) \, ,
       \label{eqn:B5}
\end{equation}
where we note that
\begin{equation}
    B_\phi(r,z_0) = \frac{\mu_0 I_{\rm M}}{2\pi r} \, .
    \label{eqn:B6}
\end{equation}
Combining equations~\ref{eqn:B1}, \ref{eqn:B4}, \ref{eqn:B5} and \ref{eqn:B6}  gives the general result (assuming axisymmetry )
\begin{equation}
    B_\phi(r,z) = \frac{\mu_0 I_{\rm M}}{2\pi r} \left(1- \frac{I_{\rm r}}{I_{\rm M}} \frac{j_{\rm ret}(z_{\rm c})}{\bar{j}_{\rm ret}} \frac{z-z_0}{z_{\rm T} - z_0} \right) \, .
     \label{eqn:B7}
\end{equation}

It follows that
\begin{equation}
    B_\phi(r,z_{\rm T}) = \frac{\mu_0 I_{\rm M}}{2\pi r}  \left(1- \frac{I_{\rm r}}{I_{\rm M}} \right) \, ,
     \label{eqn:B8}
\end{equation}
since $j_{\rm ret}(z_{\rm c}) = \bar{j}_{\rm ret}$ when $z_{\rm c} \in [z_0,z_{\rm T}]$. 

If $I_{\rm r} = I_{\rm M}$ then $B_\phi(z_{\rm T}) = 0$.

Finally, if we make the approximations that $I_{\rm r} \approx I_{\rm M}$  and $j_{\rm ret}(z_{\rm c}) \approx \bar{j}_{\rm ret} \forall z_{\rm c} \in [z_0,z]$ then equation~\ref{eqn:B7}  has the simplified form
\begin{equation}
    B_\phi(r,z) \approx \frac{\mu_0I_{\rm M}}{2\pi r} \left(1-\frac{z-z_0}{z_{\rm T}-z_0}\right) \, .
    \label{eqn:B9}
\end{equation}

\subsection{JET FORCE}

Within the acceleration region, the Lorentz force per unit volume, $\mathbf{f}_{\rm l}$, is 
\begin{equation}
    \mathbf{f}_{\rm l}(z) = j_{\rm ret}(z) B_\phi(r,z) \hat{\mathbf{z}} \, .
      \label{eqn:B10}
\end{equation}
Using equation~\ref{eqn:B7}, \ref{eqn:B10} becomes
\begin{equation}
    \mathbf{f}_{\rm l}(r,z) = \frac{j_{\rm ret}(z) \mu_0 I_{\rm M}}{2\pi r} \left(1-\frac{I_{\rm r}}{I_{\rm M}}\frac{j_{\rm ret}(z_{\rm c})}{\bar{j}_{\rm ret}}\frac{z-z_0}{z_{\rm T}-z_0} \right) \hat{\mathbf{z}} \, , 
    \label{eqn:B11}
\end{equation}
with $z_{\rm c} \in [z_0,z]$.

The magnitude of the total Lorentz force, $F_{\rm l}$, generated in the acceleration region is
\begin{equation}
    F_{\rm l} = \int_{r_{\rm i}}^{r_0} dr \int_{z_0}^{z_{\rm T}} dz \int_{0}^{2\pi} r f_{\rm l} d\theta \approx \frac{\mu_0 I_{\rm M} I_{\rm r}}{4\pi} \ln{\frac{r_0}{r_{\rm i}}}\left(2-\frac{I_{\rm r}}{I_{\rm M}}\right) \, ,
       \label{eqn:B12}
\end{equation}
where we note that in deriving equation~\ref{eqn:B12}, we have made the approximation that
\begin{align}
    \int_{r_{\rm i}}^{r_0} dr & \int_{z_0}^{z_{\rm T}} dz j_{\rm ret}(z) \mu_0 I_{\rm M} \left(1-\frac{I_{\rm r}}{I_{\rm M}}\frac{j_{\rm ret}(z_{\rm c})}{\bar{j}_{\rm ret}}\frac{z-z_0}{z_{\rm T}-z_0} \right) \approx \nonumber \\
     &\int_{r_{\rm i}}^{r_0} dr \bar{j}_{\rm ret} \mu_0 I_{\rm M} \int_{z_0}^{z_{\rm T}} dz \left(1-\frac{I_{\rm r}}{I_{\rm M}}\frac{z-z_0}{z_{\rm T}-z_0} \right) \, . 
      \label{eqn:B13}
\end{align}
If this approximation is true, then the total force from the acceleration region is independent of the $z$  behaviour of $j_{\rm ret}$.

For the case where $I_{\rm r} = I_{\rm M}$ then equation~\ref{eqn:B12}  has the form
\begin{equation}
    F_{\rm l} \approx \frac{\mu_0 I^2_{\rm M}}{4 \pi} \ln{\left(\frac{r_0}{r_{\rm i}}\right)} \, .
     \label{eqn:B14}
\end{equation}
So the total driving force is dependent on the central current flow and the radial size of the propulsion region.

\subsection{APPROXIMATE JET SPEED}

It would be useful to obtain an approximate equation for the flow speed of this magnetic jet system. An intuitive idea of how this system behaves can be obtained by solving a simplified momentum equation by ignoring gravity and pressure gradient:
\begin{equation}
    \rho \frac{d\varv}{dt} = \frac{\rho}{2}\frac{d(\varv^2)}{dz} = f_{\rm l}(r,z) 
     \label{eqn:B15}
\end{equation}
\begin{align}
    \Rightarrow \int_{z_0}^{z_{\rm T}} \frac{\rho}{2}\frac{d(\varv^2)}{dz}dz &= \frac{\rho(z_{\rm c2})}{2} \int_{z_0}^{z_{\rm T}}\frac{d(\varv^2)}{dz}dz \nonumber \\
    &= \frac{\rho(z_{\rm c2})}{2}(\varv^2(z_{\rm T})-\varv^2(z_0)) \, , \\
    \int_{z_0}^{z_{\rm T}} f_{\rm l}(r,z) dz &\approx \frac{j_{\rm ret}(z_{\rm c3})\mu_0 I_{\rm M}}{2\pi r} (z_{\rm T} - z_0) \left(1-\frac{I_{\rm r}}{2I_{\rm M}}\right) \, ,
\end{align}
where we have used the Mean Value Theorem with $z_{\rm c2}$ and $z_{\rm c3} \in [z_0,z_{\rm T}]$.

Putting all this together gives:
\begin{equation}
    \varv^2(r,z_{\rm T}) \approx \varv^2(r,z_0) + \frac{j_{\rm ret}(z_{\rm c3})\mu_0 I_{\rm M}}{\rho(z_{\rm c2})\pi r} (z_{\rm T}-z_0) \left(1-\frac{I_{\rm r}}{2I_{\rm M}}\right) \, .
       \label{eqn:B18}
\end{equation}
Making the approximations that
\begin{align}
    \varv^2(r,z_0) & \ll \varv^2(r,z_{\rm T}) \, , \\
    j_{\rm ret}(r,z_{\rm c3})&\approx \bar{j}_{\rm ret} = \frac{I_{\rm r}}{2\pi r(z_{\rm T}-z_0)} \,, \quad {\rm and} \\
    \rho(z_{\rm c3}) &\approx \bar{\rho} = \frac{\int_{z_0}^{z_{\rm T}} \rho dz}{(z_{\rm T}-z_0)} \, ,
\end{align}
then equation~\ref{eqn:B18} becomes
\begin{equation}
    \varv^2(r,z_{\rm T}) \approx \frac{\mu_0 I_{\rm r} I_{\rm M}}{2\pi^2 r^2 \bar{\rho}} \left(1-\frac{I_{\rm r}}{2I_{\rm M}}\right) \, .
\end{equation}

If we suppose that all the magnetospheric current between the star and the disc is converted into radial current, i.e., $I_{\rm r} = I_{\rm M}$ then the exit speed of the gas from the acceleration region is
\begin{align}
    \varv(r,z_{\rm T}) &\approx \sqrt{\frac{\mu_0}{\bar{\rho}}} \frac{I_{\rm M}}{2\pi r} = \frac{|B_\phi(r,z_0)|}{\sqrt{\mu_0 \bar{\rho}}} \nonumber \\ &=\sqrt{\frac{\mu_0}{\bar{\rho}}} \sigma_{\rm D}(r) r z_0 |\Omega_\star - \Omega_{\rm K}(r)| \, |B_{\rm z}(r)| \, ,
     \label{eqn:B23}
\end{align}
where we have used equations~\ref{eqn:B6}  and \ref{eqn:14} to obtain the right hand side of equation~\ref{eqn:B23}. It is useful to rewrite equation~\ref{eqn:B23}  as
\begin{equation}
    \varv(r,z_{\rm T}) \approx \sqrt{\frac{\mu_0}{\bar{\rho}}} 
    \sigma_{\rm D}(r) r z_0 \Omega_\star |1-(R_{\rm co}/r)^{3/2}| \, |B_{\rm z}(r)| \, .
\end{equation}
We note that the above derivation has ignored gravity as we have implicitly assumed that the jet propulsion occurs at or near the disc surface and that $z \ll r$. This assumption may be incorrect, but if jet flows are produced from or near the disc surface then the observed outflow speed will not be the same as the value given by equation~\ref{eqn:B23} as the jet flow will have had to overcome the gravitational potential of the star. To obtain an approximate value for the final flow speed, one can use a Bernoulli-like equation which includes gravity and angular velocity, e.g., equation 47 of \citet{1997MNRAS.290..629L}.



\section{PARTICLE MOTION}
\label{Particle_Motion}

\subsection{EQUATIONS OF MOTION}

Suppose that dust particles are, initially, in a circular Keplerian orbit at or near the inner truncation radius of the disc. As discussed in the \S~\ref{subsec:jet_exhaust_speed}, the accretional inflow and/or the protostellar jet flow gives the particles an initial `boost' velocity that is assumed to be primarily in the $z$ direction as this is perpendicular to the disc midplane. If we assume that the self-gravity of the disc is negligible compared to the gravity of the protostar then the equations of motion for a particle in the cylindrical coordinate $r,\phi$ and $z$ directions are:
\begin{equation}
    m_{\rm d} \ddot{r}_{\rm p} = m_{\rm d} r_{\rm p} \dot{\phi}^2_{\rm p} - \frac{GM_\star m_{\rm d} r_{\rm p}}{(r^2_{\rm p} + z^2_{\rm p})^{3/2}} - \frac{C_{\rm D}}{2}\rho_{\rm g} \pi a^2_{\rm d} \varv^2_{\rm pg}\hat{\mathbf{v}}_{\rm pg} \cdot  \hat{\mathbf{r}} \, ,
    \label{eqn:39} 
\end{equation}
\begin{equation}
    m_{\rm d}(r_{\rm p} \ddot{\phi}_{\rm p} + 2 \dot{r}_{\rm p} \dot{\phi}_{\rm p}) = -\frac{C_{\rm d}}{2}\rho_{\rm g} \pi a^2_{\rm d} \varv^2_{\rm pg}\hat{\mathbf{v}}_{\rm pg} \cdot  \hat{\mathbf{\phi}} \, , \label{eqn:40} 
\end{equation}
\begin{equation}
    m_{\rm d} \ddot{z}_{\rm p} = -\frac{GM_\star m_{\rm d} z_{\rm p}}{(r^2_{\rm p} + z^2_{\rm p})^{3/2}} - \frac{C_{\rm D}}{2}\rho_{\rm g} \pi a^2_{\rm d} \varv^2_{\rm pg}\hat{\mathbf{v}}_{\rm pg} \cdot  \hat{\mathbf{z}} \, ,
\end{equation}
where $r_{\rm p}, \phi_{\rm p}$ and $z_{\rm p}$ are the cylindrical coordinates of the dust particle, $\rho_{\rm g}$ the average mass density of the gas, $C_{\rm D}$ the drag coefficient for the interaction between the gas and the dust particle, and $\mathbf{v}_{\rm g}$ and $\mathbf{v}_{\rm p}$  are the gas flow velocity and dust velocity, respectively, with $\mathbf{v}_{\rm pg} = \mathbf{v}_{\rm p} - \mathbf{v}_{\rm g} $. Symbols with a caret and tilde are unit vectors, while $m_{\rm d}$ is the mass of an individual, approximately spherical, dust grain, so 
\begin{equation}
    m_{\rm d} \approx \frac{4}{3} \pi a^3_{\rm d} \rho_{\rm d} \, ,
\end{equation}
with $a_{\rm d}$ the average dust grain radius and $\rho_{\rm d}$ the average mass density of the dust grain. The azimuthal gas speed is $\varv_{{\rm g}\phi} \approx r\Omega_\star$, as we assume that the gas flow arises at the inner truncation radius and the gas flow is coupled to the protostellar magnetosphere, which is, to a first approximation, co-rotating with the protostar. 

We can normalise the above equations by setting $r'_{\rm p} = r_{\rm p} / r_0$, $z'_{\rm p} = z_{\rm p} / r_0$ and $t'_{\rm p} = t / P_0$, where $r_0$ is the initial value of $r_{\rm p}$ for the particle and $P_0$ is the orbital period of an object with a circular orbit of radius $r_0$:
\begin{equation}
    P_0 = 2\pi \sqrt{\frac{r^3_0}{GM_\star}}
\end{equation}

Dropping the primes on the main non-dimensional variables, the equations of motion become:
\begin{align}
    \ddot{r}_{\rm p} &= r_{\rm p} \dot{\phi}^2_{\rm p} - \frac{4\pi^2 r}{(r^2_{\rm p} + z^2_{\rm p})^{3/2}} - \frac{3C_{\rm D} \rho_{\rm g} r_0}{8a_{\rm d} \rho_{\rm d}} {\varv}^2_{\rm pg} \hat{\mathbf{v}}_{\rm pg} \cdot \hat{\mathbf{r}} \\
    r_{\rm p} \ddot{\phi}_{\rm p}  &= - 2 \dot{r}_{\rm p} \dot{\phi}_{\rm p} -\frac{3C_{\rm D} \rho_{\rm g} r_0}{8a_{\rm d} \rho_{\rm d}} {\varv}^2_{\rm pg} \hat{\mathbf{v}}_{\rm pg} \cdot \hat{\mathbf{\phi}} \\
    \ddot{z}_{\rm p} &= - \frac{4 \pi^2 z_{\rm p}}{(r^2_{\rm p} + z^2_{\rm p})^{3/2}} -\frac{3C_{\rm D} \rho_{\rm g} r_0}{8a_{\rm d} \rho_{\rm d}} {\varv}^2_{\rm pg} \hat{\mathbf{v}}_{\rm pg} \cdot \hat{\mathbf{z}}
\end{align}
where $\varv'_{\rm gr} = (P_0 / r_0) \varv_{\rm gr}$, $\varv'_{\rm gz} = (P_0 / r_0) \varv_{\rm gz}$, and $\Omega'_\star = P_0 \Omega_\star$.

The drag coefficient, $C_{\rm D}$, is given by 
\begin{equation}
    C_{\rm D}(s) = \frac{2}{3s}\sqrt{\frac{\pi T_{\rm p}}{T_{\rm g}}} + \frac{2s^2+1}{\sqrt{\pi}s^3}\exp(-s^2)+\frac{4s^4+4s^2-1}{2s^4}{\rm erf}(s) \, ,
    \label{eq:C_D}
\end{equation}
\citep{hayes_probstein_1959,1968Probstein} where $T_{\rm p}$ is the temperature of the particle, $\mathrm{erf}$ the error function, $\exp$ the exponential function, and $s$ is the thermal Mach number:
\begin{equation}
  s = | \mathbf{v}_{\rm pg} | / \varv_{\rm T} \, 
\end{equation}
with the thermal gas speed:
\begin{equation}
    \varv_{\rm T} = \sqrt{2k_{\rm B} T_{\rm g} / \bar{m}} \, .
\end{equation}

To compute the velocity and mass density of the gas flow, there are at least two scenarios that could be considered: accretional mass flow from the disc onto the star and/or an outflow that is ejecting material from the disc. Of course, it is possible that outflows and accretional inflows are manifestations of the same phenomena. As such, we will consider the case of accretional flow onto the star.

At the truncation radius, the infalling gas and dust will initially tend to flow along the stellar field lines with a gas velocity, $\mathbf{v}_{\rm g}$ , in the $z$ direction in an (assumed) axisymmetric channel of initial width $\Delta$ (Figure~\ref{fig:figure13}). Several authors have developed detailed and elegant flow models for the velocity and density of the infalling gas, e.g., \citet{2012ApJ...744...55A} and references therein.  However, for our purposes, we have adopted the standard boundary layer value for $\Delta$, where the stellar magnetosphere at $R_{\rm t}$ replaces the surface of a compact object (e.g., equation (6.10) in \citet{2002apa..book.....F}):
\begin{equation}
    \Delta \approx \frac{h(R_{\rm t})^2}{R_{\rm t}} = 2\times 10^{-5}\, {\rm au} \frac{(h/0.001\, {\rm au})^2}{(R_{\rm t}/0.05\,{\rm au})} \, .
    \label{eqn:52}
\end{equation}
So, we can write for the mass flow rate in the channel by using the conservation of mass
\begin{equation}
    \dot{M}_{\rm a}/2 = 2\pi R_{\rm t} \Delta \rho_{\rm g} \varv_{\rm g} \, .
    \label{eqn:53}
\end{equation}
Combining equations~\ref{eqn:52} and \ref{eqn:53} gives
\begin{equation}
    \rho_{\rm g} = \frac{\dot{M}_{\rm a}}{4\pi h(R_{\rm t})^2 \varv_{\rm g}} \, .
     \label{eqn:54}
\end{equation}

\subsection{NUMERICAL RESULTS FOR LRLL~31 \& ANALYTIC TESTS}
\label{subsec:analytic_tests}

\begin{figure}
     \begin{subfigure}{\linewidth}\centering
    \includegraphics[width=\linewidth]
    {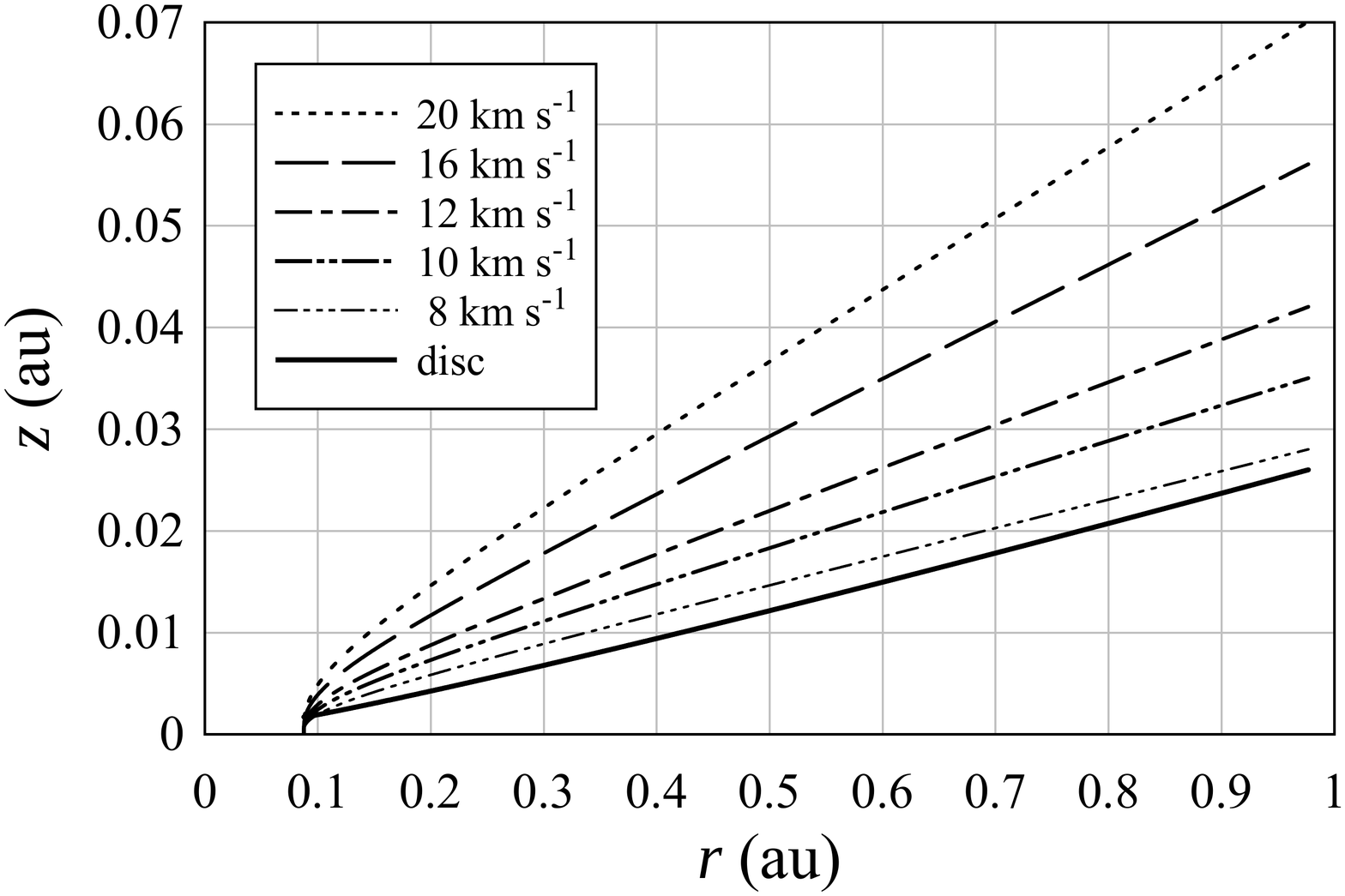}\caption{Far field}
    \end{subfigure}
    \begin{subfigure}{\linewidth}\centering
    \includegraphics[width=\linewidth]
    {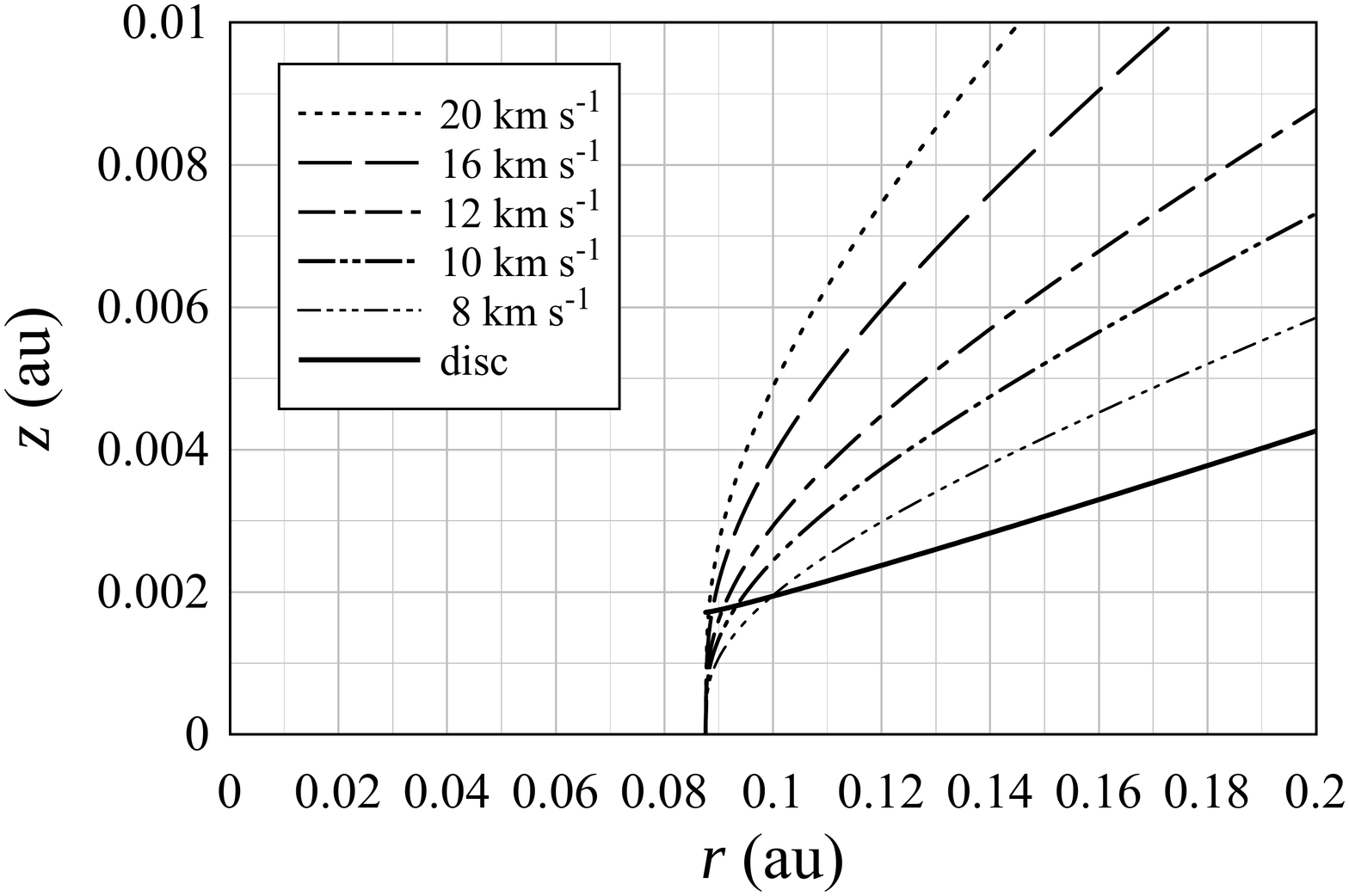}\caption{Near field}
    \end{subfigure}
    \caption{Trajectories for dust particles  in a thin gas outflow (where the outflow is not shown) in (a) the far field and (b)  near field views. The flow is located at the inner edge of the disc and is flowing perpendicular to the midplane ($z = 0$). The particle speeds, $\varv_{\rm pz}$, in the $z$ direction are shown in the insert, with the resulting particle trajectories shown relative to the scale height of the accretion disc. The particles are initially placed at the inner edge of the outflow. The particles must travel through the full width of the flow before they can be ejected and move across the face of the disc. 
    }
   \label{fig:figurecatpult}
\end{figure}

This system of equations can be solved via standard techniques and as an example, we assumed that the particles and accretional gas flow initially started at or near the midplane ($z = 0$) of the inner edge of the disc, i.e., at the truncation radius ($R_{\rm t} \approx 0.09$~au), with a corresponding mass accretion rate of $\dot{M}_{\rm a} \sim 1.6\times 10^{-8}$~M$_\odot$yr$^{-1}$. The resulting width of the channel was $\Delta \approx 5000$~km $\approx 3.3 \times 10^{-5}$~au and the dust particles were placed at the inner edge of the gas flow, so they had to travel through the entire width of the accretional flow before they could escape the flow. We set the particle drag to zero, once the dust particles left the initial gas flow. The speed of the $z$ component of the initial gas flow, $\varv_{\rm gz}$, was set as a free parameter, which, in turn, determined the mass density of the gas flow via equation~\ref{eqn:54}. The other gas velocity components were: $\varv_{\rm gr} = 0$ and $\varv_{{\rm g}\phi}=r\Omega_\star$. The results for different particle ejection speeds are shown in Figure~\ref{fig:figurecatpult}.

The dust particle parameters used were $a_{\rm d} = 0.5 \mu$m, $\rho_{\rm d} = 3$~g cm$^{-3}$, with the initial velocity components: $\varv_{\rm pr}=0, \varv_{{\rm p}\phi}=V_{\rm K}(R_{\rm t})$, (the Keplerian speed at $R_{\rm t}$) and $\varv_{\rm pz} = \varv_{\rm gz}$. As the launching, radial distance $R_{\rm t} \approx 0.09$~au $> R_{\rm co} \approx 0.05$~au then $\varv_{{\rm g}\phi} \approx R_{\rm t} \Omega_\star \approx 290$ kms$^{-1}$, which is around 2.2 times the Keplerian speed at $R_{\rm t}$. Therefore, the dust particles are subject to radial acceleration away from the star due to centrifugal force derived from the gas flow. 

\begin{figure}
	\includegraphics[width=\columnwidth]{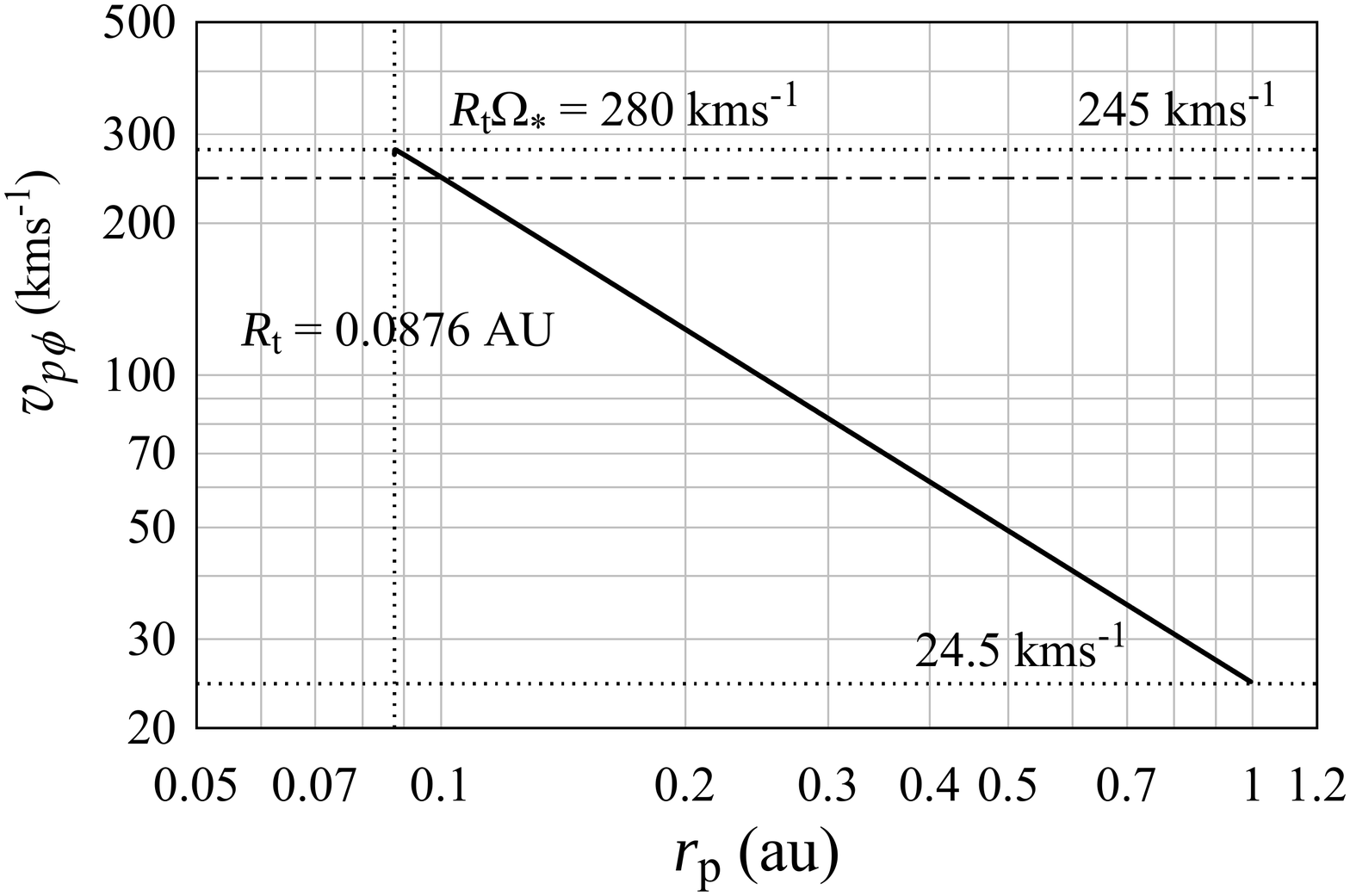}
    \caption{The particle azimuthal velocity, $\varv_{{\rm p}\phi}$, as a function of distance from the LRLL~31 protostar. The dust particle has an initial azimuthal velocity equal to the Keplerian speed of 127 km s$^{-1}$, but is quickly spun up by the mass flow from the accretion disc at the inner truncation radius, $R_{\rm t}$, to the stellar corotational speed, $R_{\rm t} \Omega_\star$ , of 280 km s$^{-1}$. Once the particle is free from the gas flow, in an assumed gas drag free environment, its angular momentum is constant and its azimuthal velocity decreases as $r_{\rm p}^{-1}$, as predicted from equation~\ref{eqn:57}. This can be seen from the azimuthal velocities at 0.1 and 1~au, which decrease from 245 km s$^{-1}$ to 24.5 km s$^{-1}$.}
    \label{fig:figure14}
\end{figure}

To check that the numerical solver was working correctly, we cross checked our numerical results with some available analytic solutions for the particle paths. For example, when the particle is free of the gas flow then equation~\ref{eqn:40} has the form
\begin{equation}
    r_{\rm p} \ddot{\phi}_{\rm p} + 2 \dot{r}_{\rm p} \dot{\phi}_{\rm p} = 0
\end{equation}
This equation has the solution
\begin{equation}
    r_{\rm p}^2 \dot{\phi}_{\rm p} = {\rm constant} = \ell 
\end{equation}
i.e., the specific angular momentum of the particle, $\ell$, when it is not subject to a torque, is a constant. It follows that
\begin{equation}
    \varv_{{\rm p}\phi} = r_{\rm p} \dot{\phi}_{\rm p} = \frac{r_0^2 \dot{\phi}_0}{r_{\rm p}} = \frac{\ell_0}{r_{\rm p}} \, ,
    \label{eqn:57}
\end{equation}
where, in this case, $r_0=R_{\rm t}$ and $\dot{\phi}_0 = \Omega_\star$. Note that even though the dust particle started with Keplerian azimuthal velocity, gas drag accelerated the particle to the co-rotational azimuthal gas velocity (Figure~\ref{fig:figure14}). The numerical calculation reproduced the values from the analytic solution: equation~\ref{eqn:57} with $\dot{\phi}_0 = \Omega_\star$. 
%

Other, approximate, analytic solutions are available when one considers the radial equation of motion, equation~\ref{eqn:39}, without gas drag:
\begin{equation}
    \ddot{r}_{\rm p} = \varv_{\rm pr}\frac{d\varv_{\rm pr}}{dr} = r_{\rm p} \dot{\phi}_{\rm p}^2 - \frac{GM_\star r_{\rm p}}{(r_{\rm p}^2 + z_{\rm p}^2)^{3/2}} = \frac{\ell^2_0}{r_{\rm p}^3} - \frac{GM_\star r_{\rm p}}{(r_{\rm p}^2 + z_{\rm p}^2)^{3/2}} \, .
\end{equation}
Suppose the particle is given a significant boost velocity in the $z$ direction such that  $z \rightarrow \infty$ (or $z$ becomes comparable to $r$) then
\begin{equation}
    \ddot{r}_{\rm p} \approx \frac{\ell_0^2}{r^3_{\rm p}} > 0 \, ,
\end{equation}
and the particle starts to accelerate in the radial direction with the subsequent radial speed 
\begin{equation}
    \varv_{\rm pr} \approx \ell_0 \sqrt{\frac{1}{r_0^2} - \frac{1}{r^2_{\rm p}}}
    \label{eqn:60}
\end{equation}
So, in this scenario, the particle increases in radial speed and we have
\begin{equation}
    \varv_{\rm pr} \rightarrow \frac{\ell_0}{r_0} = r_0 \dot{\phi}_0 \approx \sqrt{\frac{GM_\star}{r_0}} \, {\rm as } \ r_{\rm p} \rightarrow \infty \, .
\end{equation}
It could be argued that setting $z\rightarrow \infty$ is slightly unrealistic. Let us assume that the boost in the $z$ direction is small and that $z \ll r$. For such a case, another approximate analytic solution is available when one considers the radial equation of motion, equation~\ref{eqn:39}, without dust drag:
\begin{equation}
    \ddot{r}_{\rm p} = \varv_{\rm pr}\frac{d\varv_{\rm pr}}{dr} = r_{\rm p} \dot{\phi}^2_{\rm p} - \frac{GM_\star}{(r_{\rm p}^2 + z_{\rm p}^2)^{3/2}} \approx \frac{r_0^4 \dot{\phi}_0^2}{r_{\rm p}^3} -\frac{GM_\star}{r^2_{\rm p}} \, ,
    \label{eqn:62}
\end{equation}
where we have used equation~\ref{eqn:57} and assumed $z\ll r$. Equation~\ref{eqn:62} has the solution
\begin{equation}
    \varv_{\rm pr} \approx \sqrt{r_0^2 \dot{\phi}_0^2 \left(1-\left(\frac{r_0}{r_{\rm p}}\right)^2\right) - \frac{2GM_\star}{r_0}\left(1-\frac{r_0}{r_{\rm p}}\right)} \,  .
    \label{eqn:63}
\end{equation}
When $r_{\rm p} \rightarrow \infty$ then
\begin{equation}
    \varv_{\rm pr} \rightarrow \varv_{{\rm pr}\infty} \approx \sqrt{r_0^2 \dot{\phi}_0^2-\frac{2GM_\star}{r_0}} \, .
    \label{eqn:64}
\end{equation}

\begin{figure}
	\includegraphics[width=\columnwidth]{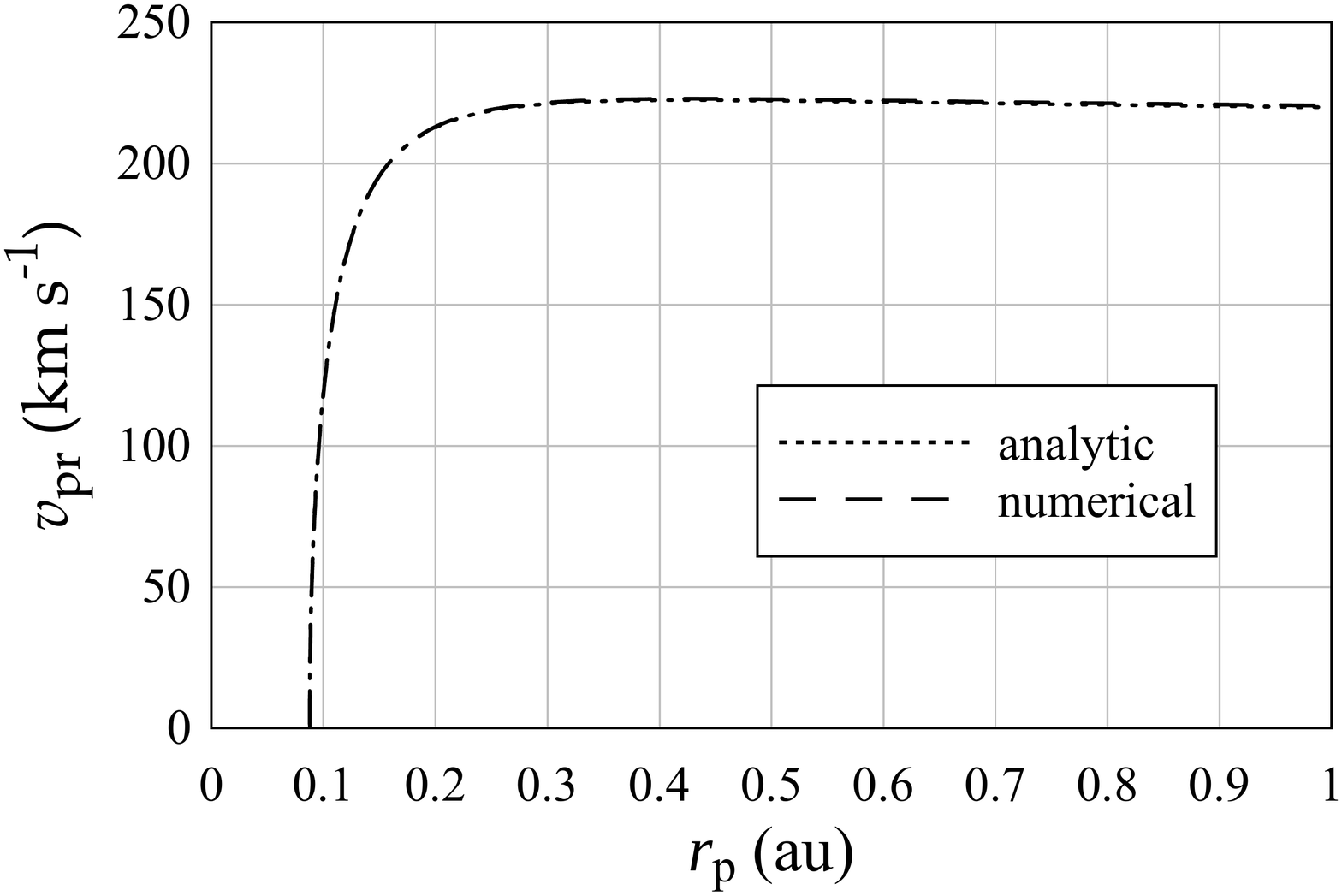}
    \caption{comparison between the numerical solution of equation~\ref{eqn:39} for the radial velocity of a particle ejected from the inner region of the LRLL~31 accretion disc relative to an analytic approximation: equation~\ref{eqn:63}. To the resolution of the graph, the two solutions are almost identical.}
    \label{fig:figure15}
\end{figure}

In Figure~\ref{fig:figure15}, we compare these analytic equations with the numerical solutions, where it can be seen that there is very little difference between the numerical solution of equation~\ref{eqn:39} for the radial velocity of a particle ejected from the inner region of the LRLL~31 accretion disc relative to an analytic approximation given by equation~\ref{eqn:63}. For this case, the expected asymptotic radial speed for large $r_{\rm p}$, as obtained from equation~\ref{eqn:64} is $\varv_{{\rm pr}\infty} \approx 215$ km s$^{-1}$. 
%


\bsp	
\label{lastpage}
\end{document}